\documentclass[sigconf,balance=false]{acmart}

\setcopyright{none}

\acmConference[]{}{}{}

\usepackage{natbib}
\usepackage{tabularray}
\usepackage{adjustbox}
\usepackage{amssymb}
\usepackage{amsmath,amsfonts}
\usepackage{algorithmic}
\usepackage{graphicx}
\usepackage{textcomp}
\usepackage{xcolor}
\usepackage{amsthm}
\usepackage{url}
\usepackage{tikz}
\usepackage{mathenv}
\usepackage{framed}
\usepackage{mdframed}  
\usepackage{newfloat}
\usepackage{xspace}
\usepackage{makecell}
\usepackage{caption}
\usepackage{enumerate}
\usepackage[shortlabels]{enumitem}
\usepackage{booktabs}
\usepackage{outlines}
\usepackage{array}
\usepackage[english]{babel}
\usepackage{blindtext}
\usepackage{afterpage}
\usepackage{multirow}
\usepackage{rotating}

\renewcommand\footnotetextcopyrightpermission[1]{} 
\DeclareFloatingEnvironment[placement={!ht}, name=Vignette]{vignette}  
\newcommand{\vignetteboxbegin}{ \begin{mdframed}[backgroundcolor=gray!10,shadow=true,shadowsize=0.5pt,shadowcolor=black!80,roundcorner=4pt] }
\newcommand{\vignetteboxend}{ \end{mdframed} \vskip -1em }

\DeclareFloatingEnvironment[placement={!ht}, name=Vig]{vig}
\newcommand{\vigbegin}{
\begin{mdframed}[backgroundcolor=gray!40,shadow=true,shadowsize=0.5pt,shadowcolor=black!80,roundcorner=4pt] }
\newcommand{\vigend}{\end{mdframed} \vskip -1em}

\usepackage{enumitem}
\setitemize{noitemsep,align=parleft,left=0pt..1em,topsep=0pt,parsep=0pt,partopsep=0pt}

\definecolor{shadecolor}{RGB}{180,180,180}

\newcommand*\halfcirc[1][.7ex]{%
 \begin{tikzpicture}
 \draw[fill] (0,0)-- (90:#1) arc (90:270:#1) -- cycle ;
 \draw (0,0) circle (#1);
 \end{tikzpicture}}
 
\newcommand*\fullcirc[1][.7ex]{%
 \begin{tikzpicture}
 \draw[black,fill=black] (0,0) circle (.7ex);
 \end{tikzpicture}}

\newcommand*\emptycirc[1][.7ex]{%
 \begin{tikzpicture}
 \draw[black,fill=none] (0,0) circle (.7ex);
 \end{tikzpicture}}
 
 \newcommand*\halfsq[1][1ex]{%
 \begin{tikzpicture}
 \draw[path picture={\fill[black] (path picture bounding box.south west)
  rectangle (path picture bounding box.east);}] (0,0) rectangle  ++ (1.5ex,1.5ex);
 \end{tikzpicture}}
 
\newcommand*\fullsq[1][1ex]{%
 \begin{tikzpicture}
 \draw[black,fill=black] (0,0) rectangle (1.5ex,1.5ex);
 \end{tikzpicture}}

 \newcommand*\emptysq[1][1ex]{%
 \begin{tikzpicture}
 \draw[black,fill=none] (0,0) rectangle (1.5ex,1.5ex);
 \end{tikzpicture}}
\newcommand{\tsref}[1]{\textsection\ref{#1}\xspace}

\def\BibTeX{{\rm B\kern-.05em{\sc i\kern-.025em b}\kern-.08em
    T\kern-.1667em\lower.7ex\hbox{E}\kern-.125emX}}
\newtheorem{theorem}{Theorem}
\newtheorem{definition}{Definition}
\graphicspath{ {./figures/} }
    \makeatletter

\let\OLDthebibliography\thebibliography
\renewcommand\thebibliography[1]{
  \OLDthebibliography{#1}
  \setlength{\parskip}{0pt}
  \setlength{\itemsep}{0pt plus 0.3ex}
}
\begin{document}

\title{SoK: The Ghost Trilemma}

\author{
Sulagna Mukherjee}
\affiliation{%
\institution{University of Southern California}
\country{}}
 \email{sulagna@usc.edu}
\author{Srivatsan Ravi}
\affiliation{%
 \institution{University of Southern California}
 \country{}}
\email{srivatsr@usc.edu}
 \author{Paul Schmitt}
\affiliation{%
 \institution{University of Hawaii}
 \country{}}
\email{pschmitt@hawaii.edu}
\author{Barath Raghavan}
 \affiliation{%
\institution{University of Southern California}
\country{}}
\email{barathra@usc.edu}

%\maketitle
\begin{abstract}
Trolls, bots, and sybils distort online discourse and compromise the security of networked platforms. User identity is central to the vectors of attack and manipulation employed in these contexts. However it has long seemed that, try as it might, the security community has been unable to stem the rising tide of such problems.
We posit the Ghost Trilemma, that there are three key properties of identity---\textbf{sentience}, \textbf{location}, and  \textbf{uniqueness}---that cannot be simultaneously verified in a fully-decentralized setting. 
Many fully-decentralized systems---whether for communication  or social coordination---grapple with this trilemma in some way, perhaps unknowingly.
In this Systematization of Knowledge (SoK) paper, we examine the design space, use cases, problems with prior approaches, and possible paths forward. We sketch a proof of this trilemma   
and outline options for practical, incrementally deployable schemes to achieve an acceptable tradeoff of trust in centralized trust anchors, decentralized
operation, and an ability to withstand a range of attacks, while protecting user privacy.
\end{abstract}
%\keywords{disinformation, centralization, user-privacy}
\maketitle

\section{Introduction}
\begin{quote}
What Orwell feared were those who would ban books. What Huxley feared was that there would be no reason to ban a book, for there would be no one who wanted to read one. Orwell feared those who would deprive us of information. Huxley feared those who would give us so much that we would be reduced to passivity and egoism. Orwell feared that the truth would be concealed from us. Huxley feared the truth would be drowned in a sea of irrelevance.

\hfill \emph{Amusing Ourselves to Death}, Postman\\
\end{quote}

We are drowning in a sea of irrelevance: so much information that we can no longer make sense of it. In a world of decontextualized shards of content, each with an increasingly short shelf life, what does free and open communication mean? To grapple with this, one common heuristic is to look to the \emph{source} of information as a guide as to its importance or veracity. But in an Internet without information gatekeepers---for good \emph{and} ill---this heuristic itself has a key flaw: who is the gatekeeper of the gatekeepers in a fully-decentralized online ecosystem?

Misinformation and disinformation campaigns are now the norm, with social media research indicating that a significant fraction of purported engagement is driven by ``sock puppet'' accounts. While some of these accounts are operated by humans and some are fully-automated bots, neither are who they claim to be. To make matters worse, social media providers are unable or unwilling to completely rid their platforms of such ghost accounts~\cite{moore2023fake}. Additionally, with the advent of AI tools that are nearing human-like language ability~\cite{diar_2023}, the boundary between what's real and what's not is blurred. 

Given the dismal status quo, voters are rightly concerned that foreign governments are having a direct effect on elections~\cite{senatereport}, while scientists are concerned that lies about their work are spreading regarding vaccines~\cite{timmer_2020} or climate change~\cite{timberg_romm_2019}. Similarly, technologists have long discussed the idea of moving important civic functions online, whether courtrooms~\cite{scigliano_2021} or elections~\cite{ford_2019,specter2020ballot,ruan2017receipt}.  The problem is truly global~\cite{goldhill2019,shu2020}. Yet inevitably, in a fully-decentralized setting, it has seemed difficult or impossible to achieve the necessary security and privacy requirements associated with such functions. So, how can we stem the rising tide of ghost accounts that poison online discourse or coordination?

Simply differentiating bots and non-bots is insufficient; the challenge in this context
goes well beyond any classic Sybil defense~\cite{al2017sybil, alvisi2013sok, viswanath2010analysis, yu2008sybilguard}. Indeed, many disinformation campaigns
are run by trained human operatives who ``step into'' the identity of others
(sometimes even stealing the identity of real, deceased
activists~\cite{stolenid}). Furthermore, platforms themselves (such as Facebook,
Twitter, and others) have thus far failed to effectively weed out accounts that
make false identity claims, a significant issue as identity is used by many
users as a proxy for credibility~\cite{sterret2018shared,grabner2015trust}.

We posit that beneath all of these challenges is what we term the \textit{Ghost Trilemma}: that it is impossible, in a decentralized setting, for  legitimate users to prove that they themselves are real humans
(\textbf{sentience}), physically where they claim to be (\textbf{location}), and
not sharing or faking credentials (\textbf{uniqueness}).  Indeed, these three
properties are what we intuitively expect from of a normal user with whom we
interact online. In addition, we wish to maintain location privacy (e.g., hiding
exact locations) and, if appropriate, anonymity (i.e., verification of these
properties need not reveal real-world identifying information such as full
names).

Both security researchers and practitioners have grappled with these challenges for decades, but to our knowledge the properties themselves and their relationship and hardness has not been highlighted previously. Our goal is to identify where the Ghost Trilemma has silently been present in prior systems and problem domains in order to call attention to the need for alternative approaches.
This paper makes the following contributions spanning both practical and theoretical dimensions of this problem space: \\ [0.5ex]
\noindent \textbf{Reframing and featurization of problem space.} We analysed a range of papers across several related and seemingly-unrelated areas in the literature (\tsref{sec:tracinginlit}). Our analysis identifies the trilemma properties lurking underneath unsolved issues present in such prior work. \\ [0.5ex]
\noindent \textbf{Framework.} We define the properties, and discuss both the implicit and explicit tradeoffs involved in designing a verification system for them. We examine how different assumptions in light of these properties create more definitive and solvable problem structures to bypass the impossibility we discuss later in \tsref{sec:introproperties} and \tsref{sec:ghostdesign}.  This enables a re-evaluation of prior work in light of the Ghost Trilemma.
%An empty line
\\ [0.5ex]
\noindent \textbf{Formulation.} We describe the Ghost Trilemma---a formulation of the problem setting of verifying sentience, location, and uniqueness---that captures the core issues at hand.  We sketch a proof of the trilemma, though fully proving it may be difficult due to its human elements, and besides the point for practical work.\\ [0.5ex]
As the Ghost Trilemma places a hard bound on how well any possible system might be able to address the fundamental challenge at hand, we can only hope for schemes that trade off various assumptions, costs, and practicalities.

\section{Background} \label{sec:background}
The identity properties central to the Ghost Trilemma show up in many unexpected contexts, but are best illustrated in the context of social media platforms, where questions of trust arise in dynamic, open, and global settings.

\subsection{Context}\label{sec:contextthreat} 
That today's online platforms are rife with misinformation and
disinformation is well known, as are the techniques used by attackers, whose
targets expand daily~\cite{NCRI_insights, covidbotsYoung}. Here we briefly outline
current scenarios and challenges. We examine a short case study about identifying trolls online to call attention to the underlying nuance that makes it a hard problem for both humans and machines.

\subsubsection{Misinformation and disinformation}
Misinformation campaigns are being used as a tool to affect
political~\cite{shu2020,goldhill2019,Motorevska2019,Alluri2019} and
apolitical~\cite{stilgherrian2020,covidbotsYoung} decisions alike, on both
domestic and international scales. At least 70 countries, including elections in the US, Ukraine~\cite{Motorevska2019}, Taiwan~\cite{shu2020}, and UK~\cite{goldhill2019} have been affected by misinformation and disinformation campaigns~\cite{bradshaw2019}. Across the globe, disinformation campaigns organized and often supported by governments~\cite{Alluri2019}, effectively cause information blackouts without shutting the internet down completely. There are some centralized efforts to counter the spread.\footnote{A dedicated team from the European Union studied the complexity of misinformation across the internet, and helped reframe the definition of misinformation and the context in which it is spread~\cite{claesson}. However, scholars~\cite{funke2018} also point out that while many nations are publicly solidifying their stance against misinformation spread by media literacy campaigns, passing laws, reforms and bills criminalizing the activity and so on, it is not being very effective.} 
A key vector of propagation is the unsuspecting, genuine human user following disguised troll accounts. While we cannot remove all such human agents of misinformation, we aim
to provide the user with a means of performing intelligent filtering based upon
the provenance of information: by ensuring that the content that they accept as
valid originates from users who are, more or less, who they purport to be.

\subsubsection{Growth of troll farms} Troll farms operated by nation states or
private enterprises have capitalized on the vagaries of culture and politics to
spread disinformation, encouraging division and undermining democratic
institutions~\cite{elder2012,vaas2018}. One of the most prominent recent
examples has been the ``Internet Research Agency'' (IRA), based in Russia.\footnote{IRA's precursor, a 2012 Kremlin-backed youth movement Nashi~\cite{elder2012}, can be traced back to the time when the Snow Revolution was gaining momentum~\cite{2011_senatereport,osborne2011,domnovosti2012}.}

A report by the US Senate Select Committee on Intelligence~\cite{senatereport} detailed the alleged involvement of IRA in the 2016 US presidential election. The report
states: ``the information warfare campaign was broad in scope and entailed
objectives beyond the result of the 2016 presidential election'' such as ``to
sow discord in American politics and society.'' Perhaps most concerning is that
the attacks required relatively few resources; the IRA reportedly spent only
about $\$100,000$ over two years on Facebook ads, and the organization
incurs operating costs of about $\$1.25$ million per month. 
The IRA's content burst increased after the 2016 election cycle. This was indicative of a continuous effort to maintain the spread of misinformation online which often directly impacted human lives. The IRA, even though controlled by Russian oligarchs, has been proven to have close ties with their  government.  They  have been known to target the African-American population the most among all racial groups. Even after suspension of several discovered troll accounts, the remaining  80\% of the accounts continue to publish content. Alongside the automated troll content that  is generated from these accounts, it is also backed by the rather large group of 400 human IRA employees who would  lend  an authentic human touch to the tweets and posts generated everyday. Besides the US, troll farms like IRA have also promoted divisive opinions such as anti-Muslim hashtags after the Brussels terror attacks, a pro-leave hashtag on the day of Britain’s Brexit referendum and leaks targeting French President Emmanuel Macron before his election~\cite{vaas2018}. 

\subsubsection{Identifying trolls} \label{subsec:spotthetroll}
However, due to increasing technological sophistication and the human touch of the troll farm employees, separating troll accounts from real ones are harder than ever, and likely to worsen still as large language models become widespread. To illustrate the difficulties, we compare examples from a website~\cite{spotthetroll} compiled by researchers for this very purpose in \tsref{para:compare}. It involved an inherent deduction around the identities and thought processes of the account holder as depicted in the account's content. 
For uniqueness, most people rely (when we can) on available resources which are often centrally maintained. For establishing location and sentience of the account holder, the contents of the account are often the only available data. Sentience can be vaguely correlated to the complexity of thoughts and contexts expressed in the contents. The location can be also be sometimes similarly deduced. 

This distinction is however impossible to see for most accounts on social media owing to several factors including user privacy. It is also very dependent on the sophistication of the trolls and bots involved. Looking at the vast bot identification oriented literature, we see both the mathematical reasoning of machine learning algorithms and abstract reasoning of the human brain fails to provide accurate classification results. 
%\section{Case Study: Bots for ``Reopening America''} 

\subsection{Case Study: Identifying Trolls}\label{para:compare} 
We discuss the difficulties of identifying trolls in the real world and look at how the three properties factor into this challenge through this case study. We expand on the definitions and why we pick these three properties in \tsref{sec:intuition} and \tsref{sec:introproperties}. To illustrate, we will compare examples from a website~\cite{spotthetroll} compiled by disinformation researchers for this very purpose. We find that there are only a few marked differences between human user accounts and known troll accounts. For instance, the sentience of any user is often demonstrated by the complexity of their thoughts and behavior displayed online, a variant of the Turing test. Location is often implicitly expressed through context and content that users post but otherwise unverifiable. Lastly, uniqueness verification is largely dependent on the personal information they choose to share, if any. Troll accounts are indicated by lighter shaded boxes, while real user accounts are represented by the darker shade. 

\begin{vignette}
\small
\vignetteboxbegin
Chloe Evans - Student, Atlanta: \newline \textbf{Location.} Covers fake news at
national level, nothing to corroborate location mentioned on profile. \newline
\textbf{Sentience.} No personalized experiences to leverage. \newline
\textbf{Uniqueness.} No defined self identified details that can be used for
verification.
\vignetteboxend
\label{v:troll-case1}
\end{vignette}

\begin{vignette}
\small
\vignetteboxbegin
Harmony Anderson, Ankeny, Iowa:  \newline
\textbf{Location.} Only political content, all unrelated to Iowa. \newline
\textbf{Sentience.} No personal (sentience based) details. \newline
\textbf{Uniqueness.} No self identifying data.
\vignetteboxend
\label{v:troll-case2}
\end{vignette}

The tweets by Chloe Evans are based on hoaxes that can be easily fact checked.
The content in Harmony Anderson's posts represent a different version of reality
that makes the reader perceive the account as inauthentic. Both the accounts use
pictures of young women and post popular content to gain followers; they connect
to communities while lacking any personal information. All of these behaviors
are markers of possible troll accounts. 

\begin{vignette}
\small
\vignetteboxbegin
 power\_to\_women\_ :
\newline
\textbf{Location.} No location given. \newline
\textbf{Sentience.} No detail. \newline
\textbf{Uniqueness.} No one listed as administrator of the group.
\vignetteboxend
\label{v:troll-case3}
\end{vignette}
A common trope for trolls is to use affinity groups like power-to-women-,
which gain followers easily. A lack of any group admin and a strict divisive
for-or-against attitude are troll markers. 

\begin{vignette}
\small
\vignetteboxbegin
Amy G, New York, NY: \newline
\textbf{Location.} No indicators to support the mentioned location. \newline
\textbf{Sentience.} No personal (sentience based) details. \newline
\textbf{Uniqueness.} No self-identification.
\vignetteboxend
\label{v:troll-case4}
\end{vignette}

Besides the usual troll markers discussed above, Amy G discusses real issues,
but not constructively, and often mimics popular users from the minority
community they pose to be a part of.

\begin{vignette}
\small
\vignetteboxbegin
Nevada Peace Officers: \newline
\textbf{Location.} No content focusing meaningfully on local issues of Nevada. \newline
\textbf{Sentience.} No indication of sentience. \newline
\textbf{Uniqueness.} No one listed as administrator of the group.
\vignetteboxend
\label{v:troll-case5}
\end{vignette}

While the Nevada Peace Officers account is associated with a swing state, there
are no clearly-verifiable local experiences, issues, or events discussed. It
also exhibits divisive ideological content and focuses on  catchy content to gain followers.

\begin{vig}
\small
\vigbegin
Christopher Worrick, Columbia City, Indiana: \newline
\textbf{Location.} Tweets addressing neighbors, not just a national audience -- personal experience of being in the location claimed in profile. \newline
\textbf{Sentience.} Was present in the experiences detailed in the profile, visibly + known to researchers. \newline
\textbf{Uniqueness.} Personally known to researchers.
\vigend
\label{v:human-case1}
\end{vig}

\begin{vig}
\small
\vigbegin
Chenjerai Kumanyika, Philadelphia, PA: \newline
\textbf{Location.} Discusses issues in the context of the specifically-mentioned region and community. \newline
\textbf{Sentience.} Mentions real life experiences, sometimes verifiable. \newline
\textbf{Uniqueness.} Verifiable identity as a scholar affiliated with an organization.
\vigend
\label{v:human-case2}
\end{vig}

Unlike unidimensionally-opinionated troll accounts, Christopher Worrick expresses complex, multidimensional opinions. Chenjerai Kumanyika mentions identifiable information that can be easily verified and local issues are the main focus of their content. 
\begin{vig}
\small
\vigbegin
Mike Adams, Lives in Austin, Texas; From Tucson, Arizona:\newline
\textbf{Location.} Known previously, no particular indicator in the profile. \newline
\textbf{Sentience.} A real identifiable person, with websites and addresses associated (maybe verifiable). \newline
\textbf{Uniqueness.} A real person who peddles misinformation, content banned by platforms, but can be traced back to an active personal account.
\vigend
\label{v:human-case3}
\end{vig}

A significant challenge is that real users
like Mike Adams, unless known previously, can be hard to distinguish from trolls. Without any definitive and verifiable human markers presented publicly, it is hard to distinguish between trolls and real accounts. 

From this case study we see the importance of these three properties with respect to verifying the realness of a user on social media. The challenge, however, is that without any explicit mechanisms for property verification, we (just like ordinary users and social media researchers) are left sifting through textual clues to perform this validation.

\section{Design Space} \label{sec:tracinginlit}

\subsection{Grounding the Intuition}\label{sec:intuition} 
Proving these three properties simultaneously while relying on a trusted central verifier is challenging, but without the centralizing influence of a ``trust anchor'' it is likely to be much harder. Here we sketch a rough background around each property to understand and identify them in the context of other research areas. 
We find that the verification of these individual properties, while not easy, is often explicit or implicit in a number of well established research areas (\tsref{subsec:voting} onwards). 
\subsubsection{Sentience} A sentient being is one that is alive and conscious of
its environment. Consciously experiencing the environment involves being capable of
engagement with multiple subject domains and contexts, rather than just a singularly directed one~\cite{duplexai}. Usually, when sentience is verified at a specific moment, it can be verified using CAPTCHAs~\cite{zeljkovic2016system, cowan2016liveness,
uzun2018rtcaptcha}.
Users are always engaged in judging the sentience continuously of others
by their behavior, even if only subconsciously.
However, content-based differentiation in itself is complex, as there are endless
perspectives and subtleties that make bots difficult to distinguish from humans.
Moreover, with dedicated resources and improving technology, verification of sentience will remain an arms race.
\subsubsection{Location} Location verification requires confirmation of a real
person being in a physical location at the time of verification. Here lies the
problem with proving location with sentience:
\begin{itemize}
    \item Common approaches to verifying location \emph{depend upon} sentience analysis; and
    \item Non-sentience-based location verification suffers from verifying simultaneity.
\end{itemize}   
First, as we saw from the troll case studies in \tsref{para:compare}, users verify others' locations through an implicit verification of their sentience and their location-specific knowledge. The knowledge that might be needed to express this is unenumerable, and as a result is difficult to formulate into an automated test (and were it to be automated, would quickly become easy to game). 

Second, a non-sentience-based location verifier, such as~\cite{abdou2015cpv,wang2011towards}, would likely build upon physical properties of the world that are difficult to fake, such as
network latency to a set of vantage points. While verifying location in this
manner may be straightforward, doing so while simultaneously verifying
sentience is challenging; how is one to verify that the real human one is
verifying is the user purporting to be in a specific location? Human
verification is necessarily latency-insensitive (both because humans cannot
reply on machine timescales and because the tasks must have some degree of
cognitive complexity); this thwarts simultaneous latency-based location
verification.

\subsubsection{Uniqueness} As we have discussed before, uniqueness by itself is
hard to establish without the cooperation of a centralized authority and the
user. For example, 
\begin{itemize}
    \item Users can be issued hardware security tokens from a trusted
vendor. These, by virtue of their uncloneability, can help a user prove that
their credentials are not being shared.
   \item A similar alternative is physically unclonable functions (PUFs) which can be used to generate unique identifier for a device by the virtue of its hardware~\cite{gassend2002silicon,8949407}.
\end{itemize}
However, today, most services rely upon user discretion and have no direct
means to verify that users have not shared their credentials. 

In a decentralized setting, users could rely upon cryptocurrency-based verification. For example, a service could bind user credentials to one or more coins, such that the
revelation of user credentials will also result in the revelation (and thus monetary loss) of those coins, which carries with it a monetary cost. However, implementing a decentralized cryptocurrency still requires agreement on the distributed state of the
individual participants (\`a la blockchain). Thus, verifying uniqueness and
location of the user by tying it to cryptocurrency-based verification and
real-world identity respectively is also hard in a decentralized system.

With this intuition in place, we can use Table~\ref{tab:comparelit1} as a guide for the following subsections and the relationship of the Ghost Trilemma properties to a wide range of research areas that have grappled with them.

 \begin{table*}
%\resizebox{14cm}{!}{
\begin{adjustbox}{width=\linewidth,center}
\def\arraystretch{1.3}
\begin{tblr}{width =1.2\linewidth,
  colspec = {|X[c,m]|X[4,m]|X[10,m]|X[c]|X[c]|X[c]|X[c,c]|},
rowsep = 4pt,
hlines = {1pt},
vlines = {1pt},}
\SetCell[r=2]{c}\textbf{Area} &\SetCell[r=2]{c}\textbf{Topic} & \SetCell[r=2]{c}\textbf{Description} &\SetCell[c=3]{c}\textbf{Relevant Properties} & & & \SetCell[r=2]{c}\textbf{Trust} \\
& & & \textit{\textbf{L}} &\textit{\textbf{S}} &\textit{\textbf{U}} & \\
\SetCell[r=3]{}\rotatebox{90}{\parbox{3cm}{\centering {\textbf{Digitizing democracy(\tsref{subsec:voting})}}}} & Swiss e-voting solution \cite{swissA, swissB,ford_2019,gjosteen2011norwegian,ford2022auditing,merino2022trip}& A case of mismatching expectations and security loopholes in the theoretical vs real-world operation of the system at scale. &  & \halfcirc & \halfcirc & \fullsq \\

 &  Estonian ID card \& e-voting system \cite{morgan2017using,evotereport,evoteinsecure} & The systems pose practical security and privacy risks, for all stakeholders. This is majorly caused by inconsistencies in cryptographic proofs and policies.
    & & \halfcirc & \halfcirc & \fullsq \\ 
    
 & Liquid democratic procedures \cite{follow2021,beauro23,fishkin2005experimenting,ford2020identity}& A requirement of occasional anonymity with transparency and a unique id for each participant calls for an individual and unique digital identity~\cite{ford2020identity}.  & \halfcirc & \halfcirc & \fullcirc &\halfsq\\ 
%\hline
 \SetCell[r=3]{}\rotatebox{90}{\parbox{3cm}{\centering {\textbf{Detecting social media trolls(\tsref{subsec:socialmedia})}}}} &
  General ML based approaches \cite{bhatia2017soc2seq,li2016persona,li2016deep,sordoni2015neural, kudugunta2018deep,sayyadiharikandeh2020detection, bradshaw2017troops,addawood2019linguistic,yang2022botometer}
 & To counter the dynamic adversarial nature of the online environment, regular re-training and patching is required as newer, more sophisticated bots are generated. &\emptycirc&\halfcirc &\emptycirc & \fullsq \\   
 & Adversarial ML based approaches \cite{cresci2019better,grimme2018changing,iliou2021web}
 & Both sides armed with GANs may eventually reach a stalemate where near perfect adversarial examples~\cite{diar_2023,duplexai} are in play. &\emptycirc&\halfcirc &\emptycirc & \fullsq\\ 
 
 & Human participation with AI Tools \cite{varol2017online, davis2016botornot,kudugunta2018deep,yang2019arming,yang2022botometer}
& Vigilant human feedback help improve the tool performances while the popularized tools in turn promote awareness. &\emptycirc &\halfcirc &\emptycirc & \fullsq \\
%\hline
\SetCell[r=3]{}\rotatebox{90}{\parbox{3cm}{\centering {\textbf{Establishing remote client location(\tsref{subsec:userlocation})}}}}&
Sattelite imaging \cite{gupta1998commercial,yeh2020using} & Never used in a location verification context but with continuing progress, using this intertwined with the client's sentient identity may be feasible. & \emptycirc&\emptycirc & & \fullsq\\   
&  Client presence verification \cite{abdou2014location, abdou2015cpv,abdou2018internet,abdouevasion, huseynov2019physical}
& Using client device geolocation as the proxy for the client's presence is not sufficient without any way to link it to the actual client's sentience. &\fullcirc& \emptycirc &  & \fullsq  \\  
& Close range communication (NFC, RFID)            \cite{coskun2013survey,cope2017investigation,sastry2003secure,talasila2010link, nosouhi2020blockchain} & The client's device is required to be in a close proximity of the verifier. Despite longstanding security concerns, they may help verify both location and sentience. &\fullcirc&\halfcirc&&\halfsq\\
%\hline
\SetCell[r=3]{}\rotatebox{90} {\parbox{3cm}{\centering {\textbf{Multi-factor authentication(\tsref{subsec:validfraud})}}}} &
 Location based multi-factor authentication \cite{kleberger2011security,ramatsakane2017pick,koved2015usable} & It is necessary to link the geolocation source to the client's physical identity. This requires full centralization. &\fullcirc&\fullcirc&\fullcirc&\fullsq  \\  
& Biometric multi-factor authentication \cite{aley2013device,chetty2006multi,hoyos2016system, kollreider2007real,10.1145/1854229.1854270,cowan2016liveness}
 & Centralized uniqueness+sentience verification. However, depending on what marker is used, there is always a chance of spoofing attacks. &&\fullcirc&\fullcirc&\fullsq   \\ 
 & Mixed multi-factor authentication \cite{zeljkovic2016system,feng2019system, aley2013device,chetty2006multi, rahman2015movee,livenessinFD}
 &Supporting a biometric marker with an additional audio, visual, non-biometric or a different type of biometric factor, helps improve robustness, reduces spoofability. & &\fullcirc&\fullcirc&\fullsq  \\
%\hline
\SetCell[r=3]{}\rotatebox{90}{\parbox{3cm}{\centering{\textbf{Open cryptocurrency systems(\tsref{subsec:pop})}}}} &
 Alternatives to  Proof of Investment (PoI) \cite{ford2008offline,borge2017proof,sanchez2019zero,individuality} & To facilitate more inclusive bitcoin crowd, Proof of Personhood (PoP) would be a good replacement for PoI. Other options include proofs of identity and individuality.& &\halfcirc&\fullcirc&\halfsq \\  
 &  Scalability of the PoP token \cite{guicciardi2022scalability,ford2020identity}
 & \cite{guicciardi2022scalability} shows PoP tokens are securely scalable to a bitcoin company clientele. However, scaling it up to include all social media users is a much harder challenge. &&\halfcirc&\fullcirc&\halfsq  \\  
& Verifiability of verifiers \cite{ford2020identity,siddarth2020watches} &Decentralization efforts require both ends to be held accountable which is more challenging with an always expanding client base. &\emptycirc&\halfcirc &\fullcirc &\halfsq\\
%\hline
\SetCell[r=2]{}\rotatebox{90}{\parbox{2.5cm}{\centering {\textbf{Open distributed systems(\tsref{subsec:distributed})}}}} &
 Byzantine-Altruistic-Rational model \cite{lilorenzo,aiyer2005bar}
 & Built to work in a cooperative environment this model, despite structurally resembling the information ecosystem, is hard to map into an adversarial space with sentient parties.
& & &\emptycirc & \emptysq  \\  
 & Rational vs Byzantine participants \cite{ford2019rationality}
 & For any parent ecosystem having the open distributed system, any rational participant's best interest can alter from external influences. Then they may behave as byzantine entities.  & & &\halfcirc & \emptysq
 %\hline

\end{tblr}
\end{adjustbox}
%} 
\caption{\centering{Tracing the presence and impact of properties discussed in the Ghost Trilemma across the literature.} 
\newline 
L$\equiv$ Location; S $\equiv$ Sentience; U $\equiv$ Uniqueness \vspace{1ex}
\fullcirc : Property is verifiable (standalone and/or combined) with reasonable guarantee, 
\halfcirc : Property is verifiable with partial guarantee,
\emptycirc : Property might need verification but is not addressed in the literature.
\vspace{1ex} \newline
\fullsq : Fully Centralized, \halfsq : Partially Decentralized, i.e., initially was aimed at full decentralization \emptysq : Fully Decentralized 
\newline 
Informally, Ghost Trilemma $\simeq$ \fullcirc \space \fullcirc \space \fullcirc \space \emptysq \space is impossible to attain.}
\label{tab:comparelit1}
\end{table*}

\subsection{Digitizing Democracy} \label{subsec:voting} 
Internet-based voting systems have been widely discussed and even implemented in some elections~\cite{jefferson2004analyzing,springall2014security,gjosteen2011norwegian,adida2008helios,haines2020not}. They have seen limited success and are often critiqued for a variety of weaknesses. Such systems aim to offer a reliable and private mechanism for casting and counting votes, but there is often little trust between the client (voter) and the server (voting network). 

Designing a reasonably secure and accurately verifiable online voting system has been long established as a difficult-to-solve problem. Few governments across the world have tried to implement online e-voting locally and nationally. Outside of the US, Norway~\cite{gjosteen2011norwegian}, Switzerland, Australia~\cite{evotereport}, and Estonia~\cite{springall2014security} are some notable examples. The Estonian system of national ID card comes close\footnote{Estonia’s Internet-based voting system has leveraged national ID cards to secure elections~\cite{springall2014security}. Estonian national ID cards are smart cards which can guarantee both vote privacy and voter uniqueness in their elections theoretically. This can also guarantee sentience, as the ID card is required to be physically inserted into a card reader once the voting process begins. Work in \cite{clark2011selections} also implements an in-person only registration process before the election date in which each voter obtains their respective credentials thus verifying sentience.}, but so far no system can simultaneously guarantee all three properties (sentience, location, and uniqueness).\footnote{The Estonian ID card has also been shown to be practically vulnerable to attacks~\cite{morgan2017using}} However, several nations like Estonia, the US, and the UAE are continuing with e-voting systems, particularly for local elections~\cite{evoteincountries}.

Several issues arise when such a system is actually implemented in practice on a national scale.  Appel~\cite{evotereport} describes the current state of e-voting and its subtle but severe issues, emphasizing the necessity of transparency around the whole process. In another article~\cite{evoteinsecure}, Appel observes that “[T]he clear consensus of computer scientists and cybersecurity experts is that paperless voting systems cannot be made sufficiently secure for use in public elections.” The inherent insecurity of e-voting comes from some persistent factors (like insecure client computers, insecure servers, and a lack of universal digital credentials), many of which cannot always be resolved practically. 

Complete transparency and offline communication channels contribute significantly towards security of such systems which operate through several dynamic, interdependent parts. Appel~\cite{evotereport} discusses how e-voting systems should ideally be evolved using expert perspectives over multiple iterations. They should also be completely transparent for audits and evaluation. For example, the Swiss government launched an e-voting system, suffered a setback, and as a result put it on hold pending further study.  Other countries like Australia, France, or several states of the US, who implement e-voting in real life are faced with the same issues (or worse) as those with the Swiss e-voting solution jointly operated by Scytl and Swisspost~\cite{haines2020not}. Transparency on part of the Swiss Post has helped identify problems encountered in Australian election administration using the same software. The out-of-band communication feature (a sheet of paper sent through the mail, to the voter) enables enhanced security guarantees in this system.

Theoretical proofs often cannot provide practical protection, especially when we overlook assumptions used in proofs that in reality end up creating security loopholes. We look at some security concerns attributable to this and addressed in the five part report~\cite{krahenbuhl2022swiss,lewis2019not}. The attacker can invalidate votes from users and skew the result to one side. Based on the assumptions used in the formal proof for the establishing the security of the system, \cite{lewis2019not} points out that a dishonest mixer can forge decryption proofs and create (dis)information that it will pass verification in the formal sense.\footnote{Likewise, in the real world of fake news and disinformation, we see similar convictions around what can be actually verified as true versus what exactly is the ground truth, if any.}

There might arise requirements for voter's location verification but ideally, it should neither invade personal privacy nor should adversaries be able to gainfully leverage it. Targeted denial of service (DoS) attacks against the Swiss system used a similar mechanism that was used by targeted Facebook ads (during the 2016 USA presidential election~\cite{senatereport}), leveraging the voter's location data. This, while unrelated to the component of location verification for any entity, does bring up the issue of people’s location privacy. The report also discusses the effectiveness of BGP hijacking attacks in context of the proximity of the attacker to the system. For remote users located far from the voting system, it is much easier for the adversary to interfere with and redirect the traffic. Similarly, within our problem context, it will be hard to incorporate features to accommodate remote out-of-country users into the system without creating exploitable vulnerabilities. This directly affects the key advantage of having broad accessibility through e-voting.

The final solution to all DNS-based attacks possible within the Swiss voting system~\cite{krahenbuhl2022swiss} is to always use end-to-end authenticated naming data. However, the DNSSEC chain of trust validation is usually outsourced to trusted third party operated resolvers due to large overheads. Delegating it to third parties creates an additional but often unavoidable loophole for attackers to leverage~\cite{schmid2021thirty}.\footnote{This is similar to how we see uniqueness verification of users which ideally has to be delegated often to trusted third parties to avoid increasing the complexity of an already cumbersome system.}

However, despite these often unavoidable circumstances, rigorous patching, documentation, and expert auditing are shown to aid in the system's evolution: “As imperfect as the current system might be when judged against a nonexistent ideal, the current system generally appears to achieve its stated goals, under the corresponding assumptions and the specific threat model around which it was designed”~\cite{appel_2022}. Though it is hard to design flawless and elegant solutions to such issues, transparent discussions in the community around practicable and iterative improvements may one day yield a acceptable system, one that leverages some degree of centralized trust.

While we have only discussed issues around large scale e-voting measures and systems, we also want to touch upon the general requirement of anonymous identity verification for participants. Such a requirement is critical to many open democratic processes, such as the selection of jury members~\cite{beauro23,fishkin2005experimenting}, and other sortition-based participation and delegative voting processes~\cite{landemore2021open,landemore2020open}. In situations which demand a fair and inclusive sampling from a limited set of participants, we do not have a secure way to establish their claim's credibility without investigating their identities at some stage.
Due to the nature of this problem, sentience and location may not be explicitly verified since the participant's presence during the process can confirm both. However, the process needs to tie each individual to the verification of their uniqueness. The PoP tokens~\cite{borge2017proof,ford2020identity} discussed in \tsref{subsec:pop}, may offer a solution with a reliable centralized mechanism for generating and distributing the tokens. However, scaling this concept for an environment (like the information ecosystem) that is as dynamic, fraught, and always increasing in size and complexity, will be challenging as described by Ford~\cite{ford2020identity}.

\subsection{Detecting Social Media Trolls} \label{subsec:socialmedia}
Experts have noted that ``scholars and administrators are constantly one step behind of malicious account developers''~\cite{cresci2020decade}. This consistent lag is very visible in social media today: ``[d]espite the increasing number of existing detection techniques, the influence of bots and other bad actors on our online discussions did not seem to decrease''~\cite{cresci2020decade}. Due to its inherently adversarial nature, the problem of identifying bots and bad actors online is better addressed than before through advances in adversarial machine learning. Before using these adversarial techniques~\cite{cresci2019better,grimme2018changing}, the majority of the other machine learning (ML) algorithms assumed a stationary and neutral environment. Hence, over time as bots have evolved, prior assumptions have failed to hold. The community hopes that with advancements in various forms of adversarial ML techniques and GANs~\cite{wu2020using,goodfellow2018making} with time~\cite{cresci2020decade} we might see better detection results. However, there is a possibility that with both sides competitively armed, we may eventually reach a stalemate with near perfect adversarial examples~\cite{diar_2023,duplexai} in play, creating uncertainty and chaos.

We see throughout the literature on bot detection using ML techniques that the accuracy of detection is heavily dependent on feature selection~\cite{varol2018feature,yang2020scalable, mbona2022feature}. Features are selected based upon how indicative they are of the anomalous behavior of a bot versus a human account. 
Extracted features (both on account and content levels) often collectively tend to point towards sentience of the user by establishing a behavioral pattern and/or social network structure around each user account. While this is very effective for accounts which showcases a distinctive bot-like behavior, it will likely miss out on the more sophisticated variations of trolls that are more ``human'' but not who they claim to be. 
The lag between the problem and the issue \cite{cresci2020decade} is actually unbridgeable because of the Ghost Trilemma. In other words, perfecting how to be a human on social media is an ideal adversarial case that cannot be decisively defeated. Hence, the solutions will still fail to identify some troll accounts and wrongly classify other legitimate accounts. We can partially also credit the focus on sentience (e.g., behavioral aspects) to the community, and less so the uniqueness and location of the account holders.\footnote{This is also because publicly-available data from social media services do not currently provide strong location or uniqueness proofs.}
Sentience verification has not yet been perfected, but it is the most developed and widely used form of verification. Hence, to effectively counter the adversary who is always a step ahead, we might need more than just sentience to further narrow the gap. 

Public participation keeps the available partial solutions updated with the current and latest generation of trolls. People are always adjusting to the various AI tools available online. To popularize these tools scholars~\cite{yang2019arming} suggest that technical solutions should properly consider how humans interact with them. They flag ``limited public awareness'' and ``unwillingness to adopt sophisticated tools to combat bots'' as bottleneck factors impacting the efficacy of the tools. Scholars~\cite{yang2019arming} identify Bot-o-meter~\cite{varol2017online, davis2016botornot} as a popular and easy to use tool. Bot-o-meter is consistently updated as the researchers refine their mechanisms using more complex deep neural network (DNN)-based architectures~\cite{kudugunta2018deep}, trying to match the increasing sophistication of bot creators with their own improving models. DNNs open up a pathway to detect bots but the same holds for troll farmers and bot creators as well~\cite{bhatia2017soc2seq,li2016persona,li2016deep,sordoni2015neural}. 

Thus, constant vigilance about regular updates for tools and their corresponding datasets is recommended to keep up with the ever changing landscape of social bots and trolls~\cite{yang2022botometer}. As scholars~\cite{yang2022botometer,cresci2020decade} often point out that in conclusion, there will always be some distribution of data where each method fails to identify bots correctly. Similarly, no one monolithic system will suffice for a problem this complex and nuanced, especially with the added constraints of a distributed setup. 

\subsection{Establishing Remote Client Location} \label{subsec:userlocation} 
Remote client verification is a vast subject, so to focus we initially discuss work on Internet location verification and client presence verification~\cite{abdou2014location, abdou2015cpv,abdou2018internet,abdouevasion, huseynov2019physical}. In most of the literature on remote client presence verification, the user device is the primary client and is nearly always used as a proxy for the user's presence at a location. However, this complete detachment of the actual human user's identity from the client device gives rise to several spoofing opportunities for adversaries, and can sometimes lead to a more severe attack \cite{kleberger2011security}. Knowing a device's geolocation is not sufficient to accept the associated person's location claim as well. Establishing trust in the actual user's location is a requirement for many of the use cases we discuss in this paper. Beyond just the location of a device as a proxy, we need both sentience and uniqueness properties here to establish trust in the user's identity. 

Prior work such as~\cite{sastry2003secure,talasila2010link, nosouhi2020blockchain} focused on secure verification of client location within the same local region as the verifier. They use close-range communication (like bluetooth or NFC) to ensure presence within a certain radius. Close-range communications, despite their security vulnerabilities~\cite{coskun2013survey,cope2017investigation}, provide the best alternative for estimating the client's location and is also easy to couple with a sentience verifier. Works such as \cite{nosouhi2020blockchain} apply unique features of a blockchain to lower the possibility of collusion (prover-prover or prover-witness) occurring within the region. However, the proxy of the user's presence is still the device they own for all this prior work. Combining sentience (often established through biometric features) and location is more commonly seen in the user validation and fraud detection literature (\tsref{subsec:validfraud}). Furthermore, the verifier in these contexts must themself be a trusted party, or part of a web of trust.

Using satellite imaging and remote sensing to establish patterns in climate changes, predict disasters, and study the impact of environments is well established~\cite{yeh2020using}. Satellite imaging has been used in a context similar to  location verification~\cite{gupta1997investigating,gupta1998commercial} for a Comprehensive Test Ban Treaty (CTBT) verification process at a nuclear power site. With continuing advancements in satellite imaging and commercialization of satellites, such remote sensing may present one avenue to verify a user's sentience and location.

\subsection{Multi-Factor Authentication} \label{subsec:validfraud}
A survey of multi-factor authentication (MFA) strategies~\cite{ometov2018multi} considered the types of MFA sensors used, including geolocation mechanisms on device, fingerprint and hand geometry, voice biometrics, ocular parameters, facial recognition, physical tokens, passwords , and more. They identify a key challenge in MFA research as the ``absence of correlation between the user identity
and the identities of smart sensors within the electronic device/system''~\cite{kleberger2011security}. This resonates with our observations in \tsref{subsec:userlocation}. 

Location-based MFA~\cite{ramatsakane2017pick,kleberger2011security} is required to tie a user's sentient identity (e.g., biometric markers) to location data (e.g., GPS, cell towers, etc.) using the same device. Together they could indicate presence of a human with the device within a definite bounding box. Using biometric markers offers the added advantage of establishing an exact and unique user identity as well which is inherently tied to the user's location and sentience. For applications in user validation, authentication, and fraud detection, the number of users will be of a limited scale (e.g., employees of some organization) with a strong trusted central authority. In the broader context we consider in this paper, the number of users requiring verification will be, potentially, all Internet users, which is a much larger and more widespread population without any natural central trusted party. 

Additionally, we focus on verification and not authentication. Verification will limit repeat users from generating unlimited unique identifiers while authentication is used to establish user access. Hence the accuracy and spoofability of the signals used demand careful consideration.\footnote{We require high accuracy signals with low spoofability because otherwise adding a new user could turn into allowing a duplicate identity into the system.}
MFA systems use three factor types, i.e., knowledge based, biometric based, and ownership based~\cite{ometov2018multi}. As a parallel to our problem context, an intelligent MFA system is likely to couple these three types of factors which in turn could interlink sentience and uniqueness and sometimes even location for the users of the system.

Liveness or sentience verification through CAPTCHAs and biometric markers in MFA systems is a thoroughly researched area. A variety of methods have been developed to detect liveness by using multi stage voice prints~\cite{aley2013device,chetty2006multi}, blinking eyes~\cite{ionita2016methods}, live facial recognition~\cite{hoyos2016system, kollreider2007real,10.1145/1854229.1854270}, body vitals measurements~\cite{cowan2016liveness}, voice challenges in multiple
stages~\cite{zeljkovic2016system,feng2019system}, and video based movement sensors~\cite{rahman2015movee}. Biometric markers especially can be used to verify both liveness and uniqueness. Although it often remains open to spoofing attacks~\cite{costa2019challenges,197225,jia2019spoofing,Boulkena7961798,GHIANI2017110}, multi-factor checks~\cite{zeljkovic2016system,feng2019system, aley2013device,chetty2006multi, rahman2015movee} and countermeasures~\cite{livenessinFD,gross2013captcha, gross2013system} can ensure a more robust performance\footnote{Companies market commercial solutions~\cite{livenessinFD} which combine liveness and uniqueness (through face biometrics) verifications to validate users to organizations globally. They apply liveness verification of the users to strengthen the security of face biometric technology against spoofing through deepfakes, masks and videos.}.  
However, all of these essentially require a fully centralized implementation with a trusted authority and a limited set of users.
\subsection{Cryptocurrency Tokens} \label{subsec:pop}
Popular cryptocurrency tokens with Sybil resistance are another domain of study in which uniqueness and sentience verification have been pursued in a distributed environment. Previously, such mechanisms utilized Proof of Investment (e.g. proof of work and proof of stake~\cite{platt2021sybil}) despite the resulting inefficiencies (i.e., high energy and environmental costs, capital hoarding, and low transaction volume). The usability of the system becomes exclusive (despite being ``permissionless'') and dependent on individual investment. 

However, given that the original aim of Bitcoin's permissionless consensus mechanisms was to allow ``anyone'' to participate by mining, permissionless cryptocurrencies should have a better capacity for inclusion. A fundamental rethinking of countering sybils while focusing on inclusivity gave way to mechanisms that leverage the user's existence and identity. Proof of Personhood (PoP)~\cite{borge2017proof}, individuality~\cite{individuality} and identity~\cite{sanchez2019zero} are strides in this direction. The goal is ideally to shift away from the one-dollar-one-vote paradigm supported by proof of investment schemes and towards a more equitable paradigm of one-person-one-vote, while managing the security and privacy risks. 

PoP~\cite{ford2020identity} satisfies the goals for establishing a digital identity online without divulging any personal identifying information. It protects and validates the digital personhood of real human Internet users. This approach addresses uniqueness and sentience validation (i.e., a real human user who abides by the one-person-one-vote standard strictly) but does not factor in the user's location within the problem context. The implementation is planned through pseudonym parties organized by a responsible party where other interested individuals turn up and are registered through a system of cryptographically-secured tokens. The authors propose to use federated pseudonym parties with synchronized deadlines for scaling beyond one single geographic location. The participants need only be present at any one location of the pseudonym parties, without any location constraints. 

However, there still remain several issues that are technically hard to combat. Coordinating such gatherings across various global timezones present a challenge. Some users would likely be able to obtain multiple (but not unlimited) tokens travelling across timezones. On another end, holding the organizers of each pseudonym parties accountable to the attendees and other organizers is a challenge because of technology like deepfakes~\cite{chesney2019deep}, which can be used to forge false evidence on video. Arranging for cross-witnessing by official and surprise unofficial volunteers across these gatherings can help. It would be inconvenient  (but not impossible) for a corrupt group of organizers to create and maintain the charade of false evidence without successfully bribing a large group of participants (who might be cross-witnessing volunteers) together. 

Coercion resistance is another such challenge that is contextually highly relevant but hard to combat technically. Nothing can prevent a participant from obtaining the PoP token through the formal procedure and then selling it to another party for their use. The best option is likely to provide alternative fake tokens (that are identifiable only to the registered user) and trust in the civic responsibility of the individual.\footnote{A similar mechanism is used in the Estonian e-voter registration process.} In the context of the use case of differentiating between fake and real users online, this remains a problem: any adversary with significant resources, including but not limited to nation states, can scheme to buy out or dupe verified users to obtain their verified and credible tokens for use by fake users.

There have been some recent commercial attempts to incentivize users to go through a Proof of Personhood verification~\cite{borge2017proof}, such as  Worldcoin~\cite{Worldcoin1}. However, it is unclear whether Worldcoin can delivering its promise of equity, fairness, and privacy. It is hard to gauge and protect against various misuses of practical implementations~\cite{problemsUnseenworldcoin}, which calls for privacy \textit{and} transparency. A biometric scan to verify a person's human-ness (what we call sentience in this work) and user identity is pre-collected through a custom hardware device.  Users are expected to trust Worldcoin, inherently a centralized entity though with the trappings of a blockchain-based decentralized service, with their sensitive biometric information. However, this biometric data is reportedly now easily available on the black market in bulk, enabling someone to purchase Worldcoin identities in bulk. 

Accurately mapping human identities into a digital space is indeed challenging. Whether such a system attempting to verify/assign unique digital identities to individual humans  can ever work will only be known with time. This inherent difficulty of binding human-ness with location and uniqueness of each user in a decentralized setting, as seen with Worldcoin and other examples, is the challenge captured by the Ghost Trilemma. 

\subsection{Modeling Distributed Systems} \label{subsec:distributed}
Distributed systems are designed assuming a mixed environment, with byzantine entities attempting to corrupt the behavior of protocols employed by honest and correct elements. Despite the obvious glaring difference in context of the type of participants present (nodes are typically machines here, and not sentient), this field often faces a similar challenge in establishing unique identification for its participating nodes and weeding out Sybils in the process. 

Here we consider the \textit{Byzantine-Altruistic-Rational} (BAR) model~\cite{lilorenzo,aiyer2005bar} which separates ``rational'' nodes from the byzantine nodes, but in a cooperative environment. The participating nodes can behave as either a byzantine entity (which exhibits irrational, unpredictable, or malicious behavior), an altruistic entity (which always follows the protocols), or as a rational one (which behaves in its own best interest). From a behavioral standpoint, this resembles a online social media (albeit with some sentient and human participants). The byzantine nodes can be seen as a parallel to troll accounts, while the rational and altruistic nodes can be compared to users who consume information online and the ones who strictly verify information before consuming or propagating it, respectively. Due to the difference between these modes of participation, we abstain from discussing incentives involved in either environment.

Classic byzantine consensus requires that no more than one third of the participating nodes be faulty (i.e., exhibiting byzantine behavior). However, many distributed environments are extremely likely to have higher than one third faulty entities.  Designs such as the BAR model~\cite{lilorenzo,aiyer2005bar} and \textit{(k,t)} robustness~\cite{abraham2011distributed} work to circumvent the classical byzantine consensus requirement by lowering the number of apparent byzantine nodes with some selfish (rational) nodes who behave honestly while it suits their interest. 

It is commonly assumed that a rational node's behavior within an open distributed system is not impacted by any external (i.e., from outside the permissionless distributed setup, but within a larger parent environment) stimulus. However, this assumption does not always hold, especially when an external intervention may give the rational node a better opportunity. This can make a practically insecure system seem secure in theory~\cite{ford2019rationality}. While mapping the concepts from this space to another having sentient participants, careful considerations are necessary especially about the assumptions used in the original space and those that are made during the mapping.

This rational behavior assumption will not hold in our problem context. The setup within which our participants function does not control any majority of total economic power or any other incentive of value that will successfully prevent a rational and sentient participant from maximizing their own gain. Even with non-sentient nodes in permissionless distributed environments, which in turn is part of a larger ecosystem, rational nodes are highly likely to be influenced by external stimuli that increase their overall perceived gain~\cite{ford2019rationality}.  Additionally, \cite{lilorenzo} also assumes unique identities for each participating node within the system, i.e., Sybil identities are not allowed. This is an unreasonable assumption in the context of global online systems without a centralized authority.

Eliminating Sybils~\cite{douceur2002sybil} in a permissionless distributed environment is a hard challenge, especially with an adversary with significant resources. Usually Sybil defense~\cite{douceur2002sybil,al2017sybil} works when the mechanisms correctly classifies every node using some structural property of the network graph. A perfect Sybil attack is when the attacker converts nodes in the graph into Sybils without introducing structural changes overall within the graph and hence the Sybils remain undetected. This parallels a real-world scenario in which an externally-available incentive coerces or otherwise persuades a ``rational'' node to give up its unique identification token to another. We again find this topic of coercion resistance cannot be managed technically and will keep reappearing for PoP tokens~\cite{ford2020identity}, in the several e-voting systems~\cite{springall2014security}, and within our problem context. The only resolutions appear to include increasing the cost of coercion for the adversary and trust in the integrity of the participants.

\section{The Ghost Trilemma} \label{sec:introproperties} 
To understand how these properties interact with each other and the related tradeoffs, next we consider the properties and their interactions and then discuss what the process of threat modeling in this context may look like depending on various threat and trust assumptions, while keeping in mind that the contexts in which these properties arise are diverse and varied and no one threat model will apply universally for future work in this area.
\begin{table*}{
\begin{adjustbox}{width=0.8\linewidth,center}
\bgroup
\begin{tblr}{width =1.1\linewidth,
  colspec = {X[c,m]X[c,m]X[c,m]X[c,m]X[4,m]},
  cell{1}{1} = {c=3}{c}, % multicolumn
  rowsep = 4pt,
  hlines = {1pt},
  vlines = {1pt} % vlines can not pass through multicolumn cells
}
  Trilemma Properties & & & \SetCell[r=2]{c} Feasibility & \SetCell[r=2]{c} Discussion \\
  Location& Sentience & Uniqueness & & \\
  
  $\times$ & & & \fullcirc & Location verification techniques that work in a distributed setting suffice.  \\
   & $\times$  &  & \fullcirc & Sentience verification techniques that work in a distributed setting suffice. \\
   &  &  $\times$   & \fullcirc & Unique identity verification techniques that work in a distributed setting suffice. \\
  $\times$ & $\times$& &\halfcirc & Simultaneous binding seems difficult through the use of network latency due to the mismatching timescales of response rates of a human and a computer. \\
  $\times$&& $\times$ & \fullcirc & Verifying the unique identity of an non-sentient entity may involve knowing its network location which can then be translated into a geographic one through network latency.\\
  &$\times$&$\times$ & \halfcirc& While they can be bound together momentarily, they are decoupled after a period of time $t$. The validity of the uniqueness verification usually outlives that of sentience.\\
  $\times$& $\times$&$\times$ & \emptycirc & The combined effect of rows 4 and 6 contributes to the impossibility of Ghost Trilemma.\\
\end{tblr}
\egroup
\end{adjustbox}
\caption{Design choices and the proofs they obtain.} 
\label{tab2}
}\end{table*}

\subsection{Overview}
The Ghost Trilemma is not a mathematical trilemma that can be proved crisply without making undue assumptions about the setting, human behavior, and the properties themselves. While we do provide an initial sketch of one proof approach, we believe it to be both unnecessary and incomplete. Instead we believe the trilemma to be a rule of thumb that can guide practical design choices made in security research and practice. Before we discuss the types of threats that one might consider when in a setting that encounters the trilemma, a bit of intuition about the three properties and the difficulty of verifying them may be of value.

In Table~\ref{tab2} we list the three properties and our assessment of the feasibility of verifying them together in a decentralized setting. Given the human complexity of verifying sentience, it alone is difficult to verify and this difficulty is compounded when it must be verified simultaneously with other properties.

The three properties are nicely orthogonal: they capture fundamentally different qualities of a user's identity in a distributed system. Due to this orthogonality, they must also individually be verified using very different mechanisms; we considered many of the mechanisms used in the literature earlier. Furthermore, even as solitary properties it is difficult to provide a universal definition for what ``counts as'', say, accurate location or true sentience. For example, a voting system may only care that a user is within the country in question but cares to an extreme degree that the user is sentient and unique, whereas a sports streaming company likely cares far more about the location of a user to show localized content but cares somewhat less about sentience.

\subsection{Threat Modelling}\label{threatmodel}
We see from the case study of spotthetroll\footnote{Quiz at \url{https://spotthetroll.org/}} (\tsref{subsec:spotthetroll}) that without a publicly displayed verifiable marker present, it can be often impossible to distinguish trolls from real human users. From the literature considered in \tsref{sec:tracinginlit} we examined the breadth of contexts in which the three properties recur, amplifying the need to understand how to verify them.

Our aim is not to provide a single threat model but rather to consider the process of threat modelling in a representative context. Specifically, to ground our threat model in real-world considerations and a scenario in which the three properties are crucial to the security of a system, we examine the setting presented in the US Senate report on foreign election interference~\cite{senatereport}: a troll farm supported by a well-resourced attacker aims to pretend to be large numbers of American users to influence an election. Other similar (and less well resourced) adversaries fall under the same model.

For example, consider the adversary $\mathcal{A}$ as a troll farm that can employ humans
to operate many accounts and implement bot-like behavior. We must assume $\mathcal{A}$ will not have unlimited resources at their disposal. We cannot prevent every possible intrusion to the system when the adversary is always prepared to spend the required capital. The worst case scenario that any solution will struggle to defend against is an adversary $\mathcal{A}$ with unlimited funds paying off many people to act on its behalf. Such users may not even realize they are helping an adversary; the adversary may be masquerading with a benign front (for example, employees of a remote company could actually help a troll farm unknowingly).
    
Here one might focus on limiting the involvement of humans and bots with fake accounts from outside a given country\footnote{There are cases where users that are physically outside of a country would wish to participate in social media (e.g., refugees, military deployments, etc.). We do not wish to bar such accounts from participating in online discourse; instead, we seek to create an additional signal that would provide users with context regarding accounts' sentience, location, and uniqueness.}, such as the US, by providing an additional means of verifying whether a user is who and where they claim to be. Of course a US-based user could still spread misinformation once ``verified'', but here we are considering verification of all three properties.  Besides the initial adoption, our biggest challenge will be for users to trust the system and integrate it into their online existence, and, over time, hopefully create a large enough network of trusted sources that the influence of the adversary will decline to a very small percentage of the verified user base.\footnote{Although human influencers with large followings can be paid off to support propaganda through disinformation spread and contribute significantly to the problem, so we consider such actors to be out of scope.}

A real-world adversary could be a nation state or similar powerful actor, and
thus have significant human and computational resources within their
national borders. One cannot hope to defend against an all-powerful adversary.
We expect such an adversary to be able to bribe or trick corruptible humans into
working for them, posing as legitimate users, validating their existence without
them being physically present in a foreign country. However, every participant in the solutions we consider is potentially byzantine~\cite{LSP82} and can deviate from the correct protocol (via stopping, crashing, or malicious intent) at any point in the system execution. Note that for simplicity, we do not distinguish between an external and internal adversary, but rather treat an adversary as a byzantine participant in the system.
%with the ability to disrupt its execution.
\subsection{Tradeoffs Encountered} \label{sec:tradeoffs}
From Table \tsref{tab2} we discuss the challenges in verifying sentience, pairwise with the other two properties.\\
\noindent \textbf{Location + Sentience.}
While each property can be verified individually binding location and sentience is hard due to the mismatching timescales of response rates of a human and a computer. The mismatching timescales between the two open up the system to the possibility of external attacks. However, we hope future works can leverage something like new geographical waypoints, sattelite imaging, etc to develop a more accurately effective binding between these two in a decentralized setup.\\[1ex]
\noindent \textbf{Sentience + Uniqueness.}
Verifying sentience is a fuzzy concept. 
While they can be bound together momentarily as we see in \cite{ford2020identity}, the binding is very easily decoupled.The verified user might choose to sell off their uniqueness identifier at time period $t+1$ if the verification which binds sentience with uniqueness ends at $t$. The validity of the uniqueness verification usually outlives that of sentience.

We now discuss the practical challenges and tradeoffs of verifying each property and selecting a trust anchor. Selecting the trust anchor determines the level of centralization within our framework. Its function is to provide an anchor to the verification processes. For the purpose of this work, we choose that our trust anchors will be pre-existing, centralized third-party systems that usually work reliably for the masses.
\subsubsection{Practical Constraints and Challenges}\label{architecture} 
Individually and separately, it is possible to prove the
user's (by proxy of a device) location~\cite{wang2011towards,abdou2015cpv},
sentience ~\cite{zeljkovic2016system, cowan2016liveness,
uzun2018rtcaptcha, ford2020identity}), or uniqueness.
However, any two of these properties together becomes
harder to establish simultaneously. Together, the location and sentience of
the same user is difficult to achieve
(\tsref{sec:intuition}), as is sentience and uniqueness without explicit
dependence on verifiable information shared by the user, often requiring the involvement of third parties. Similarly, verification of location and uniqueness
together without a trusted authority is hard. Adding
the third property to any of these pairs makes it impossible to achieve
verification for all three properties in a decentralized environment. 

A real-life example where all three properties are verified is a key
signing party~\cite{ford2020identity}. It also involves an additional human web of trust component.
The location is predetermined, acting as a trust anchor which binds the sentience and uniqueness of each user to the location (with proper documentation and hashed public key fingerprints). 
However, within our context, the user base is much more widespread and rapidly
increasing. Organizing such locally distributed events at particular times and expecting the whole user base to comply would be impossible. A distributed and accessible on-demand system is necessary to influence more people to verify themselves. Content filtration will be effective only when almost all of the legitimate accounts are verified. It is a challenge to re-organize the infrastructure of existing geographically distributed frameworks to integrate a system that verifies these three properties.\\[1ex]
Next, we discuss the challenges that a system must overcome for the attesting
and verifying the three properties.\\[1ex]
\noindent \textbf{Location + Uniqueness.}
Assuming a proof of location is for a mobile device, rather than a particular human being, then associating the proof of uniqueness obtained under such a condition, i.e.,
without the involvement of a trust anchor, is unreliable. The same person could
be simultaneously providing the same proof for multiple devices and there would
be no way to ensure user uniqueness without a trust anchor. Biometric signatures
of the owner (for proof of uniqueness) recorded through the user
device may not be as accurate as the device is under the user's control. Also,
it is easier to trick an app on the user's mobile device than in the presence of
a third party trust anchor.\\[1ex]
\noindent \textbf{Sentience + Location + Uniqueness.}
With CAPTCHA solving techniques and/or other hybrid sentience testing techniques, it is possible to distinguish a human from a non-human
as the device owner. This means we can possibly establish, while allowing a
considerable margin for error, that the device providing its own standalone
proof of location~\cite{wang2011towards} has a human owner in control, without a
trust anchor in a distributed system. However, to establish and associate a
proof of uniqueness to this human in control of the mobile device, a trust
anchor is still necessary. With the assumed threat model in our case, the lack
of inherent trust in the user only compounds the unreliability of the model
without any trust anchor.\\[1ex]
\noindent \textbf{Uniqueness and Privacy Requirements.}\label{uniqueprivacy} 
Verifying uniqueness is an essential component of any system aimed at solving
the Ghost Trilemma. Ideally, there should exist an element related to the user
that can uniquely bind them to their record without revealing their identity.
This element should not be forgeable or replaceable in any way, and the user
should not be able to trade it or deny its ownership. The closest option that
fulfills these criteria is a unique and anonymized biometric signature. This property is essential in enforcing a bound on each user's capacity to
perform unlimited attestations.

The aim of storing the biometric signature is not to link the user to their
account activities, but to ensure that under no circumstances can any user
associate proofs of location and sentience to accounts under the adversary's control. The challenge is in maintaining the
privacy of each user while also curbing the adversary's ability to possibly
corrupt any actor in the system. A rate limitation component needs to be
associated to every aspect that is related not only to the proof of uniqueness,
but also leverages the fact that the adversary has limited physical presence
inside the country (e.g., the US). Combining the effects of rate limitation while keeping entirely decoupled databases could only ensure privacy to an extent if the trust anchor
maintaining the data is trusted. 

\subsubsection{Selecting Trust Anchors}\label{trustanchortradeoff} 
Next we consider trusted third parties as trust anchors.\\[1ex]
\noindent \textbf{Trust Anchor Location Verification.} \label{trustLoc} 
To have the trust anchor verify location is to attempt to pin down a location
with a timestamp for the user via a trusted third party.  If a user can prove to
be at a point, it is prudent to allow for them to submit proofs of sentience and
uniqueness at the same point. With such a model, we can pick a fixed location
that a human user must visit and verify their presence through a sentience test,
while showing proofs of uniqueness (e.g., government ID, biometric ID, or even a finger yet to be dipped in ink). This would lead to a design option which adapts to the framework of a trust anchor that can provide a dense yet locationally distributed network.\\[1ex]
\noindent \textbf{Trust Anchor Sentience Verification.} If the trust anchor is designed to
perform verification of sentience, a possible model would require a user to
interact with some human agent, proving sentience through human contact and any
technological proof (e.g., CAPTCHA) or any verifiable action (e.g., receiving a
designated package and confirming it). Here, the human agent is a representative
of the trust anchor. Additionally, the proof of location can be obtained by
tracing the agent's location while they confirm the interaction with the user.
A framework which freely provides travelling human agents, such as mobile
notaries, to cover extensive geographical areas would be an ideal choice to
incorporate as a trust anchor in this case.\\[1ex]
\noindent \textbf{Trust Anchor Uniqueness Verification.} Verifying the uniqueness of a
user is oftentimes synonymous with collecting biometric information, though it need not be~\cite{borge2017proof}. Due to its sensitive nature, biometric information must be collected under some supervision and must be verified to establish uniqueness universally. 
When a design associates a verification of uniqueness to the trust anchor, it
often supports the proof of sentience (through biometrics) as well. The user's
location can then be linked to the location where the proof of uniqueness was
obtained. Choosing to integrate with the framework of a trust anchor which
routinely obtains uniqueness proof is harder because such frameworks are rare
and exclusive (e.g., high security jobs).  
It may not be necessary to use biometrics; a trust anchor could perform marking
in some way that is not associated with the person's identity
but nevertheless ensures uniqueness.\\[1ex]
\noindent \textbf{Ensuring User Credibility.}
Verification of these three properties of a user (sentience, location, and
uniqueness) may still not be enough to lend credibility to the user and their
data. To have a guarantee that these individuals are not somehow being
manufactured (Sybils), one can imagine adding other components of the user's
background (via address, credit history, and so on). This provides an extra
hurdle for the adversary to overcome (i.e., provide a mailable address / credit
history with each new identity they plan to verify through the system), but also increases the burden and potential unfairness of the verification system. Background verification acts as an anchor that can generate confidence in the user's history. After that has been established, the user can go forth with providing proofs for the three main properties.  

Background verification can also provide an additional proof of uniqueness and
allows for stricter rate limiting rules. In this case the procedure itself is
essentially serving as an anchor, linking the three properties to the social
structure they are a part of. However, this can also introduce negative
impacts. For instance, some techniques that can function as such an anchor
might often be related to verifiable residency proofs or credit
history. Practical, if imperfect and unfair, solutions of this sort have long
been used in the space of financial services. This raises an important concern:
due to the non-universality of such approaches, a significant portion of the
potential user base could be excluded from being seen as trustworthy, though in a decentralized setting users can decide for themselves which properties are essential.

\subsection{Formalizing the Intuition} \label{sec:ghost}
\label{subsec:trilemma}
As we noted before, the trilemma does not have a natural formal definition. Nevertheless, in this section we sketch both a formalization of the problem and a possible proof (one that makes overly strong assumptions, but perhaps a starting point for future work on the topic).

We consider how a verifier $\mathcal{V}$ can establish user sentience, location,
and uniqueness assuming the existence of a trusted party $\mathcal{F}$.
\begin{itemize}
    \item The account is operated by a human user and interaction with the
    system happens with the knowledge of the associated user (\textbf{sentience
    of the user}).
    \item The identity of this person leads back to a credible profile of
    existence, i.e., we want to know that the entity claiming the account is
    indeed a person at the expected location at the time (\textbf{location of
    the user}).
    \item A way to bind the accounts being attested to with an individual
    claiming to own said accounts, thus allowing for a limit on the number of
    verified accounts per individual. We want to establish the
    \textbf{uniqueness of the user} without necessarily deciphering the actual
    identity of the individual. One person cannot be allowed to verify an
    unlimited number of accounts. If that were allowed, the adversary's
    bottleneck of limited physical presence on US soil would be removed.
\end{itemize}
\subsubsection{Trilemma Proof Sketch} \label{sec:ghostdesign}
We assume a verifier $\mathcal{V}$ can establish user sentience, attribute location and confirm uniqueness assuming a trusted party $\mathcal{F}$. 
\theoremstyle{definition}
\begin{definition}[Attributing Sentience] \label{def1:sentience} $\mathcal{V}$
 checks if the user is a sentient being who can pass the Turing test, i.e.,
 express their capability to understand and propagate multi-dimensional
 ideologies and actions at $\mathcal{F}$. For example: $\mathcal{V}$ checks if
 user can solve a CAPTCHA (a reverse Turing test) at $\mathcal{F}$.
\end{definition}
\begin{definition}[Attributing Location] \label{def2:location} $\mathcal{V}$
computes user location using existing IP localization techniques that leverage network latencies to and from landmarks~\cite{wang2011towards}. 
\end{definition}
\begin{definition}[Attributing uniqueness] \label{def3:uniqueness} $\mathcal{V}$
maps user identity with an unique governmental identifier. For a $k$-shot user
operation, the identifier is checked $k$ times for preventing Sybils. 
\end{definition}
We now show that such a verification cannot be achieved without the trusted component. 
More specifically, it is impossible for an external verifier to deterministically attribute sentience,
attest user location, and perform privacy-preserving identity verification in a
decentralized system.

\begin{theorem}[Ghost Trilemma] \label{theorem:ghost} Consider a verifier
$\mathcal{V}$ and an asynchronous decentralized system with permissionless
participation for its participants. Then, without a honest majority of
participants, it is impossible for $\mathcal{V}$ to confirm sentience, location,
and uniqueness in finite time with respect to $\mathcal{F}$. 
\end{theorem}
\vspace{-1em}
\begin{proof}
Suppose by contradiction that there exists a verifier $\mathcal{V}$ that can
attribute sentience, user location, and perform identity verification in a
finite amount of time as specified by $\mathcal{F}$. The proof argument is
derived by showing that in order for $\mathcal{V}$ to attribute sentience, user
location, and perform identity verification in a decentralized system,
\emph{consensus} (on the \emph{state} of the decentralized system) will have to
be achieved by all the participants in \emph{any} verification protocol.
Moreover, consensus must be achieved in finite time independent of the relative
speeds of the participants: so-called \emph{wait-free} consensus~\cite{FLP85}. 

We argue that verifier $\mathcal{V}$ must \emph{deterministically} reach
consensus among the participants of the verification protocol to attribute all
three: sentience, location, and privacy-preserving uniqueness in an asynchronous
decentralized system. We build a layered argument to make the case that
$\mathcal{V}$ must take inputs from an arbitrary number of \emph{witnesses},
some of which may be byzantine, to resolve the Ghost Trilemma. Without loss of
generality, assume that there exist $n \in \mathbb{N}; n>1$ participants in the
system at any point in time. Specifically, consider an execution (at time $t$)
of a verification protocol in which:
\begin{enumerate}[wide, labelwidth=!, itemindent=!, labelindent=0pt]
\item Verifying sentience without $\mathcal{F}$ requires $\mathcal{V}$ to
implement a decentralized byzantine-resilient CAPTCHA protocol in a wait-free
manner. Any correct CAPTCHA must have a sequential specification that addresses
invariant recognition, segmentation, and parsing capabilities for a
user~\cite{captcha}. Enforcing this sequential specification in a consistent
manner across an asynchronous decentralized environment involving distributed
participants implementing the CAPTCHA protocol necessitates \emph{agreement}
among the participants. However, if one of the participants in the CAPTCHA
protocol is byzantine, it is possible to \emph{equivocate} on output of the
CAPTCHA responses thus causing valid user responses to be invalidated.

\item Now assume that $\mathcal{V}$ is able to verify sentience from the
participants in the sentience verification protocol. Let $u_0$ be the response
of the sentience verification protocol returned by the participants of the
uniqueness verification protocol. Consider any protocol for $\mathcal{V}$ to
verify uniqueness without leaking private information about the user to other
participants. Such a protocol employed by $\mathcal{V}$ must distinguish an
adversary creating multiple identities for the same application, in this case,
the identity for the sentient user $u_0$. This requires verifying the sentience
of each of the potential Sybil users. Consequently, verification involving
participants in the uniqueness protocol must validate the output $u_0$ and then
reach agreement on whether $u_0$ is unique. However, distinguishing the
execution in which $u_0$ is unique and $u_0$ contains multiple Sybils requires
the protocol participants to reach agreement.

\item Verifying location for $\mathcal{V}$ constitutes the verification of
whether the associated geographical landmark is actually at the given location.
Without a centralized trusted functionality $\mathcal{F}$, this data will need
to be gathered from disparate sources, then applied with the verification
techniques for sentience and uniqueness identified above.  
\end{enumerate}
Independent of whether the protocol participants are mutually disjoint,
$\mathcal{V}$ must reach consensus on the overall responses with the
participants in the decentralized system to attribute user sentience, location,
and uniqueness. However, reaching consensus in finite time when some of the
participants are byzantine requires a \emph{wait-free} agreement
protocol---contradiction to the impossibility of consensus~\cite{FLP85}.
\end{proof}

\section{Prototype Framework Design} \label{sec:designexample} 

Given that the Ghost Trilemma cannot be solved in a decentralized setting, we
now briefly present an example system design that leverages point-of-sale (POS) devices
used in a physically distributed chain of businesses (e.g., Starbucks,
McDonalds, etc.), grounding the verification of location in space and enabling
appropriate binding of the three properties.
\footnote{We implemented the key components of this system to verify their practicality, but \emph{do not} evaluate them in a conventional manner as the design choices we made are more relevant to the questions at hand.}
Our aim was to align the verification of the three properties with daily
routines common among ordinary users, such as going to buy coffee or fast food,
and leveraging the geographically distributed nature of such businesses and the
pre-existing security required (e.g., POS devices handle transactions and thus
businesses are inherently incentivized to keep them secure). 

\subsection{Entities}\label{entities} We narrow the scope of our discussion to
the verification of users within the United States that have an online presence
(i.e., valid social media logins), a verifiable identification, and at least one
physical mailing address.
The entities involved in the proposed design of \textsc{POS} scheme are as
follows:
\newline
\textbf{User:} Any US-based user who uses social media. 
\\
\textbf{Trust anchor:} A trusted third party, such as a store or a hardware
device with an immutable location (or satellite-verified coordinates), with a close range communication capability. 
\\
\textbf{Back-end server and datastores:} Responsible for performing the protocol
through the trust anchor/third party (i.e., mail service).\\
\textbf{Remote state-level adversary:} 
% The adversary can act in many ways, such
% as with corrupt human agents controlling the verifier, corrupt users who may be
% working with the adversary, or through remote activity via compromised devices
% or communication channels. 
The adversary has limited physical presence in the target country but otherwise capable of corrupting any component involved in this design.

\begin{figure}[t]
\centering
    \includegraphics[width=\columnwidth]{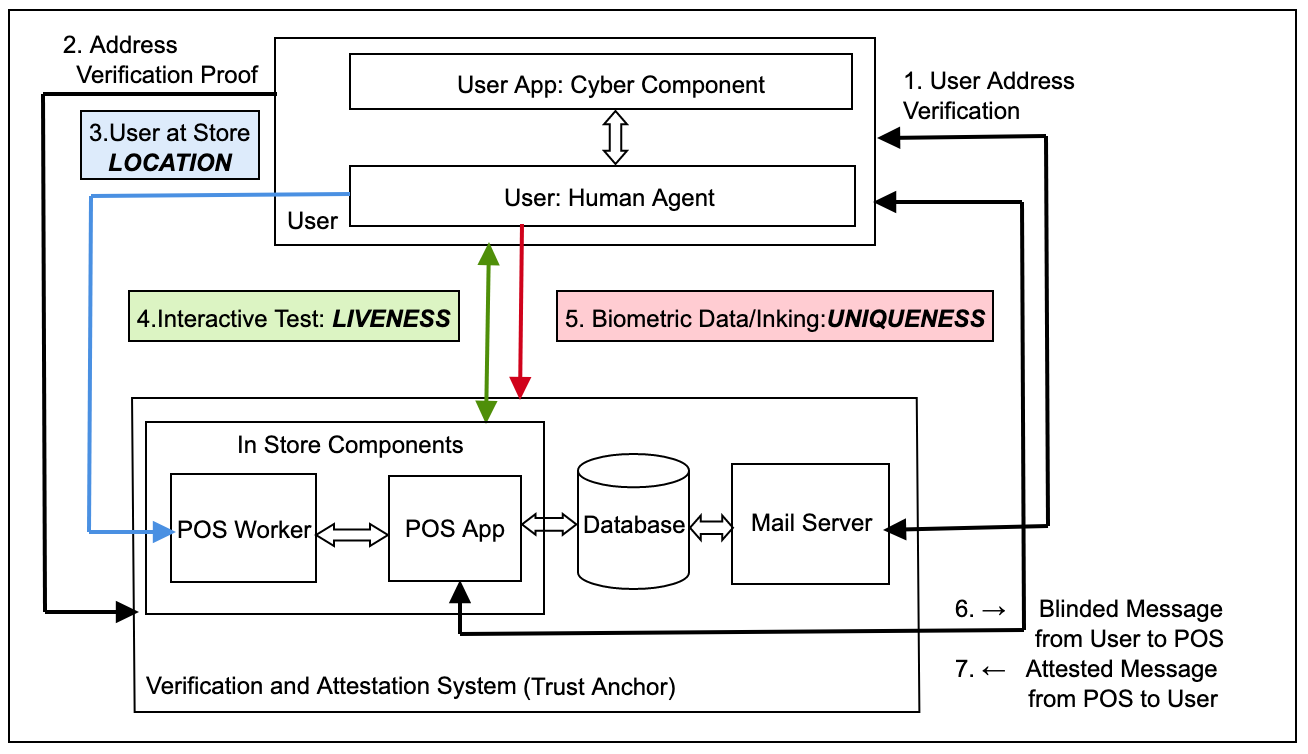}
\vskip 0em
\caption{Point-of-Sale verification design sketch.}
\label{fig:pos-design}
\vskip -1em
\end{figure}

\subsection{\textsc{POS}-Based Scheme}\label{Design1} 

Establishing sentience straddles two approaches: verification via technology and
verification via human interaction. Both are needed to avoid overdependence upon
any one. If the technological system is too complex or too simple, an adversary
could find a weak link to exploit or as we know from~\cite{duplexai,diar_2023}, artificial
intelligence has already beaten the Turing test within some conditions;
conversely, relying entirely on human-to-human verification could give way to
corrupt employees or human error. In this scheme, we use chain businesses as the
trust anchor, leveraging a Point of Sale (POS) app (controlled by a store
employee) that interacts with the user's mobile device to produce proofs of
sentience, location, and uniqueness. Figure~\tsref{fig:pos-design} shows the
high-level design. 

\paragraph{\textbf{Assumptions}}
In our example, we only consider localities that have the required verification
centers (acting as the trust anchors) which can provide a fixed location for the
user to reach (thus providing a definite proof of location). We do not consider
bringing the trust anchor to far-flung localities, though in the case of many
chain businesses this is not an issue within the United States.\footnote{For
example, it is not possible to be more than about 110 miles from a McDonalds
within the continental United States, the so-called ``McFarthest Spot.''} We
also assume that user devices have NFC capabilities and can support the
installation of social media applications. 

In this design, we rely on the trust anchor primarily to verify the location.
We considered the tradeoffs discussed in \tsref{uniqueprivacy} and \tsref{trustanchortradeoff}. One of these tradeoffs concerns the rate limiting
restrictions through the user's mailing address. This may exclude legitimate
users who also live at a particular location but may not become verified due to
a per-address limit. Here, we chose to sacrifice inclusivity in favor of
creating fewer loopholes for the adversary to exploit.

The protocol involves three stages: 1) a user's own address verification;
2) verification at a POS; and 3) intelligent content filtration via a browser
plugin. The design incorporates two different applications. One is the user
application present on the user's smartphone while the other is an application
on the POS device, controlled by a store employee. The POS acts as a
verification point, inherently providing a proof of location while obtaining and
verifying proofs of sentience and uniqueness. The user requests to verify their
physical mail address through their mobile application and brings the proof of a
successful verification to the nearest POS verification center, which is located
and operated in a chain store. The store employee verifies the proof of
sentience and uniqueness of the user through a protocol (explained in \tsref{protocol_description}), signs the social media handle(s) using a blind group
signature scheme~\cite{lysyanskaya1998group} and adds the record to a centralized data store. Users' blinded
biometric information is stored as a way to rate limit the number of
verifications and attestations allowed per person. 
The user can subsequently un-blind the signed message and publish it online (on
the specific media platform) for all other users to view. A customizable browser
plugin can be designed to facilitate this process. 

%%%%%%%%%%%%%%%%%%%%%%%%%%%%%%%%%%%%%%%%%%%%%%%%%%%%%%%%%%%%%%%%%%%%%%

\section{POS-Based Scheme}
\subsection{Protocol Description}\label{protocol_description}
In this section, we describe the POS-based scheme to understand how we verify each property.
\subsubsection{Location Verification}
The proof of location is obtained in two parts: primarily through the user's
physical presence at the POS and secondarily through the limited time validity
of the  QR code obtained from user address verification. The user can start
the address verification by placing a request through the user application. Each user
account agrees on a unique key while registering for the first time with the
system. The user application generates a random number $n$ (not visible to the user),
and signs it under its key $A$. The signed number (i.e., $(\sigma_A(n), n)$) is
received by the system along with the address which is to be verified as entered
by the user in their request. The system checks the address, signs the received
message with its own key $U$, and generates a QR code which contains the
message $(\sigma_U(m), m)$.  A physical copy of the QR code is mailed to
the address provided by the user. 

The user then scans the mailed QR code and checks its authenticity through
their application. The QR code will be valid for a limited period of time to
prevent any redirection of the mail to another address. The user must then
physically go to a store to complete the POS verification. The store
acting as the verification center is a third party with an immutable location
and hence provides an inherent proof of location when the user successfully gets
attestation from it.

\subsubsection{Sentience Verification} \label{POS verif} Next the user scans and the application
authenticates that the QR code originated from the current user account.
Only then can the user access the next stages in the application. Next, the employee
controlling the POS device (POS worker) will start the in-store verification
steps by scanning the QR code to verify the validity of the mail server's
signature. The employee then cross checks the user's proof of residency and
identity. If these are correct, the following steps are taken to obtain a proof
of sentience from the user:
\begin{itemize}
    \item The POS application starts the process by generating a random number $n$ and
    signs it with its private key $P$, producing $(\sigma_P(n), n)$; it
    then pushes this to the user application  via NFC. The number $n$ is not
    accessible by or visible to the POS worker.
    \item The user application receives the signed value via NFC, and decodes the
    message to show $n$. The user reveals the number $n$ which the POS worker
    then enters into the POS application. If it matches the original number, the POS application
    considers it as verified. If numbers do not match, the process will be
    restarted with a new random number. 
\end{itemize}
This real time short-range communication based interaction between the user
device and the POS in the presence of a human agent indicates that the device is
being controlled by a live human being, thus providing the proof of sentience.
The next step, which is intended for obtaining proof of uniqueness through
biometric data from a user, can also serve as an additional proof of sentience. 

\subsubsection{Uniqueness Verification} \label{POS verif-unique} The next step is
biometric data verification. The user is prompted by the store employee to
either have their fingerprint scanned or get an invisible/indelible ink
stamp.\footnote{We consider both options as ink stamps allow users to get
verified; we anticipate some users will be wary of providing such personal data
to any system.} If the fingerprint is used, it is hashed and stored in the centralized database along with a
timestamp and location. If the user chooses ink, then the POS application is prompted to
generate a unique ID for the user and a corresponding QR code is given to the
user to present during their next verification. The user is marked on skin with
the ink.\footnote{This can be done in a safe and cheap manner, and does not
require esoteric or toxic chemicals to be applied to the skin. For example,
banana plants produce sap that appears clear when wet but dries an indelible and
long-lasting dark brown while being hypoallergenic.} The user record is updated
accordingly to reflect this information. Together, the biometric data and address verification records provide a proof of \textbf{uniqueness} of
the user. The POS (\textbf{location}) can attest to the fact that a real human
operating a certain account (\textbf{sentience}) was present at specified
location and at a particular time. 

The user can sign one message for each social media platform during one round of
verification, through separate NFC transactions. If the user wants to verify
more than one account on the same platform, the whole verification process
address verification and POS verification will have to be
repeated. The user passes the blinded message ($<$account user-handle$>$),
signed by the user application, to the POS application via NFC. The POS application appends
($<$location$>$, $<$timestamp$>$) to the message and signs the whole message and
transmits it back to the user application via NFC. The user's record is updated
accordingly in the database. The user application stores the signed message(s) and
publishes them online (after unblinding the message) from the application into the
respective accounts. When the user wishes to re-verify the same account, the application
will overwrite the previously signed message with an updated time and location.
However, if the user wants to verify a second account on the same platform, they will
have to create a separate message and repeat the address verification and
POS verification steps.

\subsubsection {Filtering Content}
Once the POS verification is completed successfully, the user receives the
attested copy of the message. They can then unblind the message and publish it
with the digital signature on social media. The message contains a username,
location, and the time of attestation. A browser extension is
also required to complete the process. It reads the published signed information
from the user's social media account and verifies the signature to calculate the
time of last verification from the published timestamp. In this way, users will
be able to filter content they will see in their feed. Depending on the age of
attestation of different accounts, users can pick and choose time ranges whose
content they will see.

\subsubsection{Underlying Design Details} 
Both POS and user applications should be designed such that they cease to function immediately if they detect a compromised device. This theoretically rules out the possibility of the adversary gaining remote control of the devices involved in the verification protocol. Bypassing this restriction will allow the adversary to dictate the protocol operations entirely while the user or POS operator remains oblivious. 
  
As a part of the uniqueness proof, the scheme obtains biometric data from the user. A downside of using government issued IDs instead is the ease with which anyone can obtain almost any type of fake IDs online. A fake ID and a separate mailing address would allow people to conjure up multiple identities and avoid the rate limiting altogether.  However, this would be more difficult when biometric data is involved, thus providing a better guarantee of rate limitation. IDs and address proofs are verified by the human employee at the store during POS verification.

Each verification center is immediately added to a group as soon as it is registered. Upon installation, the POS device will be setup with keys according to the blind group signature protocol~\cite{chaum1983blind} that has been selected.  Ideally, the store proprietor confirms the location coordinates of the store upon installation which is recorded as the home location of the POS device. A background process constantly checks if that location ever changes, in which case it throws an error.  
Upon installation of the user application, the user's device is registered and a unique key (Universally Unique randomly generated ID (UUID) \cite{AndroidDev2019}) for the device is set up and shared with the system. The user will require to sign into all the social media accounts that they want to verify through the application. The user application allows addition of only two accounts per platform per user. The message corresponding to one particular user account, which will be signed at the POS end, can only be generated upon a successful login to that account on the user device. 
Through one NFC tap exchange of information, only one account can be verified. 
Once a particular platform's account has been attested to, the whole procedure needs to be repeated (from requesting address verification to the final in store verification) to verify a second account on the same platform. In one round (starting from address verification request to in-store signing of the message) one account from each type of social media platform supported by the application can be attested. This is done to keep a count of how many account of each type is associated to each individual's record.

\begin{table*}[t]
\begin{adjustbox}{width=0.9\linewidth,center}
\def\arraystretch{1.5}
\begin{tblr}{|m{0.42\linewidth}|m{0.2\linewidth}|c|c|c|}
\hline
\SetCell[r=2]{c}\textbf{Scheme Description}&\SetCell[r=2]{c}\textbf{Name}& \SetCell[c=3]{c}\textbf{Attributes Accounted For} & &  \\
\cline{3-5} 
 && \textbf{\textit{Location}}& \textbf{\textit{Sentience}}& \textbf{\textit{ Uniqueness}} \\
\hline
 POS based verification at a fixed store location with address verification involved through a mail server &  \textsc{pos scheme}&  \fullcirc & \fullcirc &  \\
\hline
Biometric identification implemented at a fixed location (trusted third party collects the information)& \textsc{biometric fixed}& & &\fullcirc  \\
\hline
Satellite imaging & \textsc{sat image} &\fullcirc&& \\
\hline
Address verification via USPS+ID card verification &\textsc{usps} & \halfcirc& \halfcirc & \halfcirc \\
\hline
NFC or other close range communication involving user-verifier in-person interactions & \textsc{close range comms} &\fullcirc&\fullcirc\\
\hline
Credit card verification + Residency proof and ID &\textsc{credit card}  &&&\halfcirc\\
\hline
Any ID card, license based validation &\textsc{card validation} &&&\halfcirc \\
\hline
National ID cards, e-passport based validation at a fixed location & \textsc{RFID scanning validation}& \fullcirc & \fullcirc& \halfcirc\\
\hline

\end{tblr}
\end{adjustbox}
\caption{\centering{Design choice alternatives for the POS Scheme and the proofs they obtain.}} \vspace{-2ex}
\fullcirc: Fully satisfies the requirement by itself, \halfcirc : May partially satisfy the requirement 
\label{tabPOSdesignchoices}
% \vskip -2em
\end{table*}

\subsection{Alternative Design Choices}\label{Design2} 

Our proposed POS-based design is not easily deployable for large scale testing
as it would require the cooperation of several large organizations. However, it
is modular with respect to the three properties. In this section we look at
possible alternative solutions that could be used to replace parts of the scheme
or be integrated with it to strengthen its functionality. Table~\tsref{tabPOSdesignchoices}
lists some of the possible alternative design choices we consider. 

\subsubsection{Location and Sentience}
Instead of depending on interaction with the POS application, we can also consider the
use of RSA secureID hardware tokens (e.g., one token could be registered to the
group to which the POS store would belong for the group blind signature scheme)
for proof of location. The hardware token will not be susceptible to attacks
aiming to alter its location as is possible in case of the POS application through its
network connectivity. When considered along with an implementation of
\textsc{Biometric Fixed} in Table~\tsref{tabPOSdesignchoices} (e.g., voice recording), this
could provide a binding between the user's sentience and location. The recording
could be matched against previously recorded information by the user (if any) to
verify uniqueness. It will also match the timestamp of the recording done at the
store/hardware ID location and match the time when the code from the hardware
RSA token was entered, allowing them to be simultaneously verified. The location
of an user can be verified through multiple means ranging from satellite imaging
(by being at some predefined location at a specific time) to in-person receival
of registered mails via \textsc{USPS}. The design in~\tsref{POS verif} uses a
simple sentience check coupled with the usage of \textsc{close range comms}
(Table~\tsref{tabPOSdesignchoices}). However, there are many other CAPTCHA designs to use. Any
scheme involving RFID scanning of national IDs or e-passports although satisfies
all three properties, has a more exclusive user base and is by itself a
completely centralized solution both in terms of issuance and
verification.

\subsubsection{Uniqueness}
To verify uniqueness, another option would be to use post offices rather than
chain stores, which guarantees that a third-party verifier (i.e., USPS) will be
present in every neighbourhood in the country. Using the restricted and
registered mail delivery service from \textsc{USPS}, we can ensure the mail is
received  and signed by the addressee. While it will be easier to rely on an
already widespread and established mailing system, it will have inclusivity
issues (e.g., a signer must be at least 21 years old). We can consider another
form of address verification through \textsc{credit card}s. In the US, when
credit cards are issued to a user by a US based bank, it involves a background
check and is linked to a social security number as well as an address. When
combined with an identification check of the user at the store's POS counter and
the user's biometric data/inking (following the design choices of the
proposed design example) this could replace the complex address
verification process. This may be unavailable to people without access to a
credit card or who have separate billing and residential addresses. In addition,
it is easy to obtain stolen credit cards and IDs online from deep-web
marketplaces for as little as $\$25$ each. This makes bypassing the verification
system and getting unrestricted number of accounts attested easier. Using
multiple options in combination (like \textsc{card validation} and
\textsc{credit card} from Table~\tsref{tab:costAdversary2}) could increase the cost to the
adversary, but it can still be bypassed.  

\subsubsection{Other Biometric Options} Although, the most efficient alternative is to use the biometric based option, it could feel intrusive to some users. An alternative discussed in \tsref{protocol_description} was using indelible ink stamps or invisible ink stamps on user's skin. There are, of course, ways to bypass the ink based check, which is not as reliable as the biometrics. However, it is an inherently privacy preserving option. 

Another biometric information based option, besides fingerprints, would be using voice based authentication. It might exclude a significant number of users who suffer from variations of speech based problems. However, it could be used in conjunction with fingerprints or permanent ink to implement the  rate limiting  constraint. Using a user's voice as an identification suffers from certain defects like accuracy of matching. It is dependent on the user's behavior while recording, can be affected by background and environmental noises and is  also dependent on an individual's voice changing significantly with time. Works like \cite{li2017deep} show that using neural networks and deep learning (compared to previously used architectures involving hidden Markov models) has been useful in reducing the error in speaker recognition systems. However, we see the currently available best (accuracy-wise) implementation of speaker verification techniques using DNN embedding~\cite{snyder2018x} has the best recorded accuracy of only $91\%$.

%\section{Evaluation}
\section{Analysis}\label{sec:analysis} 

We now examine the features and design choices in the prototype POS scheme
within the framework of the Ghost Trilemma and its possible attack vectors. We
then perform a cost analysis for an attacker given our design.

\subsection{Trilemma Analysis}

By itself, any one of the three trilemma properties is insufficient to defend
and filter out a determined adversary. Location can easily be faked, and without
sentience being bound to a unique (human) user, the system can provide no
guarantees. CAPTCHAs have been widely used to prevent bot infiltration, but many
troll farms employ a large human network to support their troll accounts. Even
if a CAPTCHA can perfectly thwart all bots, human employees can validate their
sentience. Hence, sentience, location, and uniqueness proofs for a user must be
bound simultaneously.

\paragraph{Location.}
Location consistency is assured by the POS applications's real-time monitoring (checking
that its location matches pre-defined coordinates). If there is an attempt to
change it to (and thus attest to) a different location, if the application does not
recognize a match, it raises an alert flag, prompting immediate action from the
controller. If the device is moved, re-calibration and re-registration of the
device is required.

Specific interactions with such a POS terminal will provide a proof of location.
The timed validity of QR codes received through address verification
also adds to the proof of location by ensuring that the QR code was read at a
certain location. 

\paragraph{Uniqueness.}
Rate limiting the ability of an user to verify accounts is accomplished by
checking uniqueness during the verification process. Address verification
is troublesome since it is difficult to find a large number of permanent mailing
addresses while requiring a user to provide proof of residency. However, as we
will discuss in \tsref{adverCost}, this increases the cost to the adversary.
Biometric information, combined with address verification, should provide a
workable system that binds each person's ability to verify accounts. The
adversary will need to create or steal identities and also arrange for
corresponding biometric data. This would ensure an increased level of
difficulty for the adversary. Adding a human verifier component through store
employees checking the QR code will add an interesting dynamic, as they would
cross-verify the address at which the QR code is received with the ID and
proof of residency presented by the user along with the QR code. While employees
can be bribed to overlook certain discrepancies, it also induces a lower
incentive to cheat on the user's side due to the unpredictability of the
employee's behavior and the public setting.

\paragraph{Sentience.}
The use of NFC for transferring messages between the POS application and the user application
strengthens the claim that the user was present in the store physically near the
POS terminal. The sentience test does not allow the operator to know the
code generated without involving the user. Thus a store employee (human)
confirms that there is indeed a person standing in front of the POS terminal
going through the verification process. This sociotechnical process yields a
proof of sentience. The honesty of the operator is not assumed in this state,
and they could behave corruptly to verify on the behalf of a friend. They can do
so within the limit of accounts they can have verified and attested. Any more
than that will not be permitted because of the biometric restrictions ensuring
uniqueness. From the user's perspective, a design where the QR code will
only be working for the device that the request originated from (within a
limited time window) will contribute to the sentience and location contexts. The
limited time window after the QR code's scanning will only allow the user
to verify within a certain local range.

The signed message includes the location and timestamp to allow the account to
be tied to a general location, independent from the platform's own features. The
attestation token available for publication is non transferable, which
contributes to the proof of location. The blind group signature on the part of
the POS terminal application aims to achieve two different types of privacy. Blindness
helps to disassociate the user from their online identity (which can be
exploited by the adversary) and the group signature does not pinpoint the exact
location of the store (and thus that of the user). Any verifier should have
access to check against all the user records and biometric data to prevent
a dishonest user from attacking the geographical bounds in hopes of
circumventing the uniqueness property. The anonymization (to the extent possible
while meeting the required level of performance of the system) of the
biometric data is necessary on several counts, primarily to prevent
possible linkage attacks that would associate the person's online identity with
the offline one. We leave studying the feasibility of this attack to future
work.

\begin{table*} [htb!]
\begin{center}
\resizebox{16cm}{!} {
\def\arraystretch{1.5}
\begin{tabular}{|m{0.42\linewidth}|m{0.2\linewidth}|m{0.42\linewidth}|}
% \begin{longtable}{|p{0.5\linewidth}|p{0.5\linewidth}|} 
% {|c|c|c|c|c|c|}
\hline
\makecell[c]{\textbf{Feature}}&\centering{\textbf{Associated Property}}&\makecell[c]{\textbf{Relation}} \\
%\cline{3-6} 
\hline \hline
A consistent check on location in POS application (run in background) &\textsc{location} & Prevents unnoticed alteration in location of the POS \\
\hline

Limited number of accounts can be verified per address verification request & Rate limiting, \textsc{uniqueness} & 
Mandatory requirement of traceable background information for each identity (Sybil or not)\\
\hline
POS verification: The code during sentience verification is only available to POS employees through user interaction  &\textsc{sentience} & Ensuring proof of sentience via human-human interaction  \\
\hline
POS verification: Verifying QR code is signed by mail server and has been validated by User application at POS &\textsc{sentience}, Rate Limiting & 
Ensuring the registered address of the user’s device is a match with the address attested by the QR code \\
\hline
Near Field Communication (NFC) is always used for communications between the POS and the User application &\textsc{sentience} & 
Ensures physical presence of the device handled by a human at the POS \\
\hline
Once the User application validates a QR code, a limited time window is available for the user to do a POS verification &\textsc{sentience}
 & 
This ensures the device requesting the verification is essentially validating it's request in front of another human (a store employee)\\
\hline
Storing user biometric data
&\textsc{uniqueness} & Provides proof of uniqueness. Linkage attacks might still be possible despite using privacy preserving measures 
\\
\hline
Alternative to biometric data: Indelible/invisible ink mark on skin &\textsc{uniqueness} & A less intrusive but a much weaker alternative to biometric data based option\\
\hline
A carefully synced in real time data management and storage system, especially for biometric data
&\textsc{uniqueness} & Prevents usage of the same Sybil identity across geographically separated group boundaries \\
\hline
\end{tabular}
}
\caption{Linking Features in \textsc{POS} Scheme to Location, Sentience or Uniqueness}
\label{tabLocLivUniFeature}
\end{center}
\end{table*}
%%%%%%%%%%%%%%%%%%%%%%%%%%%%%%%%%%%%%%%%%%%%%%%%%%%%%%%%%%%%%%%%%%%%%%%%%%%
%%%%%%%%%%%%%%%%%%%%%%%%%%%%%%%%%%%%%%%%%%%%%%%%%%%%%%%%%%%%%%%%%%%%%%%%%%

\begin{table*}[htb!]

\begin{center}
\resizebox{16cm}{!} {
\def\arraystretch{1.5}
\begin{tabular}{|m{0.4\linewidth}|m{0.2\linewidth}|m{0.4\linewidth}|}
% \begin{longtable}{|p{0.5\linewidth}|p{0.5\linewidth}|} 
% {|c|c|c|c|c|c|}
\hline
\makecell[c]{\textbf{Feature}}&\centering{\textbf{Associated Property}}&\makecell[c]{\textbf{Relation}} \\
%\cline{3-6} 
\hline \hline
Either application should not work on any rooted device & 
\textsc{Location, Sentience, Uniqueness} & Prevents the chance of remote interference \\
\hline
Address verification: Limiting the validity period of QR code received through mail & \textsc{location} & 
Prevents using outdated QR code, indirectly ties the location of the user to a locality \\
\hline
Address verification: 
    Verification of physical address by mail server and capping the number of verification requests per physical address
& Rate Limiting, \textsc{uniqueness}
  & Lends more strength to the uniqueness proof since its troublesome to gather unlimited number of verifiable physical addresses.
  \\
\hline
Verification of user’s ID and proof of residency (for address verified via QR code) by human store employees &  Rate limiting, \textsc{uniqueness} & Additional check to ensure the consistency in the background information of users\\
\hline
Moving to the next stage to start POS verification is only allowed on a device once the QR code is verified by the User application on the same device  & None & 
Ensures the device storing the messages is the one that generated the original verification request \\
\hline
Signing-in to the social media accounts is the only way to generate the message(s) & None & Ensuring the user accepts the ownership of the accounts being verified \\
\hline
The message must include the user handle of the account along with the timestamp and a general location  &  \textsc{uniqueness} & The attestation cannot be used for any other account \\
\hline

Failure at any stage of the address verification will result in starting over & None &
To enforce proper address verification, hence indirectly relevant to proof of uniqueness \\
\hline
POS uses blind group signatures \cite{lysyanskaya1998group} for attesting the user's messages after adding the location (generalized) and timestamp to it & Anonymity, \textsc{Location} & Blindness property ensures user privacy by dissociating the user's online and offline identities; group signature allows a generalized location stamp \\
\hline

\end{tabular}}
\caption{Other Features in \textsc{POS} Scheme}
\label{tabOtherFeature}
\end{center}
\end{table*}

\paragraph{Vulnerabilities.} There are some issues which arise from the proposed design's
intricacies. The usage of biometric information in our design is unlike the common use-case of biometric information. Each fingerprint needs to be compared multiple times to ensure a low probability of error in matching. This can be a hassle in some situations. When the user opts for the inking option, the address verification part of the protocol would become a significant component in enforcing rate limitation. However, the inking option places an inherent trust on the user. Thus it can create a weak link in the chain for the adversary to target. Another area this framework severely lacks in is ensuring user privacy. This relates back to the trade offs discussed in \tsref{uniqueprivacy}. 
Although with the stored biometric information, one should not be able to reconstruct original identities of the users, there is still a definite possibility of exploiting the user data through successful linkage attacks. However, this is a speculative guess since we did not try to launch and verify the success of such an attack in this case. As seen from the real world examples discussed in \tsref{sec:tracinginlit}, it is a huge responsibility for any centralized agency to be entrusted with such a bulk of sensitive user information.

Ultimately, the adversary still remains in a position to create false
identities and employ people to assume said identities with enough information
to satisfy the requirements of the \textsc{POS} scheme discussed above. Next, we analyze cost estimates for based on our prototype design.

\subsection{Cost Analysis} \label{eval} 

Here, we provide a simple estimate of how much it would cost the adversary to
maintain the expected volume of traffic.

\subsubsection{User-facing}\label{userCost} 

The user must maintain a verifiable mailing address and should be willing to
travel to the nearest in-person verification location and be in possession of a
phone.  
Based on the chosen protocol design, if the user is required to send any
mail(s), a corresponding charge may apply. If the user does not have valid ID or
proof of residency, the cost of legally obtaining those may involve yet another
expense. An integral part of this protocol requires the user to have a smart
phone that can install and operate the social media applications and user application for
verification. 
We assume every user participating will be in possession of a smartphone with
NFC capabilities.  

For any user with a generic smartphone and a valid physical mailing address with
proof of residency, the process should impose little extra cost or effort.
Unfortunately, this is not true for all legitimate and rightful users who may wish to be verified
(\tsref{trustanchortradeoff}).

\subsubsection{Adversary-facing}\label{adverCost} 

In estimating the cost for an adversary who wants to circumvent the system, we
make the following assumptions:
\begin{itemize}
    \item We consider a linear model: the adversary will have to spend the same
    for each user it attempts to validate. 
    %This would most likely not hold in a real world scenario.
    \item We do not consider infrastructural breaches that could reveal data
    directly to adversaries without cost.
\end{itemize}
Thus far, we are yet to devise a method suitable for simulating such nonlinear
real-world cost to an adversary. It is hard to predict the steps the adversary
may take or the possible weaknesses in the protocol that they may detect and
exploit. Further, any end-to-end solution will involve several components, any
of which could be broken differently. There is also the possibility of the
adversary discovering a weak link in the protocol that could decrease their cost
exponentially. However, since all these are unpredictable, we focus on a linear
model of cost calculation.

From the Senate report on IRA related investigations, we know the IRA's
operational cost to be around $\$1.25$M USD per month. They spent a meager
$\$100,000$ USD over advertisements on Facebook over two years (as reported by
Facebook). The free traffic (not ad-based), involved about 61,500 Facebook
posts, 116,000 Instagram posts, and 10.4 million tweets over two years. This
turned out to be more harmful than the ad campaigns on the platforms. 

The average Twitter user tweets twice a month; a more prolific user has 138
tweets per month; we assume 70 tweets per month for any user. 
%A user can verify a maximum of 2 accounts per social media platform. 
We limit accounts to 2 per individual user (a business account and a personal
account) for any given social media platform. Let us consider a single platform
for the rest of the analysis, Twitter. Going by the above restrictions, to hit
the goal of 0.43 million tweets a month (i.e., about 10.4 million tweets in two
years~\cite{senatereport}):
\begin{enumerate} 
    \item \textbf{Case 1:} % should use either 6190 or 6200 but not both
    The adversary would need to employ about 6200 individuals. Here we assume
    that each individual only verifies one Twitter account for the adversary,
    and uses the other for their own personal account.
    \item \textbf{Case 2:} The adversary would need to employ about 3100
    individuals. This assumes the adversary has created false identities from
    scratch and that these new users will use both their accounts for furthering
    the adversary's goals.
\end{enumerate}

\paragraph{Case 1.} The adversary may look to lure legitimate individuals (who
have valid ID and proof of residency) looking to make money. Users who have a
single account verified would likely be targeted by such ads. Targeted ads
towards this particular set of users would cost about \$5000 USD per month
(going by previous ad budgets reported by Facebook). We will consider a payment
of \$1000 per verification for each individual based off on average salaries of
people in the US (Table~\tsref{tabCost}). This would amount to \$4000 USD per
user per month with the verification frequency at once per week. This
cumulatively adds up to about \$25 million USD per month for the expected number
of users. The only challenge for the adversary is in persuading enough
legitimate users to collude with them. The variations in verification techniques
as discussed next do not have any impact on the cost estimates mentioned in this
case, though our numbers are rough. Here the adversary essentially piggybacks on
legitimate users of the system. 

\paragraph{Case 2.} We assume these are otherwise legitimate users given a new
identity and address/credit history (Sybils).\\[0.5ex]
\textbf{Baseline:} 
The baseline case of verification is when only sentience, location, and
uniqueness (via biometrics) are verified. The user ID is only checked at the
counter where some steps analogous to POS verification occurs. The
adversary only needs the same number of biometric identities as is indicated for
case 2.  Since we don't associate the biometrics with user identities, the
biometric ID of anyone in the world can be used. This may or may not involve any
cost on the adversary's part. The only measurable base expense would be
procuring the IDs once for about \$0.6 million USD for the required number of
users. The rest will have a cost (like payment to the agents who would go
through the verification process, but we cannot estimate the number of such
agents needed); this depends on the resourcefulness of the adversary.\\[0.5ex]
\textbf{Credit Card Verification:} 
The adversary must account for credit cards and correspondingly matched
documents for each of the 6200 users. This would be an estimated total of \$300
USD (\$100 [Table~\tsref{tabCost}] for the credit card and \$200
[Table~\tsref{tabCost}] for acquiring matching identification documents) one
time. The only risk here is the fake credit cards brought off the dark web might
get caught or flagged. Assuming that such a credit card will be flagged after
one transaction, the total would sum to \$1200 per month. Once a card is
flagged, the biometric identity associated with it (if any) will also be flagged
and the adversary must account for that. Considering the same incentive for the
user, as in case 1, the upper bound of the cost can be rounded to be about \$16
million USD. We ignore any possible time delays involved in cases when a card is
flagged and a new one is needed to replace it. However, if the cards take longer
to get flagged, the cost estimate will drop significantly.\\[0.5ex]
\textbf{Address Verification:} A challenging issue here is if the new user is
shown to share the same address as an original user, essentially as a
renter/sub-leaser; it would be hard to detect them as a Sybil. Rate limiting the
accounts associated per address would often be bound by the average family size
(which is just over 3 in the US, hence 4 would the ideal choice). This would
only work in favor of the adversary unless the colluding user has already
reached the limit. In order to effectively combat this, we can use something
like the census data to set the upper limit of residents for all registered
addresses. This would likely be an incomplete dataset, but having an
individualized rate limit for each address will force the adversary to associate
each identity with a new address of their own. The addresses registered to
legitimate users should be non reusable. We will assume this is the case. With
rent at \$1500 USD per month (national average is \$1494:Table~\tsref{tabCost}),
a minimum pay of \$4000 USD per employee per month (Table~\tsref{tabCost}) and a
fake ID priced at \$200 USD per employee, the cumulative price per month is
about \$17.1 to \$17.6 million USD (Table~\tsref{tab:costAdversary2}. However,
the bottleneck is more restrictive compared to case 1. Here, in addition to
locating susceptible individuals who are prepared to take on the job, the
requirement for fake identities and biometrics and new unallocated addresses is
expected to introduce more difficulties for the adversary, which cannot always
be remedied financially.\footnote{While it might be non-trivial to find the
exact number of new addresses that would fit the adversary's criteria, it is not
infeasible to find some addresses which can be systematically exploited and used
to attest multiple accounts exploiting the incompleteness of census data.}  A
combination of using both credit card and address verifications together will
increase the complexities and logistical challenges of executing such an attack.
Table~\tsref{tab:costAdversary2} compares the cost breakdowns for various
verification techniques, which takes advantage of the adversary's limited
physical presence on foreign soil. The cumulative effect can be more clearly
seen in Table~\tsref{tab:costAdversary3}, which shows the necessity of having
some form of verification.

%%%%%%%%%%%%%%%%%%%%%%%%%%%%%%%%%%%%%%%%%%%%%%%%%%%%%%%%%%%%%%%%%%%%%%%%%%
% \begin{table}[ht]
% \centering
% \begin{tabular}{cc}
% \toprule 
%  \makecell[c]{POS Scheme\\{(with Address Verification)}} & \makecell[c]{Cost per Month(USD)} \\
% \midrule 
%  \makecell[l]{\textbf{Gross for Case 1}\\- Targeted Ads\\- Incentives/User\\Total Cost/User} & \makecell[c]{ \\$\displaystyle \$5K$\\$\displaystyle \$4K$\\$\displaystyle \$4K$} \\
%  \hline
% \makecell[l]{\textbf{Gross for Case 2}\\- Rented Space/User\\- Incentives/User\\- Fake ID/User\\- Smartphone/User\\Total Cost/User} & \makecell[c]{ \\$\displaystyle \$1.5K$\\$\displaystyle \$4K$\\$\displaystyle \$200$(one time)\\$\displaystyle \$500$(one time)\\$\displaystyle \$5.5K$} \\
%  \bottomrule
% \end{tabular}
% \caption{Measurable Cost Estimate Per Month for the Adversary to Maintain the Same Traffic as has been previously seen in \cite{senatereport}}
%     \label{tab:costAdversary1}
% \end{table}
%%%%%%%%%%%%%%%%%%%%%%%%%%%%%%%%%%%%%%%%%%%%%%%%%%%%%%%%%%%%%%%%%%%%%%%%%%%%%
\begin{table}[t]
\centering
\small
\begin{tabular}{|c|c|c|}
\hline
 \makecell[c]{Verification\\{Techniques: Case 2}} & \makecell[c]{One-time \\{Cost}} & \makecell[c]{Recurring \\{Cost P.M.}}\\
\hline \hline 
\makecell[l]{\textbf{Baseline}\\- Fake IDs} & \makecell[c]{ \\ $\displaystyle \$200$}&\makecell[c]{\\-} \\
\hline
\makecell[l]{\textbf{Credit Card }\\- Credit card  ($\times 4$)\\- Incentives\\- Fake ID ($\times 4$) }  &\makecell[c]{\\-\\-\\-\\$\displaystyle$ }& \makecell[c]{ \\$ \displaystyle \$100(\times 4)$\\$\displaystyle \$4K$\\$\displaystyle \$200(\times 4)$ }\\
\hline
\makecell[l]{\textbf{User Address}\\- Rented Space\\- Incentives\\- Fake ID}  & \makecell[c]{\\- \\- \\$\displaystyle \$200$\\$\displaystyle $}& \makecell[c]{ \\$\displaystyle \$1.5K$\\$\displaystyle \$4K$\\-}\\
\hline
\end{tabular}
\caption{Measurable attack cost breakdown per user for Case 2 with different types of verification techniques to maintain the same traffic as has been previously seen in \cite{senatereport}.}
\vskip 1em
\label{tab:costAdversary2}

\end{table}
%%%%%%%%%%%%%%%%%%%%%%%%%%%%%%%%%%%%%%%%%%%%%%%%%%%%%%%%%%%%%%%%%%%%%%%%%%%%%%%%
% \begin{figure}
% %[!ht]
% \centering
% \scalebox{0.8}{
%     \centering
%     \includegraphics[scale=0.65]{plot1.png}
%     }
    
% \vspace{-1mm}
% \caption{Total monthly attack cost estimate for Case 2 with different types of verification techniques to maintain the same traffic as has been previously seen in \cite{senatereport}.}
%     \label{fig:plot1}
% %\vskip -1em
% \end{figure}
%%%%%%%%%%%%%%%%%%%%%%%%%%%%%%%%%%%%%%%%%%%%%%%%%%%%%%%%%%%%%%%%%%%%%%%%%%%%%%%%%%
\begin{table}[ht]
    \centering
    \small
    \begin{tabular}{|c|c|c|}
    \hline
    \makecell[c]{Verification\\{Techniques: Case 2}} & \makecell[c]{Initial Setup \\{(1st Month)}}& \makecell[c]{Recurring \\{Cost P.M.}}\\
    \hline \hline
         Baseline & $\displaystyle \$0.6M $ &-\\
         \hline
         Credit Card & $\displaystyle \$16.1M $ & $\displaystyle \$16.1M$\\
         \hline
         User Address  & $\displaystyle \$17.6M $ & $\displaystyle \$17.1M $  \\
         \hline
    \end{tabular}
    \caption{Total monthly attack cost estimate for Case 2 with different types of verification techniques to maintain the same traffic as has been previously seen in \cite{senatereport}.}
    \label{tab:costAdversary3}
    % \vskip -2em
\end{table}
%%%%%%%%%%%%%%%%%%%%%%%%%%%%%%%%%%%%%%%%%%%%%%%%%%%%%%%%%%%%%%%%%%%%%%%%%%%%%%%%%%

\begin{table}[ht!]
    \centering
    \resizebox{\columnwidth}{!} {
    \def\arraystretch{1.6}
    \begin{tabular}{|c|c|}
    \hline
    \textbf{Attack Vector} &\textbf{Cost}\\
    \hline \hline
         Fake IDs &$\displaystyle ~\$200 $ per ID \cite{fakeid1}\\
         \hline
         Fake credit cards & $\displaystyle \$25-\$300$ per card \cite{fakecc1} \\
         \hline
         Breaking in to POS system & \makecell[c]{$\displaystyle \$20K$ to $\displaystyle \$50K$ yearly per million\\{ in coverage of insurance Cost \cite{insurance1.3}} \\{$\displaystyle \$0.5M$ (Offensive softwares)} \\{ $\displaystyle \$3K-5K$/person/monthly \cite{phishing1, phishing2}}}  \\
         \hline
         Data Breach cost & $\displaystyle\$3.92M$ USD \cite{breach} \\
         \hline
         \makecell[c]{Fake Address (Renting \\{via shell companies etc)}}&$\displaystyle \$1494$ -$\displaystyle \$14800$ / employee /year \cite{rent}\\
         \hline
         Average Wage &$\displaystyle ~\$936$ /Week\cite{avgsal} \\
         \hline
         \makecell[c]{Stolen biometric information\\{(from deep web, breached}\\{ systems, inactive users}}& Unknown cost (non zero)\\
         \hline
    \end{tabular}}
    \caption{Cost of Attacks}
    \label{tabCost}
\end{table}

Another aggressive approach would involve buying offensive spyware and/or zero
days (cost of which would not exceed \$1M: Table~\tsref{tabCost}). Even with the
cost of hiring skilled labour to implement the attacks, overall, it could be a
much cheaper option. 

Thus even though we may succeed in raising the cost of attack from nothing to
\$16 millions to \$25 millions USD per month, it is well within the reach of a
resourceful nation-state-level adversary. Hence, it would be feasible for the
adversary to continue the process through easier implementations of the protocol
discussed above unless the rate limiting bottlenecks hold fast. It is possible
to lower the cost through social engineering hacks by the adversary. The cost
divisions are calculated based on data shown Table~\tsref{tabCost}.

\section{Conclusion}
Verifying user identities online is crucial to a wide range of services and systems today. However to date the properties required in ensuring appropriate user identity security had not be examined together. Navigating between these runs aground on the Ghost Trilemma, as we showed in this paper, but with
judicious engineering and limited trust in centralized services, we believe it
is possible to build services that enable authentic and privacy-preserving
communication for all.

%\Urlmuskip=0mu plus 1mu\relax
\small
\bibliographystyle{ACM-Reference-Format}
\bibliography{main}


\begin{thebibliography}{148}

%%% the Web address.

\ifx \showCODEN    \undefined \def \showCODEN     #1{\unskip}     \fi
\ifx \showDOI      \undefined \def \showDOI       #1{#1}\fi
\ifx \showISBNx    \undefined \def \showISBNx     #1{\unskip}     \fi
\ifx \showISBNxiii \undefined \def \showISBNxiii  #1{\unskip}     \fi
\ifx \showISSN     \undefined \def \showISSN      #1{\unskip}     \fi
\ifx \showLCCN     \undefined \def \showLCCN      #1{\unskip}     \fi
\ifx \shownote     \undefined \def \shownote      #1{#1}          \fi
\ifx \showarticletitle \undefined \def \showarticletitle #1{#1}   \fi
\ifx \showURL      \undefined \def \showURL       {\relax}        \fi
% invisible to TeX
\providecommand\bibfield[2]{#2}
\providecommand\bibinfo[2]{#2}
\providecommand\natexlab[1]{#1}
\providecommand\showeprint[2][]{arXiv:#2}

\bibitem[ind(n.\,d.)]%
        {individuality}
 \bibinfo{year}{[n.\,d.]}\natexlab{}.
\newblock \bibinfo{title}{Overview of Proof of Individuality}.
\newblock
\newblock
\urldef\tempurl%
\url{http://proofofindividuality.weebly.com/overview.html}
\showURL{%
\tempurl}


\bibitem[201(2011)]%
        {2011_senatereport}
 \bibinfo{year}{2011}\natexlab{}.
\newblock \bibinfo{title}{CENTRAL ASIA AND THE ARAB SPRING: GROWING PRESSURE
  FOR HUMAN RIGHTS?}
\newblock
\newblock
\newblock
\shownote{\url{https://www.govinfo.gov/content/pkg/CHRG-112jhrg93877/html/CHRG-112jhrg93877.htm}}.


\bibitem[fak(2015)]%
        {fakeid1}
 \bibinfo{year}{2015}\natexlab{}.
\newblock \bibinfo{title}{Fake ID Guides on the Deep Web}.
\newblock
\newblock
\newblock
\shownote{\url{https://www.reddit.com/r/deepweb/comments/3gifi3/fake\_id\_guides\_on\_the\_deep\_web/}}.


\bibitem[ren(2015)]%
        {rent}
 \bibinfo{year}{2015}\natexlab{}.
\newblock \bibinfo{title}{Here's how much your company pays to rent office
  space}.
\newblock
\newblock
\newblock
\shownote{\url{https://on.mktw.net/2UKHnU6 }}.


\bibitem[dup(2018)]%
        {duplexai}
 \bibinfo{year}{2018}\natexlab{}.
\newblock \bibinfo{booktitle}{\emph{Did Google Duplex beat the Turing Test? Yes
  and No.}}
\newblock
\newblock
\shownote{\url{http://bit.ly/3KkddR9}}.


\bibitem[phi(2018)]%
        {phishing2}
 \bibinfo{year}{2018}\natexlab{}.
\newblock \bibinfo{title}{Job Ads for Russian Troll Factory}.
\newblock
\newblock
\newblock
\shownote{\url{https://bit.ly/3e1BskS}}.


\bibitem[sto(2019)]%
        {stolenid}
 \bibinfo{year}{2019}\natexlab{}.
\newblock \bibinfo{title}{Digital Blackface: Pro-Trump Trolls Are Impersonating
  Black People on Twitter}.
\newblock
\newblock
\newblock
\shownote{\url{https://bit.ly/2XYbb1r}}.


\bibitem[avg(2019)]%
        {avgsal}
 \bibinfo{year}{2019}\natexlab{}.
\newblock \bibinfo{title}{What Is the Average Income in the U.S.?}
\newblock
\newblock
\newblock
\shownote{\url{https://www.thestreet.com/personal-finance/average-income-in-us-14852178}}.


\bibitem[bre(2020)]%
        {breach}
 \bibinfo{year}{2020}\natexlab{}.
\newblock \bibinfo{title}{Cost of Data Breaches}.
\newblock
\newblock
\newblock
\shownote{\url{https://www.scasecurity.com/cost-of-a-data-breach/ }}.


\bibitem[fak(2020)]%
        {fakecc1}
 \bibinfo{year}{2020}\natexlab{}.
\newblock \bibinfo{title}{Fake Credit Card}.
\newblock
\newblock
\newblock
\shownote{\url{https://www.deepwebsiteslinks.com/deep-web-links/}}.


\bibitem[spo(2020)]%
        {spotthetroll}
 \bibinfo{year}{2020}\natexlab{}.
\newblock \bibinfo{title}{Quiz and Resources at SpotTheTroll}.
\newblock
\newblock
\newblock
\shownote{\url{https://spotthetroll.org/start}}.


\bibitem[sen(2020)]%
        {senatereport}
 \bibinfo{year}{2020}\natexlab{}.
\newblock \bibinfo{title}{Report Of The Select Committee On Intelligence United
  States Senate On Russian Active Measures Campaigns and Interference in the
  2016 U.S. Election ' Volume 2: Russia's Use of Social Media With Additional
  Views}.
\newblock
\newblock
\newblock
\shownote{\url{http://bit.ly/37WNmK8}}.


\bibitem[ins(2021)]%
        {insurance1.3}
 \bibinfo{year}{2021}\natexlab{}.
\newblock \bibinfo{title}{Cyber Insurance}.
\newblock
\newblock
\newblock
\shownote{\url{https://foundershield.com/coverage/cyber-liability-insurance/}}.


\bibitem[phi(2021)]%
        {phishing1}
 \bibinfo{year}{2021}\natexlab{}.
\newblock \bibinfo{title}{Hacking Team Customer List}.
\newblock
\newblock
\newblock
\shownote{\url{https://en.wikipedia.org/wiki/Hacking\_Team},}.


\bibitem[fol(2021)]%
        {follow2021}
 \bibinfo{year}{2021}\natexlab{}.
\newblock \bibinfo{title}{What is a liquid democracy?}
\newblock
\newblock
\urldef\tempurl%
\url{https://followmyvote.com/liquid-democracy/}
\showURL{%
\tempurl}


\bibitem[NCR(2022)]%
        {NCRI_insights}
 \bibinfo{year}{2022}\natexlab{}.
\newblock \bibinfo{title}{New World Order Conspiracy Theories and Anti-Nato
  Rhetoric Surging on Twitter Amid Russian Invasion of Ukraine}.
\newblock
\newblock
\newblock
\shownote{\url{https://networkcontagion.us/wp-content/uploads/NCRI-Insights-SitRep-March-2022.pdf}}.


\bibitem[Abdou(2018)]%
        {abdou2018internet}
\bibfield{author}{\bibinfo{person}{AbdelRahman Abdou}.}
  \bibinfo{year}{2018}\natexlab{}.
\newblock \showarticletitle{Internet Location Verification: Challenges and
  Solutions}.
\newblock \bibinfo{journal}{\emph{arXiv preprint arXiv:1802.05169}}
  (\bibinfo{year}{2018}).
\newblock


\bibitem[Abdou et~al\mbox{.}(n.\,d.)]%
        {abdouevasion}
\bibfield{author}{\bibinfo{person}{A Abdou}, \bibinfo{person}{Ashraf Matrawy},
  {and} \bibinfo{person}{Paul~C van Oorschot}.}
  \bibinfo{year}{[n.\,d.]}\natexlab{}.
\newblock \bibinfo{booktitle}{\emph{On the Evasion of Delay-Based IP
  Geolocation}}.
\newblock \bibinfo{type}{{T}echnical {R}eport}. \bibinfo{institution}{Technical
  report, Carleton University TR-14-03, June 2014}.
\newblock


\bibitem[Abdou et~al\mbox{.}(2015)]%
        {abdou2015cpv}
\bibfield{author}{\bibinfo{person}{AbdelRahman Abdou}, \bibinfo{person}{Ashraf
  Matrawy}, {and} \bibinfo{person}{Paul~C Van~Oorschot}.}
  \bibinfo{year}{2015}\natexlab{}.
\newblock \showarticletitle{CPV: Delay-based location verification for the
  Internet}.
\newblock \bibinfo{journal}{\emph{IEEE Transactions on Dependable and Secure
  Computing}} \bibinfo{volume}{14}, \bibinfo{number}{2} (\bibinfo{year}{2015}),
  \bibinfo{pages}{130--144}.
\newblock


\bibitem[Abdou et~al\mbox{.}(2014)]%
        {abdou2014location}
\bibfield{author}{\bibinfo{person}{AbdelRahman~M Abdou},
  \bibinfo{person}{Ashraf Matrawy}, {and} \bibinfo{person}{Paul~C
  Van~Oorschot}.} \bibinfo{year}{2014}\natexlab{}.
\newblock \showarticletitle{Location verification on the Internet: Towards
  enforcing location-aware access policies over Internet clients}. In
  \bibinfo{booktitle}{\emph{2014 IEEE Conference on Communications and Network
  Security}}. IEEE, \bibinfo{pages}{175--183}.
\newblock


\bibitem[Abraham et~al\mbox{.}(2011)]%
        {abraham2011distributed}
\bibfield{author}{\bibinfo{person}{Ittai Abraham}, \bibinfo{person}{Lorenzo
  Alvisi}, {and} \bibinfo{person}{Joseph~Y Halpern}.}
  \bibinfo{year}{2011}\natexlab{}.
\newblock \showarticletitle{Distributed computing meets game theory: combining
  insights from two fields}.
\newblock \bibinfo{journal}{\emph{Acm Sigact News}} \bibinfo{volume}{42},
  \bibinfo{number}{2} (\bibinfo{year}{2011}), \bibinfo{pages}{69--76}.
\newblock


\bibitem[Addawood et~al\mbox{.}(2019)]%
        {addawood2019linguistic}
\bibfield{author}{\bibinfo{person}{Aseel Addawood}, \bibinfo{person}{Adam
  Badawy}, \bibinfo{person}{Kristina Lerman}, {and} \bibinfo{person}{Emilio
  Ferrara}.} \bibinfo{year}{2019}\natexlab{}.
\newblock \showarticletitle{Linguistic cues to deception: Identifying political
  trolls on social media}. In \bibinfo{booktitle}{\emph{Proceedings of the
  international AAAI conference on web and social media}},
  Vol.~\bibinfo{volume}{13}. \bibinfo{pages}{15--25}.
\newblock


\bibitem[Adida(2008)]%
        {adida2008helios}
\bibfield{author}{\bibinfo{person}{Ben Adida}.}
  \bibinfo{year}{2008}\natexlab{}.
\newblock \showarticletitle{Helios: Web-based Open-Audit Voting.}. In
  \bibinfo{booktitle}{\emph{USENIX security symposium}},
  Vol.~\bibinfo{volume}{17}. \bibinfo{pages}{335--348}.
\newblock


\bibitem[Aiyer et~al\mbox{.}(2005)]%
        {aiyer2005bar}
\bibfield{author}{\bibinfo{person}{Amitanand~S Aiyer}, \bibinfo{person}{Lorenzo
  Alvisi}, \bibinfo{person}{Allen Clement}, \bibinfo{person}{Mike Dahlin},
  \bibinfo{person}{Jean-Philippe Martin}, {and} \bibinfo{person}{Carl Porth}.}
  \bibinfo{year}{2005}\natexlab{}.
\newblock \showarticletitle{BAR fault tolerance for cooperative services}. In
  \bibinfo{booktitle}{\emph{Proceedings of the twentieth ACM symposium on
  Operating systems principles}}. \bibinfo{pages}{45--58}.
\newblock


\bibitem[Al-Qurishi et~al\mbox{.}(2017)]%
        {al2017sybil}
\bibfield{author}{\bibinfo{person}{Muhammad Al-Qurishi},
  \bibinfo{person}{Mabrook Al-Rakhami}, \bibinfo{person}{Atif Alamri},
  \bibinfo{person}{Majed Alrubaian}, \bibinfo{person}{Sk~Md~Mizanur Rahman},
  {and} \bibinfo{person}{M~Shamim Hossain}.} \bibinfo{year}{2017}\natexlab{}.
\newblock \showarticletitle{Sybil defense techniques in online social networks:
  a survey}.
\newblock \bibinfo{journal}{\emph{IEEE Access}}  \bibinfo{volume}{5}
  (\bibinfo{year}{2017}), \bibinfo{pages}{1200--1219}.
\newblock


\bibitem[Aley-Raz et~al\mbox{.}(2013)]%
        {aley2013device}
\bibfield{author}{\bibinfo{person}{Almog Aley-Raz}, \bibinfo{person}{Nir~Moshe
  Krause}, \bibinfo{person}{Michael~Itzhak Salmon}, {and}
  \bibinfo{person}{Ran~Yehoshua Gazit}.} \bibinfo{year}{2013}\natexlab{}.
\newblock \bibinfo{title}{Device, system, and method of liveness detection
  utilizing voice biometrics}.
\newblock
\newblock
\newblock
\shownote{US Patent 8,442,824}.


\bibitem[Alluri(2019)]%
        {Alluri2019}
\bibfield{author}{\bibinfo{person}{Aparna Alluri}.}
  \bibinfo{year}{2019}\natexlab{}.
\newblock \bibinfo{title}{WhatsApp: The 'black hole' of fake news in India's
  election}.
\newblock
\newblock
\newblock
\shownote{\url{https://www.bbc.com/news/world-asia-india-47797151}}.


\bibitem[Alvisi et~al\mbox{.}(2013)]%
        {alvisi2013sok}
\bibfield{author}{\bibinfo{person}{Lorenzo Alvisi}, \bibinfo{person}{Allen
  Clement}, \bibinfo{person}{Alessandro Epasto}, \bibinfo{person}{Silvio
  Lattanzi}, {and} \bibinfo{person}{Alessandro Panconesi}.}
  \bibinfo{year}{2013}\natexlab{}.
\newblock \showarticletitle{Sok: The evolution of sybil defense via social
  networks}. In \bibinfo{booktitle}{\emph{2013 ieee symposium on security and
  privacy}}. IEEE, \bibinfo{pages}{382--396}.
\newblock


\bibitem[Android~Developers(2019)]%
        {AndroidDev2019}
\bibfield{author}{\bibinfo{person}{. Android~Developers}.}
  \bibinfo{year}{2019}\natexlab{}.
\newblock \bibinfo{title}{UUID}.
\newblock
\newblock
\newblock
\shownote{\url{https://developer.android.com/reference/java/util/UUID}}.


\bibitem[Appel(2022a)]%
        {evotereport}
\bibfield{author}{\bibinfo{person}{Andrew Appel}.}
  \bibinfo{year}{2022}\natexlab{a}.
\newblock \bibinfo{title}{How to Assess an E-Voting System}.
\newblock
\newblock
\newblock
\shownote{\url{https://bit.ly/3QhVldv}}.


\bibitem[Appel(2022b)]%
        {evoteinsecure}
\bibfield{author}{\bibinfo{person}{Andrew Appel}.}
  \bibinfo{year}{2022}\natexlab{b}.
\newblock \bibinfo{title}{How to Assess an E-Voting System}.
\newblock
\newblock
\newblock
\shownote{\url{https://bit.ly/47dFm6f}}.


\bibitem[Appel(2022c)]%
        {appel_2022}
\bibfield{author}{\bibinfo{person}{Andrew Appel}.}
  \bibinfo{year}{2022}\natexlab{c}.
\newblock \bibinfo{title}{Switzerland's e-voting: The threat model}.
\newblock
\newblock
\urldef\tempurl%
\url{https://bit.ly/43MZqJG}
\showURL{%
\tempurl}


\bibitem[Bailey and Almond(n.\,d.)]%
        {problemsUnseenworldcoin}
\bibfield{author}{\bibinfo{person}{Andrew Bailey} {and} \bibinfo{person}{Nick
  Almond}.} \bibinfo{year}{[n.\,d.]}\natexlab{}.
\newblock \bibinfo{title}{A New Identity and Financial Network}.
\newblock
\newblock
\newblock
\shownote{\url{https://blockworks.co/news/worldcoin-privacy-concerns}}.


\bibitem[Barni et~al\mbox{.}(2010)]%
        {10.1145/1854229.1854270}
\bibfield{author}{\bibinfo{person}{Mauro Barni}, \bibinfo{person}{Tiziano
  Bianchi}, \bibinfo{person}{Dario Catalano}, \bibinfo{person}{Mario
  Di~Raimondo}, \bibinfo{person}{Ruggero Donida~Labati},
  \bibinfo{person}{Pierluigi Failla}, \bibinfo{person}{Dario Fiore},
  \bibinfo{person}{Riccardo Lazzeretti}, \bibinfo{person}{Vincenzo Piuri},
  \bibinfo{person}{Fabio Scotti}, {and} \bibinfo{person}{Alessandro Piva}.}
  \bibinfo{year}{2010}\natexlab{}.
\newblock \showarticletitle{Privacy-Preserving Fingercode Authentication}. In
  \bibinfo{booktitle}{\emph{Proceedings of the 12th ACM Workshop on Multimedia
  and Security}} (Roma, Italy) \emph{(\bibinfo{series}{MM\&amp;Sec '10})}.
  \bibinfo{publisher}{Association for Computing Machinery},
  \bibinfo{address}{New York, NY, USA}, \bibinfo{pages}{231–240}.
\newblock
\showISBNx{9781450302869}
\urldef\tempurl%
\url{https://doi.org/10.1145/1854229.1854270}
\showDOI{\tempurl}


\bibitem[Bhatia et~al\mbox{.}(2017)]%
        {bhatia2017soc2seq}
\bibfield{author}{\bibinfo{person}{Parminder Bhatia}, \bibinfo{person}{Marsal
  Gavalda}, {and} \bibinfo{person}{Arash Einolghozati}.}
  \bibinfo{year}{2017}\natexlab{}.
\newblock \showarticletitle{soc2seq: Social embedding meets conversation
  model}.
\newblock \bibinfo{journal}{\emph{arXiv preprint arXiv:1702.05512}}
  (\bibinfo{year}{2017}).
\newblock


\bibitem[Blania and Altman(n.\,d.)]%
        {Worldcoin1}
\bibfield{author}{\bibinfo{person}{Alex Blania} {and} \bibinfo{person}{Sam
  Altman}.} \bibinfo{year}{[n.\,d.]}\natexlab{}.
\newblock \bibinfo{title}{A New Identity and Financial Network}.
\newblock
\newblock
\newblock
\shownote{\url{https://whitepaper.worldcoin.org/}}.


\bibitem[Borge et~al\mbox{.}(2017)]%
        {borge2017proof}
\bibfield{author}{\bibinfo{person}{Maria Borge}, \bibinfo{person}{Eleftherios
  Kokoris-Kogias}, \bibinfo{person}{Philipp Jovanovic}, \bibinfo{person}{Linus
  Gasser}, \bibinfo{person}{Nicolas Gailly}, {and} \bibinfo{person}{Bryan
  Ford}.} \bibinfo{year}{2017}\natexlab{}.
\newblock \showarticletitle{Proof-of-personhood: Redemocratizing permissionless
  cryptocurrencies}. In \bibinfo{booktitle}{\emph{2017 IEEE European Symposium
  on Security and Privacy Workshops (EuroS\&PW)}}. IEEE,
  \bibinfo{pages}{23--26}.
\newblock


\bibitem[{Boulkenafet} et~al\mbox{.}(2017)]%
        {Boulkena7961798}
\bibfield{author}{\bibinfo{person}{Z. {Boulkenafet}}, \bibinfo{person}{J.
  {Komulainen}}, \bibinfo{person}{L. {Li}}, \bibinfo{person}{X. {Feng}}, {and}
  \bibinfo{person}{A. {Hadid}}.} \bibinfo{year}{2017}\natexlab{}.
\newblock \showarticletitle{OULU-NPU: A Mobile Face Presentation Attack
  Database with Real-World Variations}. In \bibinfo{booktitle}{\emph{2017 12th
  IEEE International Conference on Automatic Face Gesture Recognition (FG
  2017)}}. \bibinfo{pages}{612--618}.
\newblock


\bibitem[Bradshaw and Howard(2017)]%
        {bradshaw2017troops}
\bibfield{author}{\bibinfo{person}{Samantha Bradshaw} {and}
  \bibinfo{person}{Philip Howard}.} \bibinfo{year}{2017}\natexlab{}.
\newblock \showarticletitle{Troops, trolls and troublemakers: A global
  inventory of organized social media manipulation}.
\newblock  (\bibinfo{year}{2017}).
\newblock


\bibitem[Bradshaw and Howard(2019)]%
        {bradshaw2019}
\bibfield{author}{\bibinfo{person}{Samantha Bradshaw} {and}
  \bibinfo{person}{Philip~N Howard}.} \bibinfo{year}{2019}\natexlab{}.
\newblock \bibinfo{title}{The Global Disinformation Order 2019 Global Inventory
  of Organised Social Media Manipulation}.
\newblock
\newblock
\newblock
\shownote{\url{http://bit.ly/2SleUSP}}.


\bibitem[Bureau~of Democracy and Labor(2023)]%
        {beauro23}
\bibfield{author}{\bibinfo{person}{Human~Rights Bureau~of Democracy} {and}
  \bibinfo{person}{Labor}.} \bibinfo{year}{2023}\natexlab{}.
\newblock \bibinfo{title}{Supporting critical open source technologies that
  enable a free and open internet - united states department of state}.
\newblock
\newblock
\urldef\tempurl%
\url{https://tinyurl.com/2fnx7ebd}
\showURL{%
\tempurl}


\bibitem[Chaum(1983)]%
        {chaum1983blind}
\bibfield{author}{\bibinfo{person}{David Chaum}.}
  \bibinfo{year}{1983}\natexlab{}.
\newblock \showarticletitle{Blind signatures for untraceable payments}. In
  \bibinfo{booktitle}{\emph{Advances in cryptology}}. Springer,
  \bibinfo{pages}{199--203}.
\newblock


\bibitem[Chellapilla et~al\mbox{.}(2005)]%
        {captcha}
\bibfield{author}{\bibinfo{person}{Kumar Chellapilla}, \bibinfo{person}{Kevin
  Larson}, \bibinfo{person}{Patrice~Y. Simard}, {and} \bibinfo{person}{Mary
  Czerwinski}.} \bibinfo{year}{2005}\natexlab{}.
\newblock \showarticletitle{Designing human friendly human interaction proofs
  (HIPs)}. In \bibinfo{booktitle}{\emph{Proceedings of the 2005 Conference on
  Human Factors in Computing Systems, {CHI} 2005, Portland, Oregon, USA, April
  2-7, 2005}}, \bibfield{editor}{\bibinfo{person}{Gerrit~C. van~der Veer} {and}
  \bibinfo{person}{Carolyn Gale}} (Eds.). \bibinfo{publisher}{ACM},
  \bibinfo{pages}{711--720}.
\newblock


\bibitem[Chesney and Citron(2019)]%
        {chesney2019deep}
\bibfield{author}{\bibinfo{person}{Bobby Chesney} {and}
  \bibinfo{person}{Danielle Citron}.} \bibinfo{year}{2019}\natexlab{}.
\newblock \showarticletitle{Deep fakes: A looming challenge for privacy,
  democracy, and national security}.
\newblock \bibinfo{journal}{\emph{Calif. L. Rev.}}  \bibinfo{volume}{107}
  (\bibinfo{year}{2019}), \bibinfo{pages}{1753}.
\newblock


\bibitem[Chetty and Wagner(2006)]%
        {chetty2006multi}
\bibfield{author}{\bibinfo{person}{Girija Chetty} {and}
  \bibinfo{person}{Michael Wagner}.} \bibinfo{year}{2006}\natexlab{}.
\newblock \showarticletitle{Multi-level liveness verification for face-voice
  biometric authentication}. In \bibinfo{booktitle}{\emph{2006 Biometrics
  Symposium: Special Session on Research at the Biometric Consortium
  Conference}}. IEEE, \bibinfo{pages}{1--6}.
\newblock


\bibitem[Claesson(2019)]%
        {claesson}
\bibfield{author}{\bibinfo{person}{Annina Claesson}.}
  \bibinfo{year}{2019}\natexlab{}.
\newblock \bibinfo{title}{Coming Together to Fight Fake News: Lessons from the
  European Approach to Disinformation}.
\newblock
\newblock
\newblock
\shownote{\url{https://bit.ly/2XK8oYv}}.


\bibitem[Clark and Hengartner(2011)]%
        {clark2011selections}
\bibfield{author}{\bibinfo{person}{Jeremy Clark} {and} \bibinfo{person}{Urs
  Hengartner}.} \bibinfo{year}{2011}\natexlab{}.
\newblock \showarticletitle{Selections: Internet voting with over-the-shoulder
  coercion-resistance}. In \bibinfo{booktitle}{\emph{International Conference
  on Financial Cryptography and Data Security}}. Springer,
  \bibinfo{pages}{47--61}.
\newblock


\bibitem[Cope et~al\mbox{.}(2017)]%
        {cope2017investigation}
\bibfield{author}{\bibinfo{person}{Peter Cope}, \bibinfo{person}{Joseph
  Campbell}, {and} \bibinfo{person}{Thaier Hayajneh}.}
  \bibinfo{year}{2017}\natexlab{}.
\newblock \showarticletitle{An investigation of Bluetooth security
  vulnerabilities}. In \bibinfo{booktitle}{\emph{2017 IEEE 7th annual computing
  and communication workshop and conference (CCWC)}}. IEEE,
  \bibinfo{pages}{1--7}.
\newblock


\bibitem[Coskun et~al\mbox{.}(2013)]%
        {coskun2013survey}
\bibfield{author}{\bibinfo{person}{Vedat Coskun}, \bibinfo{person}{Busra
  Ozdenizci}, {and} \bibinfo{person}{Kerem Ok}.}
  \bibinfo{year}{2013}\natexlab{}.
\newblock \showarticletitle{A survey on near field communication (NFC)
  technology}.
\newblock \bibinfo{journal}{\emph{Wireless personal communications}}
  \bibinfo{volume}{71} (\bibinfo{year}{2013}), \bibinfo{pages}{2259--2294}.
\newblock


\bibitem[Costa-Pazo et~al\mbox{.}(2019)]%
        {costa2019challenges}
\bibfield{author}{\bibinfo{person}{Artur Costa-Pazo}, \bibinfo{person}{Esteban
  Vazquez-Fernandez}, \bibinfo{person}{Jos{\'e}~Luis Alba-Castro}, {and}
  \bibinfo{person}{Daniel Gonz{\'a}lez-Jim{\'e}nez}.}
  \bibinfo{year}{2019}\natexlab{}.
\newblock \showarticletitle{Challenges of face presentation attack detection in
  real scenarios}.
\newblock In \bibinfo{booktitle}{\emph{Handbook of Biometric Anti-Spoofing}}.
  \bibinfo{publisher}{Springer}, \bibinfo{pages}{247--266}.
\newblock


\bibitem[Cowan et~al\mbox{.}(2016)]%
        {cowan2016liveness}
\bibfield{author}{\bibinfo{person}{Melissa~A Cowan}, \bibinfo{person}{Ramune
  Nagisetty}, \bibinfo{person}{Jason Martin}, \bibinfo{person}{Richard~A
  Forand}, \bibinfo{person}{Conor~P Cahill}, {and} \bibinfo{person}{Bradley~A
  Jackson}.} \bibinfo{year}{2016}\natexlab{}.
\newblock \bibinfo{title}{Liveness Detection for User Authentication}.
\newblock
\newblock
\newblock
\shownote{US Patent App. 14/499,138}.


\bibitem[Cresci(2020)]%
        {cresci2020decade}
\bibfield{author}{\bibinfo{person}{Stefano Cresci}.}
  \bibinfo{year}{2020}\natexlab{}.
\newblock \showarticletitle{A decade of social bot detection}.
\newblock \bibinfo{journal}{\emph{Commun. ACM}} \bibinfo{volume}{63},
  \bibinfo{number}{10} (\bibinfo{year}{2020}), \bibinfo{pages}{72--83}.
\newblock


\bibitem[Cresci et~al\mbox{.}(2019)]%
        {cresci2019better}
\bibfield{author}{\bibinfo{person}{Stefano Cresci}, \bibinfo{person}{Marinella
  Petrocchi}, \bibinfo{person}{Angelo Spognardi}, {and}
  \bibinfo{person}{Stefano Tognazzi}.} \bibinfo{year}{2019}\natexlab{}.
\newblock \showarticletitle{Better safe than sorry: an adversarial approach to
  improve social bot detection}. In \bibinfo{booktitle}{\emph{Proceedings of
  the 10th ACM Conference on Web Science}}. \bibinfo{pages}{47--56}.
\newblock


\bibitem[D(2020)]%
        {livenessinFD}
\bibfield{author}{\bibinfo{person}{ID~R~\& D}.}
  \bibinfo{year}{2020}\natexlab{}.
\newblock \bibinfo{booktitle}{\emph{THE IMPORTANT ROLE OF LIVENESS DETECTION IN
  FACE BIOMETRIC AUTHENTICATION}}.
\newblock \bibinfo{type}{{T}echnical {R}eport}. \bibinfo{institution}{ID R \&
  D}.
\newblock
\urldef\tempurl%
\url{https://www.idrnd.ai/wp-content/uploads/2020/09/IDRD-Facial-Liveness-WHITEPAPER-Sept2020.pdf}
\showURL{%
\tempurl}


\bibitem[Davis et~al\mbox{.}(2016)]%
        {davis2016botornot}
\bibfield{author}{\bibinfo{person}{Clayton~Allen Davis}, \bibinfo{person}{Onur
  Varol}, \bibinfo{person}{Emilio Ferrara}, \bibinfo{person}{Alessandro
  Flammini}, {and} \bibinfo{person}{Filippo Menczer}.}
  \bibinfo{year}{2016}\natexlab{}.
\newblock \showarticletitle{Botornot: A system to evaluate social bots}. In
  \bibinfo{booktitle}{\emph{Proceedings of the 25th international conference
  companion on world wide web}}. \bibinfo{pages}{273--274}.
\newblock


\bibitem[de~Araujo et~al\mbox{.}(2019)]%
        {8949407}
\bibfield{author}{\bibinfo{person}{L.~Santiago de Araujo},
  \bibinfo{person}{V.~C. Patil}, \bibinfo{person}{L.~Augusto~Justen Marzulo},
  \bibinfo{person}{F.~Maia~Galvao Franca}, {and} \bibinfo{person}{S. Kundu}.}
  \bibinfo{year}{2019}\natexlab{}.
\newblock \showarticletitle{Efficient Testing of Physically Unclonable
  Functions for Uniqueness}. In \bibinfo{booktitle}{\emph{2019 IEEE 28th Asian
  Test Symposium (ATS)}}. \bibinfo{publisher}{IEEE Computer Society},
  \bibinfo{address}{Los Alamitos, CA, USA}, \bibinfo{pages}{117--1175}.
\newblock
\urldef\tempurl%
\url{https://doi.org/10.1109/ATS47505.2019.00022}
\showDOI{\tempurl}


\bibitem[Douceur(2002)]%
        {douceur2002sybil}
\bibfield{author}{\bibinfo{person}{John~R Douceur}.}
  \bibinfo{year}{2002}\natexlab{}.
\newblock \showarticletitle{The sybil attack}. In
  \bibinfo{booktitle}{\emph{International workshop on peer-to-peer systems}}.
  Springer, \bibinfo{pages}{251--260}.
\newblock


\bibitem[Elder(2012)]%
        {elder2012}
\bibfield{author}{\bibinfo{person}{Miriam Elder}.}
  \bibinfo{year}{2012}\natexlab{}.
\newblock \bibinfo{title}{Emails give insight into Kremlin youth group's
  priorities, means and concerns}.
\newblock
\newblock
\newblock
\shownote{\url{https://bit.ly/3eBc9pI}}.


\bibitem[FederalChancellery(n.\,d.a)]%
        {swissA}
\bibfield{author}{\bibinfo{person}{FCh FederalChancellery}.}
  \bibinfo{year}{[n.\,d.]}\natexlab{a}.
\newblock \showarticletitle{Redesign and relaunch of trials:Final report of the
  Steering Committee Vote Electronique}.
\newblock  (\bibinfo{year}{[n.\,d.]}).
\newblock
\newblock
\shownote{\url{shorturl.at/tuyG9}}.


\bibitem[FederalChancellery(n.\,d.b)]%
        {swissB}
\bibfield{author}{\bibinfo{person}{FCh FederalChancellery}.}
  \bibinfo{year}{[n.\,d.]}\natexlab{b}.
\newblock \showarticletitle{Summary of the expert dialog: Redesign of Internet
  Voting Trials in Switzerland 2020}.
\newblock  (\bibinfo{year}{[n.\,d.]}).
\newblock
\urldef\tempurl%
\url{https://www.newsd.admin.ch/newsd/message/attachments/61843.pdf}
\showURL{%
\tempurl}
\newblock
\shownote{Created: 11-19-2020}.


\bibitem[Feng and Wang(2019)]%
        {feng2019system}
\bibfield{author}{\bibinfo{person}{Xuetao Feng} {and} \bibinfo{person}{Yan
  Wang}.} \bibinfo{year}{2019}\natexlab{}.
\newblock \bibinfo{title}{System and method for efficient liveness detection}.
\newblock
\newblock
\newblock
\shownote{US Patent App. 16/019,955}.


\bibitem[Fischer et~al\mbox{.}(1985)]%
        {FLP85}
\bibfield{author}{\bibinfo{person}{Michael~J. Fischer},
  \bibinfo{person}{Nancy~A. Lynch}, {and} \bibinfo{person}{Michael~S.
  Paterson}.} \bibinfo{year}{1985}\natexlab{}.
\newblock \showarticletitle{Impossibility of Distributed Consensus with one
  Faulty Process}.
\newblock \bibinfo{journal}{\emph{J. ACM}} \bibinfo{volume}{32},
  \bibinfo{number}{2} (\bibinfo{date}{April} \bibinfo{year}{1985}),
  \bibinfo{pages}{374--382}.
\newblock


\bibitem[Fishkin and Luskin(2005)]%
        {fishkin2005experimenting}
\bibfield{author}{\bibinfo{person}{James~S Fishkin} {and}
  \bibinfo{person}{Robert~C Luskin}.} \bibinfo{year}{2005}\natexlab{}.
\newblock \showarticletitle{Experimenting with a democratic ideal: Deliberative
  polling and public opinion}.
\newblock \bibinfo{journal}{\emph{Acta politica}}  \bibinfo{volume}{40}
  (\bibinfo{year}{2005}), \bibinfo{pages}{284--298}.
\newblock


\bibitem[for democracy and Assistance(n.\,d.)]%
        {evoteincountries}
\bibfield{author}{\bibinfo{person}{International IDEA~Institute for democracy}
  {and} \bibinfo{person}{Electoral Assistance}.}
  \bibinfo{year}{[n.\,d.]}\natexlab{}.
\newblock \bibinfo{title}{Use of E-Voting Around the World}.
\newblock
\newblock
\newblock
\shownote{\url{https://www.idea.int/news-media/media/use-e-voting-around-world}}.


\bibitem[Ford(2019)]%
        {ford_2019}
\bibfield{author}{\bibinfo{person}{Bryan Ford}.}
  \bibinfo{year}{2019}\natexlab{}.
\newblock \bibinfo{title}{The Remote Voting Minefield: from North Carolina to
  Switzerland}.
\newblock
\newblock
\urldef\tempurl%
\url{https://bford.info/2019/02/22/voting/}
\showURL{%
\tempurl}


\bibitem[Ford(2020)]%
        {ford2020identity}
\bibfield{author}{\bibinfo{person}{Bryan Ford}.}
  \bibinfo{year}{2020}\natexlab{}.
\newblock \showarticletitle{Identity and Personhood in Digital Democracy:
  Evaluating Inclusion, Equality, Security, and Privacy in Pseudonym Parties
  and Other Proofs of Personhood}.
\newblock \bibinfo{journal}{\emph{arXiv preprint arXiv:2011.02412}}
  (\bibinfo{year}{2020}).
\newblock


\bibitem[Ford(2022)]%
        {ford2022auditing}
\bibfield{author}{\bibinfo{person}{Bryan Ford}.}
  \bibinfo{year}{2022}\natexlab{}.
\newblock \showarticletitle{Auditing the Swiss Post E-voting System: An
  Architectural Perspective}.
\newblock  (\bibinfo{year}{2022}).
\newblock


\bibitem[Ford and B{\"o}hme(2019)]%
        {ford2019rationality}
\bibfield{author}{\bibinfo{person}{Bryan Ford} {and} \bibinfo{person}{Rainer
  B{\"o}hme}.} \bibinfo{year}{2019}\natexlab{}.
\newblock \showarticletitle{Rationality is self-defeating in permissionless
  systems}.
\newblock \bibinfo{journal}{\emph{arXiv preprint arXiv:1910.08820}}
  (\bibinfo{year}{2019}).
\newblock


\bibitem[Ford and Strauss(2008)]%
        {ford2008offline}
\bibfield{author}{\bibinfo{person}{Bryan Ford} {and} \bibinfo{person}{Jacob
  Strauss}.} \bibinfo{year}{2008}\natexlab{}.
\newblock \showarticletitle{An offline foundation for online accountable
  pseudonyms}. In \bibinfo{booktitle}{\emph{Proceedings of the 1st workshop on
  Social network systems}}. \bibinfo{pages}{31--36}.
\newblock


\bibitem[Funke and Flamini(2018)]%
        {funke2018}
\bibfield{author}{\bibinfo{person}{Daniel Funke} {and} \bibinfo{person}{Daniela
  Flamini}.} \bibinfo{year}{2018}\natexlab{}.
\newblock \bibinfo{title}{A guide to anti-misinformation actions around the
  world}.
\newblock
\newblock
\newblock
\shownote{\url{https://www.poynter.org/ifcn/anti-misinformation-actions/}}.


\bibitem[Gassend et~al\mbox{.}(2002)]%
        {gassend2002silicon}
\bibfield{author}{\bibinfo{person}{Blaise Gassend}, \bibinfo{person}{Dwaine
  Clarke}, \bibinfo{person}{Marten Van~Dijk}, {and} \bibinfo{person}{Srinivas
  Devadas}.} \bibinfo{year}{2002}\natexlab{}.
\newblock \showarticletitle{Silicon physical random functions}. In
  \bibinfo{booktitle}{\emph{Proceedings of the 9th ACM conference on Computer
  and communications security}}. \bibinfo{pages}{148--160}.
\newblock


\bibitem[Gebru et~al\mbox{.}(2023)]%
        {diar_2023}
\bibfield{author}{\bibinfo{person}{Timnit Gebru}, \bibinfo{person}{Emily~M
  Bender}, \bibinfo{person}{Angelina McMillan-Major}, {and}
  \bibinfo{person}{Margaret Mitchell}.} \bibinfo{year}{2023}\natexlab{}.
\newblock \bibinfo{title}{Statement from the listed authors of Stochastic
  Parrots on the ``AI pause'' letter}.
\newblock
\newblock
\urldef\tempurl%
\url{https://www.dair-institute.org/blog/letter-statement-March2023}
\showURL{%
\tempurl}


\bibitem[Ghiani et~al\mbox{.}(2017)]%
        {GHIANI2017110}
\bibfield{author}{\bibinfo{person}{Luca Ghiani}, \bibinfo{person}{David~A.
  Yambay}, \bibinfo{person}{Valerio Mura}, \bibinfo{person}{Gian~Luca
  Marcialis}, \bibinfo{person}{Fabio Roli}, {and} \bibinfo{person}{Stephanie~A.
  Schuckers}.} \bibinfo{year}{2017}\natexlab{}.
\newblock \showarticletitle{Review of the Fingerprint Liveness Detection
  (LivDet) competition series: 2009 to 2015}.
\newblock \bibinfo{journal}{\emph{Image and Vision Computing}}
  \bibinfo{volume}{58} (\bibinfo{year}{2017}), \bibinfo{pages}{110--128}.
\newblock
\showISSN{0262-8856}
\urldef\tempurl%
\url{https://doi.org/10.1016/j.imavis.2016.07.002}
\showDOI{\tempurl}


\bibitem[Gj{\o}steen(2011)]%
        {gjosteen2011norwegian}
\bibfield{author}{\bibinfo{person}{Kristian Gj{\o}steen}.}
  \bibinfo{year}{2011}\natexlab{}.
\newblock \showarticletitle{The Norwegian internet voting protocol}. In
  \bibinfo{booktitle}{\emph{International Conference on E-Voting and
  Identity}}. Springer, \bibinfo{pages}{1--18}.
\newblock


\bibitem[Goldhill(2019)]%
        {goldhill2019}
\bibfield{author}{\bibinfo{person}{Olivia Goldhill}.}
  \bibinfo{year}{2019}\natexlab{}.
\newblock \bibinfo{title}{Politicians are embracing disinformation in the UK
  election}.
\newblock
\newblock
\newblock
\shownote{\url{https://bit.ly/2TUFrbg}}.


\bibitem[Goodfellow et~al\mbox{.}(2018)]%
        {goodfellow2018making}
\bibfield{author}{\bibinfo{person}{Ian Goodfellow}, \bibinfo{person}{Patrick
  McDaniel}, {and} \bibinfo{person}{Nicolas Papernot}.}
  \bibinfo{year}{2018}\natexlab{}.
\newblock \showarticletitle{Making machine learning robust against adversarial
  inputs}.
\newblock \bibinfo{journal}{\emph{Commun. ACM}} \bibinfo{volume}{61},
  \bibinfo{number}{7} (\bibinfo{year}{2018}), \bibinfo{pages}{56--66}.
\newblock


\bibitem[Grabner-Kr{\"a}uter and Bitter(2015)]%
        {grabner2015trust}
\bibfield{author}{\bibinfo{person}{Sonja Grabner-Kr{\"a}uter} {and}
  \bibinfo{person}{Sofie Bitter}.} \bibinfo{year}{2015}\natexlab{}.
\newblock \showarticletitle{Trust in online social networks: A multifaceted
  perspective}. In \bibinfo{booktitle}{\emph{Forum for social economics}},
  Vol.~\bibinfo{volume}{44}. Taylor And Francis, \bibinfo{pages}{48--68}.
\newblock


\bibitem[Grimme et~al\mbox{.}(2018)]%
        {grimme2018changing}
\bibfield{author}{\bibinfo{person}{Christian Grimme}, \bibinfo{person}{Dennis
  Assenmacher}, {and} \bibinfo{person}{Lena Adam}.}
  \bibinfo{year}{2018}\natexlab{}.
\newblock \showarticletitle{Changing perspectives: Is it sufficient to detect
  social bots?}. In \bibinfo{booktitle}{\emph{International conference on
  social computing and social media}}. Springer, \bibinfo{pages}{445--461}.
\newblock


\bibitem[Gross(2013a)]%
        {gross2013captcha}
\bibfield{author}{\bibinfo{person}{John~Nicholas Gross}.}
  \bibinfo{year}{2013}\natexlab{a}.
\newblock \bibinfo{title}{Captcha Using Challenges Optimized for distinguishing
  between humans and machines}.
\newblock
\newblock
\newblock
\shownote{US Patent 8,494,854}.


\bibitem[Gross(2013b)]%
        {gross2013system}
\bibfield{author}{\bibinfo{person}{John~Nicholas Gross}.}
  \bibinfo{year}{2013}\natexlab{b}.
\newblock \bibinfo{title}{System and method for generating challenge items for
  CAPTCHAs}.
\newblock
\newblock
\newblock
\shownote{US Patent 8,380,503}.


\bibitem[Guicciardi(2022)]%
        {guicciardi2022scalability}
\bibfield{author}{\bibinfo{person}{Piero Guicciardi}.}
  \bibinfo{year}{2022}\natexlab{}.
\newblock \emph{\bibinfo{title}{Scalability of Encointer-a Proof-Of-Personhood
  Cryptocurrency}}.
\newblock \bibinfo{thesistype}{Master's\ thesis}. \bibinfo{school}{ETH Zurich}.
\newblock


\bibitem[Gupta and Pabian(1997)]%
        {gupta1997investigating}
\bibfield{author}{\bibinfo{person}{Vipin Gupta} {and} \bibinfo{person}{Frank
  Pabian}.} \bibinfo{year}{1997}\natexlab{}.
\newblock \showarticletitle{Investigating the allegations of Indian nuclear
  test preparations in the Rajasthan desert: A CTB Verification Exercise Using
  Commercial Satellite Imagery}.
\newblock \bibinfo{journal}{\emph{Science \& Global Security}}
  \bibinfo{volume}{6}, \bibinfo{number}{2} (\bibinfo{year}{1997}),
  \bibinfo{pages}{101--188}.
\newblock


\bibitem[Gupta and Pabian(1998)]%
        {gupta1998commercial}
\bibfield{author}{\bibinfo{person}{Vipin Gupta} {and} \bibinfo{person}{Frank
  Pabian}.} \bibinfo{year}{1998}\natexlab{}.
\newblock \showarticletitle{Commercial satellite imagery and the CTBT
  verification process}.
\newblock \bibinfo{journal}{\emph{The Nonproliferation Review}}
  \bibinfo{volume}{5}, \bibinfo{number}{3} (\bibinfo{year}{1998}),
  \bibinfo{pages}{89--97}.
\newblock


\bibitem[Haines et~al\mbox{.}(2020)]%
        {haines2020not}
\bibfield{author}{\bibinfo{person}{Thomas Haines}, \bibinfo{person}{Sarah~Jamie
  Lewis}, \bibinfo{person}{Olivier Pereira}, {and} \bibinfo{person}{Vanessa
  Teague}.} \bibinfo{year}{2020}\natexlab{}.
\newblock \showarticletitle{How not to prove your election outcome}. In
  \bibinfo{booktitle}{\emph{2020 IEEE Symposium on Security and Privacy (SP)}}.
  IEEE, \bibinfo{pages}{644--660}.
\newblock


\bibitem[Hoyos(2016)]%
        {hoyos2016system}
\bibfield{author}{\bibinfo{person}{Hector Hoyos}.}
  \bibinfo{year}{2016}\natexlab{}.
\newblock \bibinfo{title}{System and method for determining liveness}.
\newblock
\newblock
\newblock
\shownote{US Patent 9,313,200}.


\bibitem[Huseynov and Seigneur(2019)]%
        {huseynov2019physical}
\bibfield{author}{\bibinfo{person}{Emin Huseynov} {and}
  \bibinfo{person}{Jean-Marc Seigneur}.} \bibinfo{year}{2019}\natexlab{}.
\newblock \showarticletitle{Physical presence verification using TOTP and QR
  codes}. In \bibinfo{booktitle}{\emph{34th International Conference on ICT
  Systems Security and Privacy Protection-IFIP SEC 2019}}.
\newblock


\bibitem[Iliou et~al\mbox{.}(2021)]%
        {iliou2021web}
\bibfield{author}{\bibinfo{person}{Christos Iliou}, \bibinfo{person}{Theodoros
  Kostoulas}, \bibinfo{person}{Theodora Tsikrika}, \bibinfo{person}{Vasilis
  Katos}, \bibinfo{person}{Stefanos Vrochidis}, {and} \bibinfo{person}{Ioannis
  Kompatsiaris}.} \bibinfo{year}{2021}\natexlab{}.
\newblock \showarticletitle{Web bot detection evasion using generative
  adversarial networks}. In \bibinfo{booktitle}{\emph{2021 IEEE International
  Conference on Cyber Security and Resilience (CSR)}}. IEEE,
  \bibinfo{pages}{115--120}.
\newblock


\bibitem[Ionita(2016)]%
        {ionita2016methods}
\bibfield{author}{\bibinfo{person}{Mircea Ionita}.}
  \bibinfo{year}{2016}\natexlab{}.
\newblock \bibinfo{title}{Methods and systems for determining user liveness}.
\newblock
\newblock
\newblock
\shownote{US Patent 9,305,225}.


\bibitem[Jefferson et~al\mbox{.}(2004)]%
        {jefferson2004analyzing}
\bibfield{author}{\bibinfo{person}{David Jefferson}, \bibinfo{person}{Aviel~D
  Rubin}, \bibinfo{person}{Barbara Simons}, {and} \bibinfo{person}{David
  Wagner}.} \bibinfo{year}{2004}\natexlab{}.
\newblock \showarticletitle{Analyzing internet voting security}.
\newblock \bibinfo{journal}{\emph{Commun. ACM}} \bibinfo{volume}{47},
  \bibinfo{number}{10} (\bibinfo{year}{2004}), \bibinfo{pages}{59--64}.
\newblock


\bibitem[Jia et~al\mbox{.}(2019)]%
        {jia2019spoofing}
\bibfield{author}{\bibinfo{person}{Shan Jia}, \bibinfo{person}{Xin Li},
  \bibinfo{person}{Chuanbo Hu}, {and} \bibinfo{person}{Zhengquan Xu}.}
  \bibinfo{year}{2019}\natexlab{}.
\newblock \showarticletitle{Spoofing and Anti-Spoofing with Wax Figure Faces}.
\newblock \bibinfo{journal}{\emph{arXiv preprint arXiv:1910.05457}}
  (\bibinfo{year}{2019}).
\newblock


\bibitem[Kleberger et~al\mbox{.}(2011)]%
        {kleberger2011security}
\bibfield{author}{\bibinfo{person}{Pierre Kleberger}, \bibinfo{person}{Tomas
  Olovsson}, {and} \bibinfo{person}{Erland Jonsson}.}
  \bibinfo{year}{2011}\natexlab{}.
\newblock \showarticletitle{Security aspects of the in-vehicle network in the
  connected car}. In \bibinfo{booktitle}{\emph{2011 IEEE Intelligent Vehicles
  Symposium (IV)}}. IEEE, \bibinfo{pages}{528--533}.
\newblock


\bibitem[Kollreider et~al\mbox{.}(2007)]%
        {kollreider2007real}
\bibfield{author}{\bibinfo{person}{Klaus Kollreider}, \bibinfo{person}{Hartwig
  Fronthaler}, \bibinfo{person}{Maycel~Isaac Faraj}, {and}
  \bibinfo{person}{Josef Bigun}.} \bibinfo{year}{2007}\natexlab{}.
\newblock \showarticletitle{Real-time face detection and motion analysis with
  application in ``liveness'' assessment}.
\newblock \bibinfo{journal}{\emph{IEEE Transactions on Information Forensics
  and Security}} \bibinfo{volume}{2}, \bibinfo{number}{3}
  (\bibinfo{year}{2007}), \bibinfo{pages}{548--558}.
\newblock


\bibitem[Koved(2015)]%
        {koved2015usable}
\bibfield{author}{\bibinfo{person}{Larry Koved}.}
  \bibinfo{year}{2015}\natexlab{}.
\newblock \bibinfo{booktitle}{\emph{Usable multi-factor authentication and
  risk-based authorization}}.
\newblock \bibinfo{type}{{T}echnical {R}eport}.
  \bibinfo{institution}{INTERNATIONAL BUSINESS MACHINES CORP YORKTOWN HEIGHTS
  NY}.
\newblock


\bibitem[Kr{\"a}henb{\"u}hl et~al\mbox{.}(2022)]%
        {krahenbuhl2022swiss}
\bibfield{author}{\bibinfo{person}{Cyrill Kr{\"a}henb{\"u}hl},
  \bibinfo{person}{Marc Wyss}, \bibinfo{person}{Robin Burkhard},
  \bibinfo{person}{Joel Wanner}, {and} \bibinfo{person}{Adrian Perrig}.}
  \bibinfo{year}{2022}\natexlab{}.
\newblock \showarticletitle{Swiss Post E-Voting Scope 4: Network Security
  Analysis}.
\newblock  (\bibinfo{year}{2022}).
\newblock


\bibitem[Kudugunta and Ferrara(2018)]%
        {kudugunta2018deep}
\bibfield{author}{\bibinfo{person}{Sneha Kudugunta} {and}
  \bibinfo{person}{Emilio Ferrara}.} \bibinfo{year}{2018}\natexlab{}.
\newblock \showarticletitle{Deep neural networks for bot detection}.
\newblock \bibinfo{journal}{\emph{Information Sciences}}  \bibinfo{volume}{467}
  (\bibinfo{year}{2018}), \bibinfo{pages}{312--322}.
\newblock


\bibitem[Lamport et~al\mbox{.}(1982)]%
        {LSP82}
\bibfield{author}{\bibinfo{person}{Leslie Lamport}, \bibinfo{person}{Robert
  Shostak}, {and} \bibinfo{person}{Marshall Pease}.}
  \bibinfo{year}{1982}\natexlab{}.
\newblock \showarticletitle{The Byzantine Generals Problem}.
\newblock \bibinfo{journal}{\emph{ACM Transactions on Programming Languages and
  Systems}} \bibinfo{volume}{4}, \bibinfo{number}{3} (\bibinfo{date}{July}
  \bibinfo{year}{1982}), \bibinfo{pages}{382--401}.
\newblock


\bibitem[Landemore(2020)]%
        {landemore2020open}
\bibfield{author}{\bibinfo{person}{H{\'e}l{\`e}ne Landemore}.}
  \bibinfo{year}{2020}\natexlab{}.
\newblock \bibinfo{booktitle}{\emph{Open democracy: Reinventing popular rule
  for the twenty-first century}}.
\newblock \bibinfo{publisher}{Princeton University Press}.
\newblock


\bibitem[Landemore(2021)]%
        {landemore2021open}
\bibfield{author}{\bibinfo{person}{H{\'e}l{\`e}ne Landemore}.}
  \bibinfo{year}{2021}\natexlab{}.
\newblock \showarticletitle{Open democracy and digital technologies}.
\newblock \bibinfo{journal}{\emph{Digital technology and democratic theory}}
  (\bibinfo{year}{2021}), \bibinfo{pages}{62--89}.
\newblock


\bibitem[Lewis et~al\mbox{.}(2019)]%
        {lewis2019not}
\bibfield{author}{\bibinfo{person}{Sarah~Jamie Lewis}, \bibinfo{person}{Olivier
  Pereira}, {and} \bibinfo{person}{Vanessa Teague}.}
  \bibinfo{year}{2019}\natexlab{}.
\newblock \bibinfo{booktitle}{\emph{How not to prove your election outcome}}.
\newblock \bibinfo{type}{{T}echnical {R}eport}. \bibinfo{institution}{Technical
  Report, March}.
\newblock


\bibitem[Li et~al\mbox{.}(2017)]%
        {li2017deep}
\bibfield{author}{\bibinfo{person}{Chao Li}, \bibinfo{person}{Xiaokong Ma},
  \bibinfo{person}{Bing Jiang}, \bibinfo{person}{Xiangang Li},
  \bibinfo{person}{Xuewei Zhang}, \bibinfo{person}{Xiao Liu},
  \bibinfo{person}{Ying Cao}, \bibinfo{person}{Ajay Kannan}, {and}
  \bibinfo{person}{Zhenyao Zhu}.} \bibinfo{year}{2017}\natexlab{}.
\newblock \showarticletitle{Deep speaker: an end-to-end neural speaker
  embedding system}.
\newblock \bibinfo{journal}{\emph{arXiv preprint arXiv:1705.02304}}
  (\bibinfo{year}{2017}).
\newblock


\bibitem[Li et~al\mbox{.}(n.\,d.)]%
        {lilorenzo}
\bibfield{author}{\bibinfo{person}{Harry~C Li}, \bibinfo{person}{Allen
  Clement}, \bibinfo{person}{Edmund~L Wong}, \bibinfo{person}{Jeff Napper},
  {and} \bibinfo{person}{Indrajit Roy}.} \bibinfo{year}{[n.\,d.]}\natexlab{}.
\newblock \showarticletitle{Lorenzo Alvisi, Michael Dahlin Laboratory for
  Advanced Systems Research (LASR), Dept. of Computer Sciences, The University
  of Texas at Austin}.
\newblock  (\bibinfo{year}{[n.\,d.]}).
\newblock


\bibitem[Li et~al\mbox{.}(2016a)]%
        {li2016persona}
\bibfield{author}{\bibinfo{person}{Jiwei Li}, \bibinfo{person}{Michel Galley},
  \bibinfo{person}{Chris Brockett}, \bibinfo{person}{Georgios~P Spithourakis},
  \bibinfo{person}{Jianfeng Gao}, {and} \bibinfo{person}{Bill Dolan}.}
  \bibinfo{year}{2016}\natexlab{a}.
\newblock \showarticletitle{A persona-based neural conversation model}.
\newblock \bibinfo{journal}{\emph{arXiv preprint arXiv:1603.06155}}
  (\bibinfo{year}{2016}).
\newblock


\bibitem[Li et~al\mbox{.}(2016b)]%
        {li2016deep}
\bibfield{author}{\bibinfo{person}{Jiwei Li}, \bibinfo{person}{Will Monroe},
  \bibinfo{person}{Alan Ritter}, \bibinfo{person}{Michel Galley},
  \bibinfo{person}{Jianfeng Gao}, {and} \bibinfo{person}{Dan Jurafsky}.}
  \bibinfo{year}{2016}\natexlab{b}.
\newblock \showarticletitle{Deep reinforcement learning for dialogue
  generation}.
\newblock \bibinfo{journal}{\emph{arXiv preprint arXiv:1606.01541}}
  (\bibinfo{year}{2016}).
\newblock


\bibitem[Lysyanskaya and Ramzan(1998)]%
        {lysyanskaya1998group}
\bibfield{author}{\bibinfo{person}{Anna Lysyanskaya} {and}
  \bibinfo{person}{Zulfikar Ramzan}.} \bibinfo{year}{1998}\natexlab{}.
\newblock \showarticletitle{Group blind digital signatures: A scalable solution
  to electronic cash}. In \bibinfo{booktitle}{\emph{International Conference on
  Financial Cryptography}}. Springer, \bibinfo{pages}{184--197}.
\newblock


\bibitem[Mbona and Eloff(2022)]%
        {mbona2022feature}
\bibfield{author}{\bibinfo{person}{Innocent Mbona} {and}
  \bibinfo{person}{Jan~HP Eloff}.} \bibinfo{year}{2022}\natexlab{}.
\newblock \showarticletitle{Feature selection using Benford's law to support
  detection of malicious social media bots}.
\newblock \bibinfo{journal}{\emph{Information Sciences}}  \bibinfo{volume}{582}
  (\bibinfo{year}{2022}), \bibinfo{pages}{369--381}.
\newblock


\bibitem[Merino et~al\mbox{.}(2022)]%
        {merino2022trip}
\bibfield{author}{\bibinfo{person}{Louis-Henri Merino}, \bibinfo{person}{Simone
  Colombo}, \bibinfo{person}{Jeff Allen}, \bibinfo{person}{Vero
  Estrada-Gali{\~n}anes}, {and} \bibinfo{person}{Bryan Ford}.}
  \bibinfo{year}{2022}\natexlab{}.
\newblock \showarticletitle{TRIP: Trustless Coercion-Resistant In-Person Voter
  Registration}.
\newblock \bibinfo{journal}{\emph{arXiv preprint arXiv:2202.06692}}
  (\bibinfo{year}{2022}).
\newblock


\bibitem[Moore(2023)]%
        {moore2023fake}
\bibfield{author}{\bibinfo{person}{Martin Moore}.}
  \bibinfo{year}{2023}\natexlab{}.
\newblock \showarticletitle{Fake accounts on social media, epistemic
  uncertainty and the need for an independent auditing of accounts}.
\newblock \bibinfo{journal}{\emph{Internet Policy Review}}
  \bibinfo{volume}{12}, \bibinfo{number}{1} (\bibinfo{year}{2023}).
\newblock


\bibitem[Morgan and Parsovs(2017)]%
        {morgan2017using}
\bibfield{author}{\bibinfo{person}{Danielle Morgan} {and}
  \bibinfo{person}{Arnis Parsovs}.} \bibinfo{year}{2017}\natexlab{}.
\newblock \showarticletitle{Using the Estonian Electronic Identity Card for
  Authentication to a Machine (Extended Version)}.
\newblock \bibinfo{journal}{\emph{Cryptology ePrint Archive}}
  (\bibinfo{year}{2017}).
\newblock


\bibitem[Motorevska et~al\mbox{.}(2019)]%
        {Motorevska2019}
\bibfield{author}{\bibinfo{person}{Yevheniia Motorevska},
  \bibinfo{person}{Dmytr Replianchuk}, {and} \bibinfo{person}{Vasyk Bidun}.}
  \bibinfo{year}{2019}\natexlab{}.
\newblock \bibinfo{title}{Inside a Ukrainian Troll Farm}.
\newblock
\newblock
\newblock
\shownote{\url{https://www.occrp.org/en/investigations/inside-a-ukrainian-troll-farm}}.


\bibitem[Nosouhi et~al\mbox{.}(2020)]%
        {nosouhi2020blockchain}
\bibfield{author}{\bibinfo{person}{Mohammad~Reza Nosouhi},
  \bibinfo{person}{Shui Yu}, \bibinfo{person}{Wanlei Zhou},
  \bibinfo{person}{Marthie Grobler}, {and} \bibinfo{person}{Habiba Keshtiar}.}
  \bibinfo{year}{2020}\natexlab{}.
\newblock \showarticletitle{Blockchain for secure location verification}.
\newblock \bibinfo{journal}{\emph{J. Parallel and Distrib. Comput.}}
  \bibinfo{volume}{136} (\bibinfo{year}{2020}), \bibinfo{pages}{40--51}.
\newblock


\bibitem[Novosti(2012)]%
        {domnovosti2012}
\bibfield{author}{\bibinfo{person}{RIA Novosti}.}
  \bibinfo{year}{2012}\natexlab{}.
\newblock \bibinfo{title}{Russia's Snow Revolutionaries Ponder Next Move}.
\newblock
\newblock
\urldef\tempurl%
\url{https://sputniknews.com/20120206/171180369.html}
\showURL{%
\tempurl}


\bibitem[Ometov et~al\mbox{.}(2018)]%
        {ometov2018multi}
\bibfield{author}{\bibinfo{person}{Aleksandr Ometov}, \bibinfo{person}{Sergey
  Bezzateev}, \bibinfo{person}{Niko M{\"a}kitalo}, \bibinfo{person}{Sergey
  Andreev}, \bibinfo{person}{Tommi Mikkonen}, {and} \bibinfo{person}{Yevgeni
  Koucheryavy}.} \bibinfo{year}{2018}\natexlab{}.
\newblock \showarticletitle{Multi-factor authentication: A survey}.
\newblock \bibinfo{journal}{\emph{Cryptography}} \bibinfo{volume}{2},
  \bibinfo{number}{1} (\bibinfo{year}{2018}), \bibinfo{pages}{1}.
\newblock


\bibitem[Osborne(2011)]%
        {osborne2011}
\bibfield{author}{\bibinfo{person}{Andrew Osborne}.}
  \bibinfo{year}{2011}\natexlab{}.
\newblock \bibinfo{title}{Bloggers who are changing the face of Russia as the
  Snow Revolution takes hold}.
\newblock
\newblock
\urldef\tempurl%
\url{https://bit.ly/3E4bXyU}
\showURL{%
\tempurl}


\bibitem[Platt and McBurney(2021)]%
        {platt2021sybil}
\bibfield{author}{\bibinfo{person}{Moritz Platt} {and} \bibinfo{person}{Peter
  McBurney}.} \bibinfo{year}{2021}\natexlab{}.
\newblock \showarticletitle{Sybil attacks on identity-augmented
  Proof-of-Stake}.
\newblock \bibinfo{journal}{\emph{Computer Networks}}  \bibinfo{volume}{199}
  (\bibinfo{year}{2021}), \bibinfo{pages}{108424}.
\newblock


\bibitem[Rahman et~al\mbox{.}(2015)]%
        {rahman2015movee}
\bibfield{author}{\bibinfo{person}{Mahmudur Rahman}, \bibinfo{person}{Umut
  Topkara}, {and} \bibinfo{person}{Bogdan Carbunar}.}
  \bibinfo{year}{2015}\natexlab{}.
\newblock \showarticletitle{Movee: Video liveness verification for mobile
  devices using built-in motion sensors}.
\newblock \bibinfo{journal}{\emph{IEEE Transactions on Mobile Computing}}
  \bibinfo{volume}{15}, \bibinfo{number}{5} (\bibinfo{year}{2015}),
  \bibinfo{pages}{1197--1210}.
\newblock


\bibitem[Ramatsakane and Leung(2017)]%
        {ramatsakane2017pick}
\bibfield{author}{\bibinfo{person}{Kobosa~Icconies Ramatsakane} {and}
  \bibinfo{person}{Wai~Sze Leung}.} \bibinfo{year}{2017}\natexlab{}.
\newblock \showarticletitle{Pick location security: Seamless integrated
  multi-factor authentication}. In \bibinfo{booktitle}{\emph{2017 IST-Africa
  Week Conference (IST-Africa)}}. IEEE, \bibinfo{pages}{1--10}.
\newblock


\bibitem[Ruan and Zou(2017)]%
        {ruan2017receipt}
\bibfield{author}{\bibinfo{person}{Yefeng Ruan} {and} \bibinfo{person}{Xukai
  Zou}.} \bibinfo{year}{2017}\natexlab{}.
\newblock \showarticletitle{Receipt-freeness and coercion resistance in remote
  E-voting systems}.
\newblock  (\bibinfo{year}{2017}).
\newblock


\bibitem[S{\'a}nchez(2019)]%
        {sanchez2019zero}
\bibfield{author}{\bibinfo{person}{David~Cerezo S{\'a}nchez}.}
  \bibinfo{year}{2019}\natexlab{}.
\newblock \showarticletitle{Zero-knowledge proof-of-identity: Sybil-resistant,
  anonymous authentication on permissionless blockchains and incentive
  compatible, strictly dominant cryptocurrencies}.
\newblock \bibinfo{journal}{\emph{arXiv preprint arXiv:1905.09093}}
  (\bibinfo{year}{2019}).
\newblock


\bibitem[Sastry et~al\mbox{.}(2003)]%
        {sastry2003secure}
\bibfield{author}{\bibinfo{person}{Naveen Sastry}, \bibinfo{person}{Umesh
  Shankar}, {and} \bibinfo{person}{David Wagner}.}
  \bibinfo{year}{2003}\natexlab{}.
\newblock \showarticletitle{Secure verification of location claims}. In
  \bibinfo{booktitle}{\emph{Proceedings of the 2nd ACM workshop on Wireless
  security}}. \bibinfo{pages}{1--10}.
\newblock


\bibitem[Sayyadiharikandeh et~al\mbox{.}(2020)]%
        {sayyadiharikandeh2020detection}
\bibfield{author}{\bibinfo{person}{Mohsen Sayyadiharikandeh},
  \bibinfo{person}{Onur Varol}, \bibinfo{person}{Kai-Cheng Yang},
  \bibinfo{person}{Alessandro Flammini}, {and} \bibinfo{person}{Filippo
  Menczer}.} \bibinfo{year}{2020}\natexlab{}.
\newblock \showarticletitle{Detection of novel social bots by ensembles of
  specialized classifiers}. In \bibinfo{booktitle}{\emph{Proceedings of the
  29th ACM international conference on information \& knowledge management}}.
  \bibinfo{pages}{2725--2732}.
\newblock


\bibitem[Schmid(2021)]%
        {schmid2021thirty}
\bibfield{author}{\bibinfo{person}{Giovanni Schmid}.}
  \bibinfo{year}{2021}\natexlab{}.
\newblock \showarticletitle{Thirty years of DNS insecurity: Current issues and
  perspectives}.
\newblock \bibinfo{journal}{\emph{IEEE Communications Surveys \& Tutorials}}
  \bibinfo{volume}{23}, \bibinfo{number}{4} (\bibinfo{year}{2021}),
  \bibinfo{pages}{2429--2459}.
\newblock


\bibitem[Scigliano(2021)]%
        {scigliano_2021}
\bibfield{author}{\bibinfo{person}{Eric Scigliano}.}
  \bibinfo{year}{2021}\natexlab{}.
\newblock \bibinfo{title}{Zoom Court Is Changing How Justice Is Served}.
\newblock
\newblock
\urldef\tempurl%
\url{https://www.theatlantic.com/magazine/archive/2021/05/can-justice-be-served-on-zoom/618392/}
\showURL{%
\tempurl}


\bibitem[Shu(2020)]%
        {shu2020}
\bibfield{author}{\bibinfo{person}{Catherine Shu}.}
  \bibinfo{year}{2020}\natexlab{}.
\newblock \bibinfo{title}{Why the world must pay attention to the fight against
  disinformation and fake news in Taiwan}.
\newblock
\newblock
\newblock
\shownote{\url{https://tcrn.ch/31IiScd}}.


\bibitem[Siddarth et~al\mbox{.}(2020)]%
        {siddarth2020watches}
\bibfield{author}{\bibinfo{person}{Divya Siddarth}, \bibinfo{person}{Sergey
  Ivliev}, \bibinfo{person}{Santiago Siri}, {and} \bibinfo{person}{Paula
  Berman}.} \bibinfo{year}{2020}\natexlab{}.
\newblock \showarticletitle{Who watches the watchmen? a review of subjective
  approaches for sybil-resistance in proof of personhood protocols}.
\newblock \bibinfo{journal}{\emph{Frontiers in Blockchain}}
  (\bibinfo{year}{2020}), \bibinfo{pages}{46}.
\newblock


\bibitem[Snyder et~al\mbox{.}(2018)]%
        {snyder2018x}
\bibfield{author}{\bibinfo{person}{David Snyder}, \bibinfo{person}{Daniel
  Garcia-Romero}, \bibinfo{person}{Gregory Sell}, \bibinfo{person}{Daniel
  Povey}, {and} \bibinfo{person}{Sanjeev Khudanpur}.}
  \bibinfo{year}{2018}\natexlab{}.
\newblock \showarticletitle{X-vectors: Robust dnn embeddings for speaker
  recognition}. In \bibinfo{booktitle}{\emph{2018 IEEE International Conference
  on Acoustics, Speech and Signal Processing (ICASSP)}}. IEEE,
  \bibinfo{pages}{5329--5333}.
\newblock


\bibitem[Sordoni et~al\mbox{.}(2015)]%
        {sordoni2015neural}
\bibfield{author}{\bibinfo{person}{Alessandro Sordoni}, \bibinfo{person}{Michel
  Galley}, \bibinfo{person}{Michael Auli}, \bibinfo{person}{Chris Brockett},
  \bibinfo{person}{Yangfeng Ji}, \bibinfo{person}{Margaret Mitchell},
  \bibinfo{person}{Jian-Yun Nie}, \bibinfo{person}{Jianfeng Gao}, {and}
  \bibinfo{person}{Bill Dolan}.} \bibinfo{year}{2015}\natexlab{}.
\newblock \showarticletitle{A neural network approach to context-sensitive
  generation of conversational responses}.
\newblock \bibinfo{journal}{\emph{arXiv preprint arXiv:1506.06714}}
  (\bibinfo{year}{2015}).
\newblock


\bibitem[Specter et~al\mbox{.}(2020)]%
        {specter2020ballot}
\bibfield{author}{\bibinfo{person}{Michael~A Specter}, \bibinfo{person}{James
  Koppel}, {and} \bibinfo{person}{Daniel Weitzner}.}
  \bibinfo{year}{2020}\natexlab{}.
\newblock \showarticletitle{The ballot is busted before the blockchain: A
  security analysis of voatz, the first internet voting application used in us
  federal elections}. In \bibinfo{booktitle}{\emph{29th $\{$USENIX\$\}\$
  Security Symposium (\$\{\$USENIX\$\}\$ Security 20)}}.
  \bibinfo{pages}{1535--1553}.
\newblock


\bibitem[Springall et~al\mbox{.}(2014)]%
        {springall2014security}
\bibfield{author}{\bibinfo{person}{Drew Springall}, \bibinfo{person}{Travis
  Finkenauer}, \bibinfo{person}{Zakir Durumeric}, \bibinfo{person}{Jason
  Kitcat}, \bibinfo{person}{Harri Hursti}, \bibinfo{person}{Margaret
  MacAlpine}, {and} \bibinfo{person}{J.~Alex Halderman}.}
  \bibinfo{year}{2014}\natexlab{}.
\newblock \showarticletitle{Security Analysis of the Estonian Internet Voting
  System}. In \bibinfo{booktitle}{\emph{Proceedings of the 2014 ACM SIGSAC
  Conference on Computer and Communications Security}} (Scottsdale, Arizona,
  USA) \emph{(\bibinfo{series}{CCS '14})}. \bibinfo{publisher}{Association for
  Computing Machinery}, \bibinfo{address}{New York, NY, USA},
  \bibinfo{pages}{703–715}.
\newblock
\showISBNx{9781450329576}
\urldef\tempurl%
\url{https://doi.org/10.1145/2660267.2660315}
\showDOI{\tempurl}


\bibitem[Sterret et~al\mbox{.}(2018)]%
        {sterret2018shared}
\bibfield{author}{\bibinfo{person}{D Sterret}, \bibinfo{person}{Dan Malato},
  \bibinfo{person}{Jennifer Benz}, \bibinfo{person}{Liz Kantor},
  \bibinfo{person}{Trevor Tompson}, \bibinfo{person}{Tom Rosenstiel},
  \bibinfo{person}{Jeff Sonderman}, \bibinfo{person}{Kevin Loker}, {and}
  \bibinfo{person}{Emily Swanson}.} \bibinfo{year}{2018}\natexlab{}.
\newblock \bibinfo{title}{Who shared it?: How Americans decide what news to
  trust on social media}.
\newblock
\newblock


\bibitem[Stilgherrian(2020)]%
        {stilgherrian2020}
\bibfield{author}{\bibinfo{person}{Stilgherrian}.}
  \bibinfo{year}{2020}\natexlab{}.
\newblock \bibinfo{title}{Twitter bots and trolls promote conspiracy theories
  about Australian bushfires}.
\newblock
\newblock
\newblock
\shownote{\url{https://zd.net/2Ux2jhV}}.


\bibitem[Talasila et~al\mbox{.}(2010)]%
        {talasila2010link}
\bibfield{author}{\bibinfo{person}{Manoop Talasila}, \bibinfo{person}{Reza
  Curtmola}, {and} \bibinfo{person}{Cristian Borcea}.}
  \bibinfo{year}{2010}\natexlab{}.
\newblock \showarticletitle{Link: Location verification through immediate
  neighbors knowledge}. In \bibinfo{booktitle}{\emph{International Conference
  on Mobile and Ubiquitous Systems: Computing, Networking, and Services}}.
  Springer, \bibinfo{pages}{210--223}.
\newblock


\bibitem[Timberg and Romm(2019)]%
        {timberg_romm_2019}
\bibfield{author}{\bibinfo{person}{Craig Timberg} {and} \bibinfo{person}{Tony
  Romm}.} \bibinfo{year}{2019}\natexlab{}.
\newblock \bibinfo{title}{Russian trolls sought to inflame debate over climate
  change, fracking, Dakota pipeline}.
\newblock
\newblock
\urldef\tempurl%
\url{https://www.chicagotribune.com/nation-world/ct-russian-trolls-climate-change-20180301-story.html}
\showURL{%
\tempurl}


\bibitem[Timmer(2020)]%
        {timmer_2020}
\bibfield{author}{\bibinfo{person}{John Timmer}.}
  \bibinfo{year}{2020}\natexlab{}.
\newblock \bibinfo{title}{Study looks at how Russian troll farms are
  politicizing vaccines}.
\newblock
\newblock
\urldef\tempurl%
\url{https://arstechnica.com/science/2020/04/study-looks-at-how-russian-troll-farms-are-politicizing-vaccines/}
\showURL{%
\tempurl}


\bibitem[Uzun et~al\mbox{.}(2018)]%
        {uzun2018rtcaptcha}
\bibfield{author}{\bibinfo{person}{Erkam Uzun}, \bibinfo{person}{Simon Pak~Ho
  Chung}, \bibinfo{person}{Irfan Essa}, {and} \bibinfo{person}{Wenke Lee}.}
  \bibinfo{year}{2018}\natexlab{}.
\newblock \showarticletitle{rtCaptcha: A Real-Time CAPTCHA Based Liveness
  Detection System.}. In \bibinfo{booktitle}{\emph{NDSS}}.
\newblock


\bibitem[Vaas(2018)]%
        {vaas2018}
\bibfield{author}{\bibinfo{person}{Lisa Vaas}.}
  \bibinfo{year}{2018}\natexlab{}.
\newblock \bibinfo{title}{Twitter publishes data on Iranian and Russian troll
  farms}.
\newblock
\newblock
\newblock
\shownote{\url{https://bit.ly/2Mgfaja}}.


\bibitem[Varol et~al\mbox{.}(2018)]%
        {varol2018feature}
\bibfield{author}{\bibinfo{person}{Onur Varol}, \bibinfo{person}{Clayton~A
  Davis}, \bibinfo{person}{Filippo Menczer}, {and} \bibinfo{person}{Alessandro
  Flammini}.} \bibinfo{year}{2018}\natexlab{}.
\newblock \showarticletitle{Feature engineering for social bot detection}.
\newblock In \bibinfo{booktitle}{\emph{Feature engineering for machine learning
  and data analytics}}. \bibinfo{publisher}{CRC Press},
  \bibinfo{pages}{311--334}.
\newblock


\bibitem[Varol et~al\mbox{.}(2017)]%
        {varol2017online}
\bibfield{author}{\bibinfo{person}{Onur Varol}, \bibinfo{person}{Emilio
  Ferrara}, \bibinfo{person}{Clayton Davis}, \bibinfo{person}{Filippo Menczer},
  {and} \bibinfo{person}{Alessandro Flammini}.}
  \bibinfo{year}{2017}\natexlab{}.
\newblock \showarticletitle{Online human-bot interactions: Detection,
  estimation, and characterization}. In \bibinfo{booktitle}{\emph{Proceedings
  of the international AAAI conference on web and social media}},
  Vol.~\bibinfo{volume}{11}. \bibinfo{pages}{280--289}.
\newblock


\bibitem[Viswanath et~al\mbox{.}(2010)]%
        {viswanath2010analysis}
\bibfield{author}{\bibinfo{person}{Bimal Viswanath}, \bibinfo{person}{Ansley
  Post}, \bibinfo{person}{Krishna~P Gummadi}, {and} \bibinfo{person}{Alan
  Mislove}.} \bibinfo{year}{2010}\natexlab{}.
\newblock \showarticletitle{An analysis of social network-based sybil
  defenses}.
\newblock \bibinfo{journal}{\emph{ACM SIGCOMM Computer Communication Review}}
  \bibinfo{volume}{40}, \bibinfo{number}{4} (\bibinfo{year}{2010}),
  \bibinfo{pages}{363--374}.
\newblock


\bibitem[Wang et~al\mbox{.}(2011)]%
        {wang2011towards}
\bibfield{author}{\bibinfo{person}{Yong Wang}, \bibinfo{person}{Daniel
  Burgener}, \bibinfo{person}{Marcel Flores}, \bibinfo{person}{Aleksandar
  Kuzmanovic}, {and} \bibinfo{person}{Cheng Huang}.}
  \bibinfo{year}{2011}\natexlab{}.
\newblock \showarticletitle{Towards Street-Level Client-Independent IP
  Geolocation.}. In \bibinfo{booktitle}{\emph{NSDI}},
  Vol.~\bibinfo{volume}{11}. \bibinfo{pages}{27--27}.
\newblock


\bibitem[Wu et~al\mbox{.}(2020)]%
        {wu2020using}
\bibfield{author}{\bibinfo{person}{Bin Wu}, \bibinfo{person}{Le Liu},
  \bibinfo{person}{Yanqing Yang}, \bibinfo{person}{Kangfeng Zheng}, {and}
  \bibinfo{person}{Xiujuan Wang}.} \bibinfo{year}{2020}\natexlab{}.
\newblock \showarticletitle{Using improved conditional generative adversarial
  networks to detect social bots on Twitter}.
\newblock \bibinfo{journal}{\emph{IEEE Access}}  \bibinfo{volume}{8}
  (\bibinfo{year}{2020}), \bibinfo{pages}{36664--36680}.
\newblock


\bibitem[Xu et~al\mbox{.}(2016)]%
        {197225}
\bibfield{author}{\bibinfo{person}{Yi Xu}, \bibinfo{person}{True Price},
  \bibinfo{person}{Jan-Michael Frahm}, {and} \bibinfo{person}{Fabian Monrose}.}
  \bibinfo{year}{2016}\natexlab{}.
\newblock \showarticletitle{Virtual U: Defeating Face Liveness Detection by
  Building Virtual Models from Your Public Photos}. In
  \bibinfo{booktitle}{\emph{25th {USENIX} Security Symposium ({USENIX} Security
  16)}}. \bibinfo{publisher}{{USENIX} Association}, \bibinfo{address}{Austin,
  TX}, \bibinfo{pages}{497--512}.
\newblock
\showISBNx{978-1-931971-32-4}
\urldef\tempurl%
\url{https://www.usenix.org/conference/usenixsecurity16/technical-sessions/presentation/xu}
\showURL{%
\tempurl}


\bibitem[Yang et~al\mbox{.}(2022)]%
        {yang2022botometer}
\bibfield{author}{\bibinfo{person}{Kai-Cheng Yang}, \bibinfo{person}{Emilio
  Ferrara}, {and} \bibinfo{person}{Filippo Menczer}.}
  \bibinfo{year}{2022}\natexlab{}.
\newblock \showarticletitle{Botometer 101: Social bot practicum for
  computational social scientists}.
\newblock \bibinfo{journal}{\emph{arXiv preprint arXiv:2201.01608}}
  (\bibinfo{year}{2022}).
\newblock


\bibitem[Yang et~al\mbox{.}(2019)]%
        {yang2019arming}
\bibfield{author}{\bibinfo{person}{Kai-Cheng Yang}, \bibinfo{person}{Onur
  Varol}, \bibinfo{person}{Clayton~A Davis}, \bibinfo{person}{Emilio Ferrara},
  \bibinfo{person}{Alessandro Flammini}, {and} \bibinfo{person}{Filippo
  Menczer}.} \bibinfo{year}{2019}\natexlab{}.
\newblock \showarticletitle{Arming the public with artificial intelligence to
  counter social bots}.
\newblock \bibinfo{journal}{\emph{Human Behavior and Emerging Technologies}}
  \bibinfo{volume}{1}, \bibinfo{number}{1} (\bibinfo{year}{2019}),
  \bibinfo{pages}{48--61}.
\newblock


\bibitem[Yang et~al\mbox{.}(2020)]%
        {yang2020scalable}
\bibfield{author}{\bibinfo{person}{Kai-Cheng Yang}, \bibinfo{person}{Onur
  Varol}, \bibinfo{person}{Pik-Mai Hui}, {and} \bibinfo{person}{Filippo
  Menczer}.} \bibinfo{year}{2020}\natexlab{}.
\newblock \showarticletitle{Scalable and generalizable social bot detection
  through data selection}. In \bibinfo{booktitle}{\emph{Proceedings of the AAAI
  conference on artificial intelligence}}, Vol.~\bibinfo{volume}{34}.
  \bibinfo{pages}{1096--1103}.
\newblock


\bibitem[Yeh et~al\mbox{.}(2020)]%
        {yeh2020using}
\bibfield{author}{\bibinfo{person}{Christopher Yeh}, \bibinfo{person}{Anthony
  Perez}, \bibinfo{person}{Anne Driscoll}, \bibinfo{person}{George Azzari},
  \bibinfo{person}{Zhongyi Tang}, \bibinfo{person}{David Lobell},
  \bibinfo{person}{Stefano Ermon}, {and} \bibinfo{person}{Marshall Burke}.}
  \bibinfo{year}{2020}\natexlab{}.
\newblock \showarticletitle{Using publicly available satellite imagery and deep
  learning to understand economic well-being in Africa}.
\newblock \bibinfo{journal}{\emph{Nature communications}} \bibinfo{volume}{11},
  \bibinfo{number}{1} (\bibinfo{year}{2020}), \bibinfo{pages}{2583}.
\newblock


\bibitem[Young(2020)]%
        {covidbotsYoung}
\bibfield{author}{\bibinfo{person}{Virginia~Alvino Young}.}
  \bibinfo{year}{2020}\natexlab{}.
\newblock \bibinfo{title}{Nearly Half Of The Twitter Accounts Discussing
  `Reopening America' May Be Bots}.
\newblock
\newblock
\newblock
\shownote{\url{https://bit.ly/2WX5j83}}.


\bibitem[Yu et~al\mbox{.}(2008)]%
        {yu2008sybilguard}
\bibfield{author}{\bibinfo{person}{Haifeng Yu}, \bibinfo{person}{Michael
  Kaminsky}, \bibinfo{person}{Phillip~B Gibbons}, {and}
  \bibinfo{person}{Abraham~D Flaxman}.} \bibinfo{year}{2008}\natexlab{}.
\newblock \showarticletitle{Sybilguard: defending against sybil attacks via
  social networks}.
\newblock \bibinfo{journal}{\emph{IEEE/ACM Transactions on networking}}
  \bibinfo{volume}{16}, \bibinfo{number}{3} (\bibinfo{year}{2008}),
  \bibinfo{pages}{576--589}.
\newblock


\bibitem[Zeljkovic et~al\mbox{.}(2016)]%
        {zeljkovic2016system}
\bibfield{author}{\bibinfo{person}{Ilija Zeljkovic}, \bibinfo{person}{Taniya
  Mishra}, \bibinfo{person}{Amanda Stent}, \bibinfo{person}{Ann~K Syrdal},
  {and} \bibinfo{person}{Jay Wilpon}.} \bibinfo{year}{2016}\natexlab{}.
\newblock \bibinfo{title}{System and method for generating challenge utterances
  for speaker verification}.
\newblock
\newblock
\newblock
\shownote{US Patent 9,318,114}.


\end{thebibliography}

%\vspace{12pt}
% \normalsize
% % \vspace{12pt}

\end{document}